\documentclass[iop, twocolappendix, numberedappendix]{emulateapj}
\usepackage{apjfonts}
\usepackage{graphicx}
\usepackage{natbib}
\usepackage{amsmath}
\usepackage{xcolor}
\usepackage[caption=false]{subfig}
\usepackage{subfloat}
\PassOptionsToPackage{hyphens}{url}
\usepackage{hyperref}
\hypersetup{pdfpagemode=UseNone}

\makeatletter

\newcommand{\msun}{M$_\odot$}
\newcommand{\zsun}{$Z_\odot$}
\newcommand{\tmin}{$t_\mathrm{min}^\mathrm{SNIa}$}

\newcommand{\flexce}{\texttt{flexCE}}
\newcommand{\flexceurl}{\url{https://github.com/bretthandrews/flexCE}}

\makeatother
\begin{document}

\title{Inflow, Outflow, Yields, and Stellar Population Mixing in Chemical
Evolution Models}
\author{Brett H.~Andrews\altaffilmark{1, 2, 3},
 David H.~Weinberg\altaffilmark{2, 3},
 Ralph Sch\"onrich\altaffilmark{2, 4},
 Jennifer A.~Johnson\altaffilmark{2, 3}
}

\altaffiltext{1}{PITT PACC, Department of Physics and Astronomy, University of
Pittsburgh, Pittsburgh, PA 15260, andrewsb@pitt.edu}
\altaffiltext{2}{Department of Astronomy, The Ohio State University, 140 West
18th Avenue, Columbus, OH 43210}
\altaffiltext{3}{Center for Cosmology and Astro-Particle Physics, The Ohio
State University, 191 West Woodruff Avenue, Columbus, OH 43210}
\altaffiltext{4}{Rudolf Peierls Centre for Theoretical Physics, 1 Keble Road,
Oxford OX1 3NP, GB}

\begin{abstract}

Chemical evolution models are powerful tools for interpreting stellar abundance
surveys and understanding galaxy evolution.  However, their predictions depend
heavily on the treatment of inflow, outflow, star formation efficiency (SFE),
the stellar initial mass function, the Type Ia supernova delay time
distribution, stellar yields, and stellar population mixing.  Using \flexce, a
flexible one-zone chemical evolution code, we investigate the effects of and
trade-offs between parameters.  Two critical parameters are SFE and the outflow
mass-loading parameter, which shift the knee in [O/Fe]--[Fe/H] and the
equilibrium abundances that the simulations asymptotically approach,
respectively.  One-zone models with simple star formation histories follow
narrow tracks in [O/Fe]--[Fe/H] unlike the observed bimodality (separate
high-$\alpha$ and low-$\alpha$ sequences) in this plane.  A mix of one-zone
models with inflow timescale and outflow mass-loading parameter variations,
motivated by the inside-out galaxy formation scenario with radial mixing,
reproduces the two sequences better than a one-zone model with two infall
epochs.  We present [$X$/Fe]--[Fe/H] tracks for 20 elements assuming three
different supernova yield models and find some significant discrepancies with
solar neighborhood observations, especially for elements with strongly
metallicity-dependent yields.  We apply principal component abundance analysis
(PCAA) to the simulations and existing data to reveal the main correlations
amongst abundances and quantify their contributions to variation in abundance
space.  For the stellar population mixing scenario, the abundances of
$\alpha$-elements and elements with metallicity-dependent yields dominate the
first and second principal components, respectively, and collectively explain
99\% of the variance in the model.  \flexce\ is a python package available at
\flexceurl.

\end{abstract}


\defcitealias{chieffi2004}{CL04}
\defcitealias{woosley1995}{WW95}
\defcitealias{limongi2006}{LC06}
\defcitealias{weinberg2016a}{WAF}


\section{Introduction}

Elemental abundances provide some of the strongest constraints for galaxy
evolution models because they encode information about the entire formation
history of a galaxy.  Galaxy scaling relations, such as the mass--metallicity
relation \citep{tremonti2004, andrews2013} and the mass--metallicity--SFR
relation \citep{mannucci2010, laralopez2010, andrews2013}, are useful for
understanding the evolution of global properties of galaxy populations but are
typically limited to measuring only overall enrichment with coarse spatial
resolution. The Milky Way provides a unique opportunity for studying the full
assembly and enrichment history of one galaxy because we can measure the
multi-element abundances of individual stars born during all epochs of its
formation.  The new generation of stellar abundance surveys, such as APOGEE
\citep{majewski2015}, GALAH \citep{freeman2010}, and Gaia-ESO
\citep{gilmore2012}, represents a significant leap forward in the number
of stars with abundances measured for $>$10 elements.  However, connecting the
abundances from these surveys with galaxy evolution scenarios requires chemical
evolution models. This paper examines the predictions of one-zone (fully mixed)
chemical evolution models for a wide variety of parameter choices and several
assumptions about supernova yields, including predictions for 20 different
elements that will be probed by these large surveys.

Classical chemical evolution models use prescriptions for the physical processes
that drive galaxy formation, like star formation and enrichment, to predict the
evolution of stellar abundances.  They have proven to be important tools for
understanding galaxy evolution ever since the pioneering work of
\citet{schmidt1959, schmidt1963} that introduced the most basic one-zone model
of chemical evolution, an interesting limiting case with an analytic solution
that is known as the ``Simple'' model.  The Simple model assumes a closed system
of initially pristine gas, a constant stellar initial mass function (IMF), and
complete and instantaneous mixing of the gas reservoir. The Simple model can be
further idealized by adopting the instantaneous recycling approximation
\citep{talbot1971, searle1972}, which approximates the yields of stellar
populations by just two classes: massive stars that are assumed to die and free
their yields instantly, and low-mass stars that are assumed to live and lock-up
all their mass forever.  This approximation works well for primary elements,
which are synthesized from hydrogen and helium, but not for secondary elements,
whose yields depend on the initial metal content of the star or for elements
that have a major contribution from long lived stars. Hence, the instantaneous
recycling approximation is of limited utility for comparing with multi-element
abundances. While the Simple model does not reproduce all observational
constraints, such as the metallicity distribution function (MDF) of the solar
neighborhood (a discrepancy known as the G-dwarf problem;
\citealt{vandenbergh1962, schmidt1963}), it is still frequently used as an
illustrative example.

Subsequent models have relaxed the assumptions of the Simple model.  Various
hypotheses were proposed to solve the G-dwarf problem and reproduce the
tendency for abundances to asymptote to a constant value, including
pre-enrichment \citep{schmidt1963}, variable IMF or yields \citep{truran1971b},
and inflow \citep{larson1972, tinsley1974, tinsley1976, tinsley1977}.  Other
models have sought to recreate the inside-out formation of the disk
\citep{larson1976} by representing the Milky Way as a series of concentric
annuli \citep{tinsley1980, chiosi1980, matteucci1989, chiappini1997}.
Initially, multi-zonal models evolved the annuli in isolation, though later
studies coupled the annuli by considering radial gas flows \citep{lacey1985,
goetz1992, portinari2000}.

Observations hinted that radial mixing of stars could be important for
understanding the chemical evolution of the solar neighborhood.
\citet{grenon1987, grenon1999} and \citet{francois1993} argued that the most
metal-rich stars in the solar neighborhood must have originated in the inner
Galaxy because they have higher metallicities than the current gas-phase
metallicity of the solar annulus. Increases in epicyclic amplitude
(``blurring'') were known to cause stellar orbits to reach more extreme radii.
But it was the discovery of radial migration (``churning''), a process that
changes the guiding center radii of stars via resonant scattering off of spiral
arms, by \citet{sellwood2002} that turned out to have special importance
because of its ability to transport stars over several kpc without inducing a
large ellipticity.

Several recent chemodynamical studies \citep{roskar2008, schoenrich2009a,
schoenrich2009b, minchev2013, minchev2014, kubryk2015a, kubryk2015b,
spitoni2015} have included radial mixing of stars (blurring and churning).
There is a general consensus that stellar mixing is critical for reproducing
the dispersion in the age--metallicity relation, and increasing recognition
that it may play a key role in shaping MDFs \citep{hayden2015, loebman2016,
martinezmedina2016}, though the extent of this impact remains a matter of
debate (e.g., \citealt{haywood2015}).

The complexity of these chemodynamical models can obscure the effects of and
trade-offs between model parameters and assumptions, so we adopt a
complementary approach using a one-zone model to vary the ingredients of
chemical evolution models and isolate their effects.  In particular, we focus
on understanding some novel aspects of the \citet{schoenrich2009a} model, such
as their finding that the tracks of individual annuli in [O/Fe]--[Fe/H] space
are much steeper than the thin disk sequence.  Though we devote much
of our attention to the impact of model parameters on the stellar
distribution in [O/Fe]--[Fe/H] space, another of our primary motivations
was to extend the suite of elements tracked by the \citet{schoenrich2009a}
model to include as many elements as possible, especially those with
metallicity-dependent yields.

Our goals for this paper are as follows:
\begin{enumerate}
\item Understand the trade-offs among the various ingredients of chemical
  evolution models, particularly the effects associated with inflow,
  outflow, star formation efficiency (SFE), the Type Ia supernova (SNIa) delay
  time distribution, and nucleosynthetic yields.
\item Examine predictions for a large number of elements within a
  common modeling framework, in part to identify elements for which
  current supernova yield models may be inaccurate.
\item Investigate mixing of stellar populations as a source of scatter
  and bimodality in
  [$X$/Fe]--[Fe/H] diagrams.  Scatter arises automatically in multi-zone models
  that incorporate radial mixing \citep[e.g.,][]{schoenrich2009a,
  schoenrich2009b}, but here we can study it in a simpler and somewhat more
  general context.
\item Present simple model predictions for the principal components
  of the distribution of stars in multi-element abundance space
  \citep{andrews2012, ting2012}.
\end{enumerate}
Throughout the paper we attempt to provide intuitive explanations for the
sometimes complex behavior that arises even within the one-zone framework. An
accompanying paper (\citealt{weinberg2016a}, hereafter
\citetalias{weinberg2016a}) provides analytic insights that describe several of
our key findings for MDFs and [O/Fe]--[Fe/H] evolutionary tracks.

In Section \ref{sec:model_description}, we describe the basics of \flexce, our
one-zone chemical evolution code, and the fiducial simulation.  In Section
\ref{sec:variations}, we show the effect of varying model parameters on model
tracks in [O/Fe]--[Fe/H].  In Section \ref{sec:yields_multielement}, we
compare the mean tracks for 20 elements ($X$) in [$X$/Fe]--[Fe/H] for two sets
of CCSN yields. In Section \ref{sec:scatter}, we produce a suite of model
tracks that span the scatter in [O/Fe]--[Fe/H] by simultaneously varying star
formation efficiency and outflow rate, as might result from mixing stellar
populations born at different galactocentric radii.  We apply principal
component abundance analysis (PCAA) to this suite of simulations in Section
\ref{sec:pcaa}.  Finally, Section \ref{sec:conclusions} highlights the
successes and potential uses of \flexce.


\section{Model Description}
\label{sec:model_description}

Our goal is to explore the sensitivity of chemical evolution models to
variations in individual model parameters and trade-offs between them, so we
constructed a code, \flexce, to include the major physical processes that
drive chemical evolution while retaining the simplicity of a one-zone model.

\flexce\ computes the evolution of a one-zone chemical evolution model with
inflow and outflow (i.e., open box) whose gas is instantaneously and completely
mixed.  The model is evolved for 12~Gyr with time steps every 30~Myr, which
corresponds to the lifetime of the longest-lived star that will explode as a
CCSN.  The model includes enrichment from CCSN, SNIa, and AGB stars, and the
yields of the latter two sources are staggered (i.e., not instantaneously
recycled). The model stochastically draws individual stars from the stellar
initial mass function (IMF) and stochastically explodes white dwarfs (WDs) as
SNIa.  The stochastic nature of the model enables investigations of scatter in
regimes where the IMF is not fully populated.  As the total mass increases and
the IMF becomes well-sampled, stochastic effects become unimportant.  Our
adopted inputs for \flexce\ are described in the following sections, and details
of the calculation method appear in the Appendix.

\subsection{Parameters of Fiducial Simulation}
\label{sec:fiducial}

The parameters of the fiducial simulation were chosen to broadly match the
observed solar neighborhood trend in [O/Fe]--[Fe/H], and we use this set of
parameters for all subsequent simulations unless stated otherwise.  We adopt the
net CCSN yields from \citet{chieffi2004} and \citet{limongi2006}, net AGB star
yields from \citet{karakas2010}, and the W70 SNIa yields from
\citet{iwamoto1999} (see Section \ref{sec:yields_multielement} for more
details). We normalized the abundances to the \citet{lodders2003} photospheric
solar abundances.

The simulation starts with 2$\times$10$^{10}$~\msun\ of primordial composition
gas, and an additional 3.5$\times$10$^{11}$~\msun\ primordial composition gas
flows into the galaxy with an exponentially declining time profile
($\tau$~=~6~Gyr). The final total mass of the simulation is
6.9$\times$10$^{10}$~\msun, with 3.2$\times$10$^9$~\msun\ in gas,
5.2$\times$10$^{10}$~\msun\ in stars, and 1.4$\times$10$^{10}$~\msun\ in
remnants (black holes, neutron stars, and white dwarfs).  Though we compare the
model abundances to solar neighborhood data and match the model mass to the
total mass of the Milky Way, the total mass of the model is essentially
arbitrary (except for small stochastic supernova effects). All masses scale in
proportion, so elemental abundances and gas fractions are independent of the
total mass.  Abundances across the Milky Way cannot be accurately modeled as a
single zone due to different conditions at different radii.  We present a suite
of models in Section \ref{sec:superposition} as a first step towards capturing
this diversity.

The star formation rate (SFR) is set by a constant star formation efficiency
(SFE) of 10$^{-9}$ yr$^{-1}$, converting $M_\mathrm{gas}(t)\cdot(\Delta t /
1$~Gyr) of the available gas mass into stars in a time interval $\Delta t$.  The
cold ISM is ejected in an outflow at a rate of 2.5$\times$SFR.  We adopt a
\citet{kroupa2001} IMF from 0.1--100 \msun\ and the stellar lifetime function of
\citet{renzini1986} (see Equations 3 and 4 from \citealt{padovani1993}.  The
SNIa delay time distribution (DTD) is an exponential with a timescale of 1.5 Gyr
and a minimum delay time of 150 Myr. We list the parameters in Table
\ref{table:parameters} and explain the choice of parameters in more detail in
Section \ref{sec:variations}.

\begin{deluxetable}{ll}[!t]
\tabletypesize{\small}
\tablecaption{Fiducial Model Parameters and Variations
\label{table:parameters}}
\tablewidth{0pt}
\tablehead{
\multicolumn{2}{c}{Fiducial Parameters}
}
\startdata

{}Outflow mass loading factor  &  $\eta$ = 2.5 \\
{}Star formation efficiency    &  SFE = 1.0 $\times 10^{-9}$ yr$^{-1}$ \\
{}Exponential SNIa DTD         &  $\tau_{\rm Ia}$ = 1.5 Gyr, $t_{\rm min}$ = 0.15 Gyr \\
{}Exponential inflow history   &  $\tau_{\rm inf}$ = 6 Gyr \\
{}IMF                          &  Kroupa (2001) $M$ = 0.1--100 M$_\odot$ \\

\cutinhead{Variations}

{}Inflow timescale            &  Figure \ref{fig:ofe_mdf_tau_inflow} \\
{}Inflow time history         &  Figures \ref{fig:ofe_mdf_inflow_rate}, \ref{fig:sfh} \\
{}Star formation efficiency   &  Figure \ref{fig:ofe_mdf_sfe} \\
{}Outflow mass loading        &  Figure \ref{fig:ofe_mdf_outflow} \\
{}Inflow metallicity          &  Figure \ref{fig:ofe_mdf_inflow_metallicity} \\
{}IMF                         &  Figure \ref{fig:ofe_mdf_imf} \\
{}Minimum SNIa delay time     &  Figure \ref{fig:ofe_min_snia_time} \\
{}Form of SNIa DTD            &  Figures \ref{fig:dtd}, \ref{fig:ofe_snia_dtd} \\

\enddata
\end{deluxetable}

\begin{figure}
\centerline{
\includegraphics[width=9cm]
{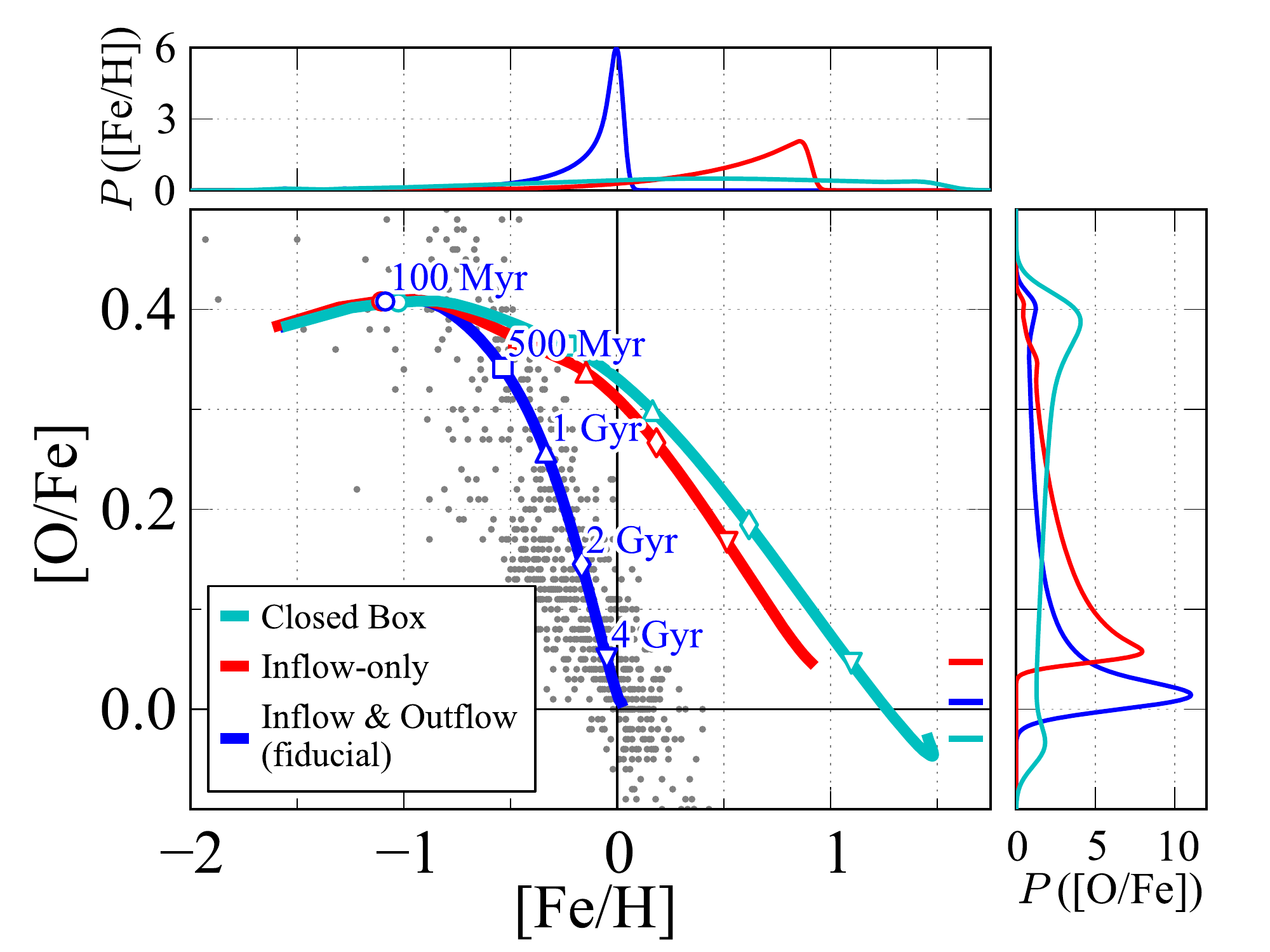}}

\caption{[O/Fe]--[Fe/H], metallicity distribution functions, and [O/Fe]
distribution functions for a closed box (no gas inflow or outflow) simulation,
an inflow only simulation, and the fiducial simulation that includes both
inflow and outflow.  Inflow dilutes the metallicity but boosts [O/Fe] because
it fuels late time star formation and CCSN enrichment, as can be seen in the
more prominent high metallicity and low [O/Fe] peaks of the inflow only and
fiducial simulations relative to the closed box simulation.  Outflow decreases
metallicity by ejecting metals and the gas that would have fueled additional
star formation. Outflow produces more strongly peaked MDFs and [O/Fe]-DFs by
speeding up the convergence to the equilibrium abundance and by significantly
reducing the equilibrium metallicity.  The black solid lines mark the solar
abundance.  Time is indicated by the colored circles (100 Myr), squares (500
Myr), upward-pointing triangles (1 Gyr), diamonds (2 Gyr), and
downward-pointing triangles (4 Gyr). The colored tick marks on the right
indicate the equilibrium [O/Fe].  The gray points show the iron abundances and
oxygen abundances derived using non-LTE analysis
of 775 local thin disk, thick disk, and
halo stars from \citet{ramirez2013}.  For display purposes, we applied a kernel
density estimate to the distribution functions that uses a Gaussian kernel with
a bandwidth of 0.03.}

\label{fig:ofe_gas_flows}
\end{figure}

\subsection{Including Inflow and Outflow}

To illustrate the importance of accretion and outflows, Figure
\ref{fig:ofe_gas_flows} shows the evolution in [O/Fe]--[Fe/H], the metallicity
distribution function (MDF), and the [O/Fe] distribution function ([O/Fe]-DF) of
the fiducial simulation, a closed box (no gas inflow or outflow) simulation, and
an inflow-only simulation. For the purpose of comparison, we have adopted the
fiducial simulation parameters for the closed box and inflow-only simulations,
except for the gas flows themselves. Parameter adjustments could improve the
agreement between these models and observations, though reproducing the observed
track in full would require substantial changes to adopted supernova yields.

The closed box simulation forms many of its stars early, as shown by the large
high-$\alpha$ peak of its [O/Fe]-DF, and then steadily forms stars across a wide
range of metallicities, producing a very broad MDF.  It reaches an unreasonably
high metallicity ([Fe/H]~$\approx$~1.5) because most of the gas has been turned
into stars ($M_\mathrm{\star}^\mathrm{final} /
M_\mathrm{gas}^\mathrm{final}$~=~200), leaving only a small amount of hydrogen
to dilute the iron synthesized by CCSN and SNIa.

The inflow-only simulation achieves a lower but still excessive metallicity of
[Fe/H]~$\approx$~0.9 and finishes at a higher [O/Fe] (0.05 vs.~$-$0.05 for the
closed box simulation), because inflow dilutes the metallicity by providing
additional hydrogen and fuels more star formation and CCSN enrichment at late
times relative to a closed box.  This late time accretion results in the more
prominent high metallicity and low-$\alpha$ peaks of the MDF and [O/Fe]-DF,
respectively.

The fiducial simulation demonstrates the effect of outflows, which decrease the
overall metallicity ([Fe/H]~$\approx$~0 at the end of the simulation by design)
by removing metals and the gas that would have fueled more star formation and
metal production. Outflows have the secondary effect of decreasing [O/Fe]
because CCSN oxygen yields slightly increase with metallicity. The MDF and
[O/Fe]-DF of the fiducial simulation are even more sharply peaked than the
inflow-only simulation.

The fiducial simulation roughly reproduces the observed abundance trend in the
data of \citet{ramirez2013}, though we caution that the detailed appearance of
the observations depends on sample selection and on the methodology for
abundance estimates.  In the fiducial simulation and all of the other
simulations that approximately reproduce observed abundance trends, the
abundance evolution slows dramatically at late times, approaching an
approximate equilibrium between sources and sinks of different elements. Below
we refer to these approximately asymptotic states as equilibrium abundances.
The emergence of equilibrium abundances and the timescales for reaching them
can be understood analytically given some simplifying assumptions, as described
by \citetalias{weinberg2016a}.

\begin{figure}
\centerline{
\includegraphics[width=9cm]
{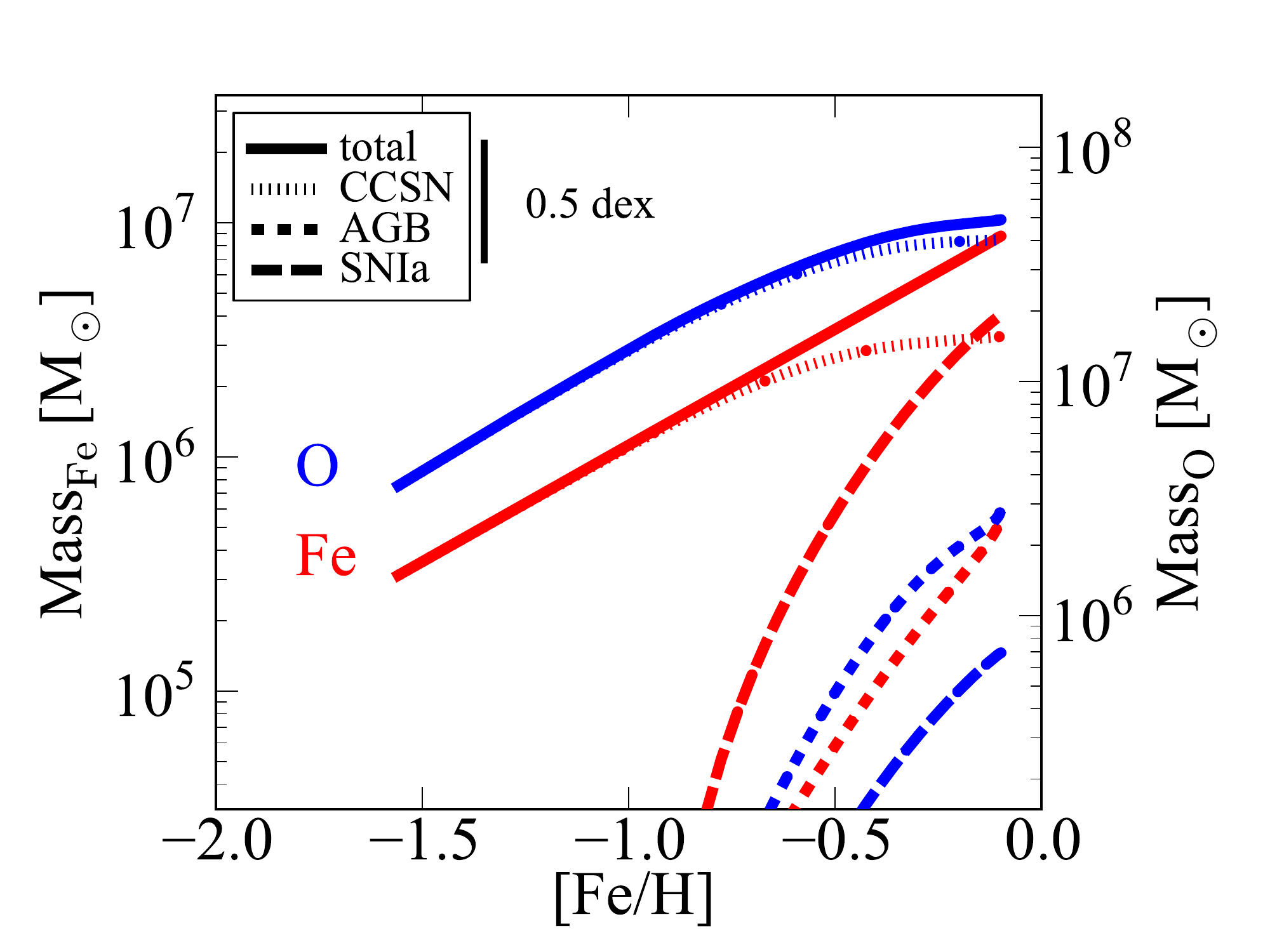}}
\caption{The mass of gas-phase oxygen and iron as a function of [Fe/H] for the
  constant SFR simulation.  The line styles indicate the contributions from
  different enrichment sources (including recycled material---i.e., atoms
  incorporated into a star at birth and returned at death).  The ordinate axes
  are aligned such that the total oxygen and total iron lines cross at solar
  [O/Fe].  The black bar indicates a 0.5~dex offset.}
\label{fig:ofe_river}
\end{figure}

\subsection{Model Calibration}

Our goal in this paper is to understand how model predictions depend on model
inputs, not to present a specific model of the Galaxy or the solar neighborhood.
It is nonetheless worth explaining our choice of fiducial model parameters in
the broader context of chemical evolution modeling, and especially our adoption
of a high outflow mass loading factor $\eta=2.5$.  Our choice of the stellar IMF
is motivated by the observational evidence summarized by \cite{kroupa2001}, and
our choice of the SNIa DTD is motivated by a combination of empirical and
theoretical considerations as described in Section \ref{sec:snia_dtd}.  With
these choices, the predicted present-day ratio of CCSN to SNIa in the fiducial
model is 3.7, consistent with the (rather loose) empirical constraint of $4.3
\pm 1.3$ found by \citet{li2011}.  Population averaged yields follow from the
nucleosynthesis calculations cited in Section \ref{sec:fiducial}. Our adoption
of a 6~Gyr e-folding timescale for gas infall is motivated by observational
estimates of the star formation history of the solar neighborhood, which suggest
slowly declining star formation over the history of the Galaxy
\citep{twarog1980, rochapinto2000}.

With these quantities set, the main adjustable parameters of the model are
$\eta$ and the star formation efficiency.  As shown in 
\S\ref{sec:variations} (see Figure~\ref{fig:ofe_mdf_multi}), the value of the
metallicity at late times is controlled mainly by $\eta$, and we choose
$\eta$~=~2.5 so that the fiducial model reaches [Fe/H]~$\approx$~0 at late
times. For fixed $\eta$, the SFE controls the location of the knee in
[O/Fe]--[Fe/H], and we select an SFE of $10^{-9}\,{\rm yr}^{-1}$ to reproduce
the observed location at [Fe/H]~$\approx -1$.

The notion that strong outflows are required to achieve solar abundance in the
ISM is common in the literature on the galaxy mass--metallicity relation (e.g.,
\citealt{finlator2008, peeples2011, zahid2012}), but it is not universal in the
Galactic chemical evolution literature. \cite{peeples2014} conclude that typical
star-forming galaxies have ejected 75--80\% of the oxygen they have produced,
and our fiducial model predicts 82\% ejection. \citetalias{weinberg2016a} show
that in the instantaneous recycling approximation the oxygen abundance evolves
to an equilibrium value $Z_{\rm O,eq} = m_{\rm O}^{\rm cc}/(1+\eta-r)$ for a
constant star formation rate (and slightly higher for a slowly declining star
formation rate).  Here $m_{\rm O}^{\rm cc}$ is the IMF-averaged oxygen yield
from CCSNe and $r$ is the mass recycling fraction, approximately 0.45 for a
Kroupa IMF. Our IMF and yield assumptions imply $m_{\rm O}^{cc} \approx 0.017$
(i.e., $1.7 M_\odot$ of oxygen produced per $100 M_\odot$ of stars formed). The
\citet{lodders2003} solar oxygen abundance, similar to that found in nearby
B-stars \citep{przybilla2008}, corresponds to $Z_{\rm O} = 0.0056$, and
reproducing it with $m_{\rm O}^{\rm cc}$ requires $\eta$~=~2--3.  We assume that
all stars from 8--100~$M_\odot$ explode as CCSN, and the oxygen yield could be
significantly lower if the more massive stars (which produce most of the oxygen)
instead collapse to black holes or if the wind loss from these stars is
overpredicted. For example, if we cut out oxygen from stars with
$M>35\,M_\odot$ then $m_{\rm O}^{\rm cc}$ drops by a factor of two, reducing
the required outflow efficiency to $\eta \approx 1$. However, one would also
need to adjust the CCSN and SNIa iron yields to reproduce solar iron abundance
with this lower efficiency.

Many historical chemical evolution models have reproduced broad characteristics
of the solar neighborhood without requiring outflows (e.g.,
\citealt{matteucci1986, chiappini1997}). Some of the earlier models effectively
treated the population-averaged supernova yield as an adjustable parameter,
concentrating on the form of the MDF rather than its normalization, and thus
omitted the main constraint that drives us to substantial outflows. Another
difference in many of these models is the use of a \citet{salpeter1955} IMF
rather than a \citet{kroupa2001} or \citet{chabrier2003} IMF.  The much larger
population of low mass stars in a Salpeter IMF reduces the IMF-averaged
supernova yield; for example, with the same supernova yields that we adopt, a
Salpeter IMF gives $m_{\rm O}^{\rm cc} = 0.011$ instead of 0.017, so the same
oxygen abundance can be achieved with a lower $\eta$. Even with a Salpeter IMF,
\citet{matteucci1986} find that ``the predicted solar iron abundance is higher
by more than a factor of two than the observed one,'' which is precisely the
mismatch that leads us to adopt strong outflows. The \cite{chiappini1997} model
incorporates both two episodes of infall and a decreased star formation
efficiency during the second phase.  It is unclear whether these changes can
explain the lower abundances relative to a one-zone model with no outflow and
fixed parameter; they may also be adopting lower yields than we do, but the
descriptions are sufficiently different that a direct comparison is difficult.
Radial gas flows \citep{spitoni2011, bilitewski2012, pezzulli2016} are another
possible mechanism for reducing metallicity by bringing in gas from the
metal-poor outer disk. Nonetheless, in the context of one-zone models with an
IMF and supernova yields similar to those adopted here, it is clear that strong
outflows are a necessary ingredient for reproducing solar abundances.

For simplicity, we have restricted our analysis to models in which the outflow
efficiency and star formation efficiency are constant in time, though either
could plausibly evolve as the disk potential grows and the gas fraction
declines. The high star formation efficiency of our fiducial model, chosen to
reproduce the location of the [$\alpha$/Fe]--[Fe/H] knee, leads to a low gas
fraction at late times, $M_{\rm gas}/M_* \approx 0.05$. This is considerably
lower than the recent estimate $M_{\rm gas}/M_* \approx 0.4$ for the solar
neighborhood by \citet{mckee2015}. For a specified star formation history,
the gas fraction is inversely proportional to the SFE, so this mismatch likely
indicates a substantially decreased SFE at late times, relative to the high
efficiency at early times when [$\alpha$/Fe] is just beginning its decline
toward solar values.

Two other solar-neighborhood constraints we do not attempt to match are the
age--metallicity relation \citep{twarog1980, edvardsson1993, casagrande2011} and
the metallicity distribution function \citep{pagel1975, nordstrom2004,
casagrande2011}. Both of these quantities are likely to be strongly shaped by
radial migration of stars, as argued by \citet{schoenrich2009a} and
\citet{hayden2015} building on suggestions by \citet{wielen1977} and
\citet{sellwood2002}. One-zone models with constant parameters predict a
one-to-one age--metallicity relation, while the observed relation shows large
scatter \citep{edvardsson1993}.  As shown below, these models generically
predict negatively skewed, sharply peaked MDFs like those seen in the inner
Galaxy \citep{hayden2015}, not the broad, roughly Gaussian (in [Fe/H]) form seen
in the solar neighborhood.  One-zone models provide a useful starting point for
inferring the impact of radial migration, which mixes populations formed with
different enrichment histories, but a single one-zone model cannot explain these
aspects of solar neighborhood chemical evolution.

\subsection{Oxygen and Iron Production}

Since we will be using [O/Fe]--[Fe/H] as the main diagnostic plot, it is
helpful to understand the enrichment sources responsible for producing oxygen
and iron.  Figure \ref{fig:ofe_river} shows the mass of gas-phase oxygen and
iron as a function of metallicity (stellar abundances are taken to be the
gas-phase abundances when a stellar generation is born). For ease of
interpretation we show the results from a constant SFR simulation because its
iron mass is linear with [Fe/H]; this simulation follows a very similar locus
in [O/Fe]--[Fe/H] to the fiducial simulation, though its final [Fe/H] is lower
and [O/Fe] is higher (see Figure \ref{fig:ofe_mdf_inflow_rate}).  The solid
line shows the total gas-phase mass of each element. The dotted, short-dashed,
and long-dashed lines indicate the contributions from CCSN, AGB stars, and
SNIa, respectively.  The ordinate axes have been offset such that the total
oxygen and iron lines cross at solar [O/Fe].

One can metaphorically regard CCSN, SNIa, and AGB stars as tributaries that
enrich the ``river'' of the main gas reservoir, which is itself depleted by
star formation and outflow.  Far ``upstream,'' at early times and low [Fe/H],
CCSN are the only significant channel and the ratio [O/Fe]~$\approx$~0.4
reflects their high-$\alpha$ yields.  As SNIa become important, contributing
iron but minimal oxygen, the [O/Fe] ratio declines toward solar values.  In the
constant SFR simulation, the transition from CCSN to SNIa dominating iron
production occurs at $t$~=~2.7~Gyr and [Fe/H]~$\approx -0.17$. CCSN remain the
dominant source of oxygen production at all times, with a much smaller
contribution from AGB stars.  Since these lines include recycled material, AGB
stars ``contribute'' iron even though they do not directly synthesize it
because they return iron incorporated at birth.  For similar plots of the full
suite of elements and more in-depth discussion of multi-element abundances see
Figure \ref{fig:river_all} and Section \ref{sec:multielement}.


\section{Varying Parameters: Model Tracks in [O/Fe]--[Fe/H]}
\label{sec:variations}

In this section, we vary the parameters of the model to understand how galaxy
evolution parameters affect the mean track in [O/Fe]--[Fe/H], the [O/Fe]-DF,
and the MDF.  The mean track in this space reflects the relative enrichment
from CCSN and SNIa ([O/Fe]) as a function of overall metallicity ([Fe/H]).
Since metallicity maps onto time nearly monotonically in the simulations, the
mean track also encodes the history of CCSN versus SNIa enrichment.

\begin{figure*}

\subfloat{
\includegraphics[width=9cm]
{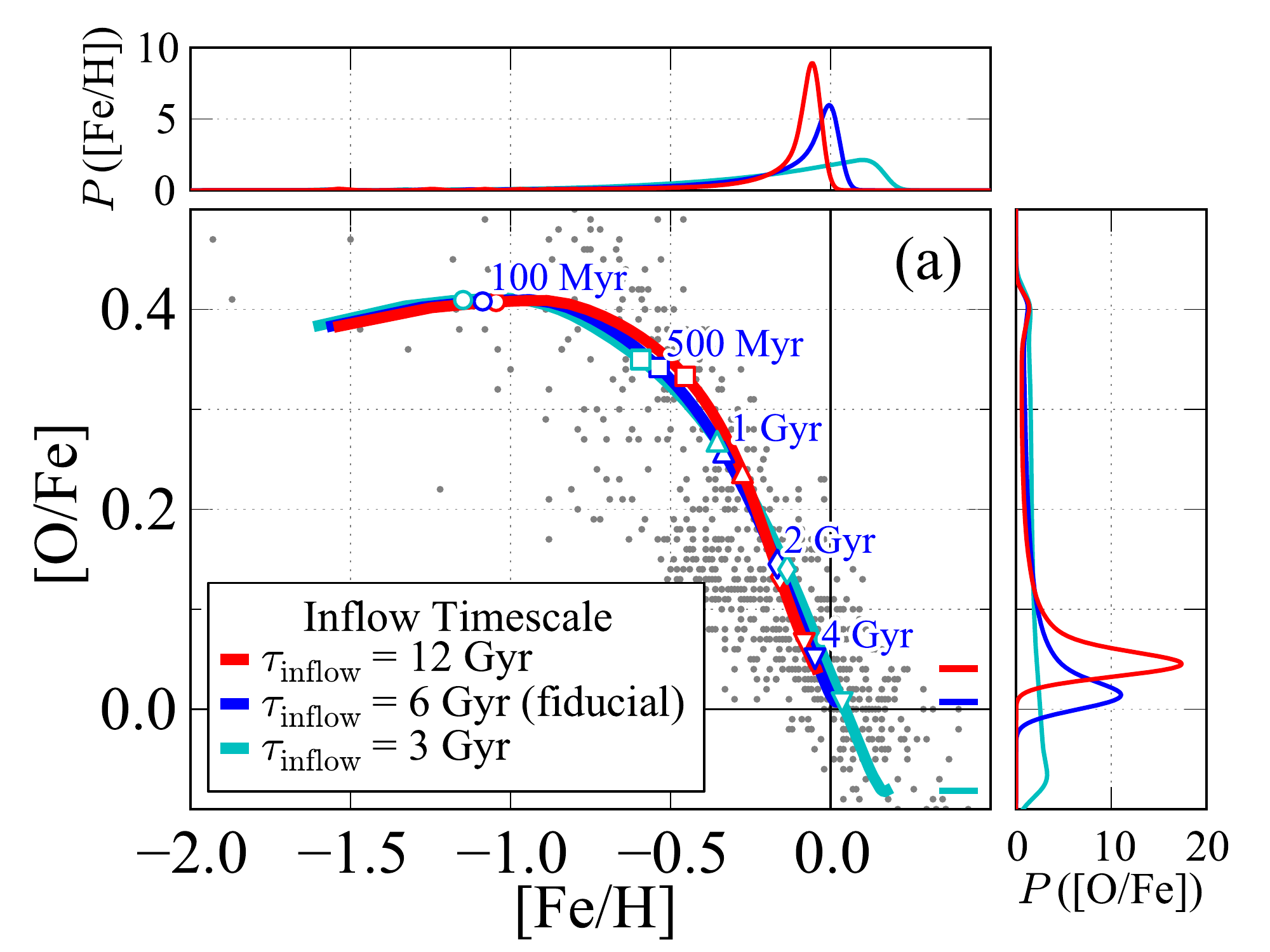}
\label{fig:ofe_mdf_tau_inflow}}
\subfloat{
\includegraphics[width=9cm]
{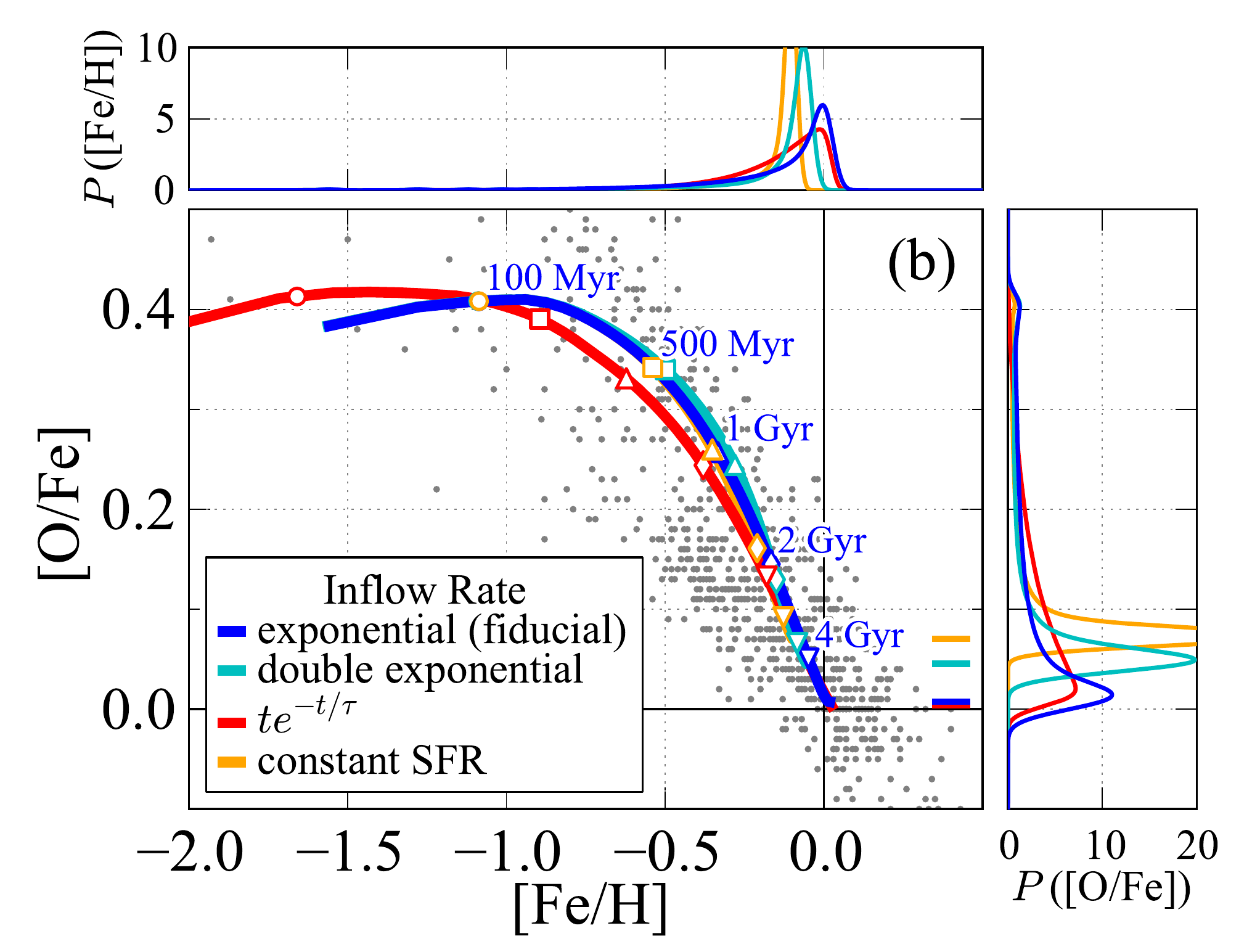}
\label{fig:ofe_mdf_inflow_rate}}

\subfloat{
\includegraphics[width=9cm]
{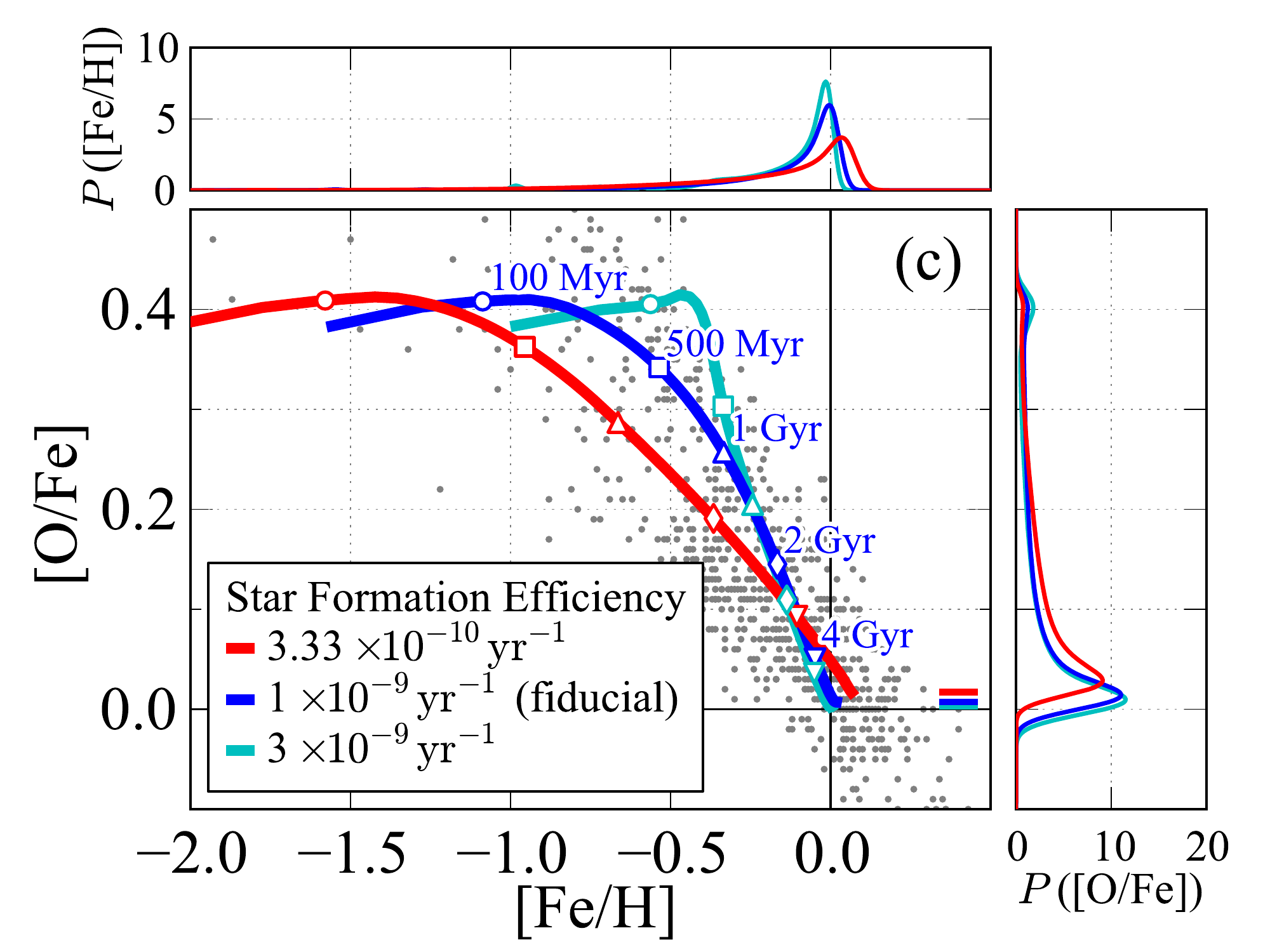}
\label{fig:ofe_mdf_sfe}}
\subfloat{
\includegraphics[width=9cm]
{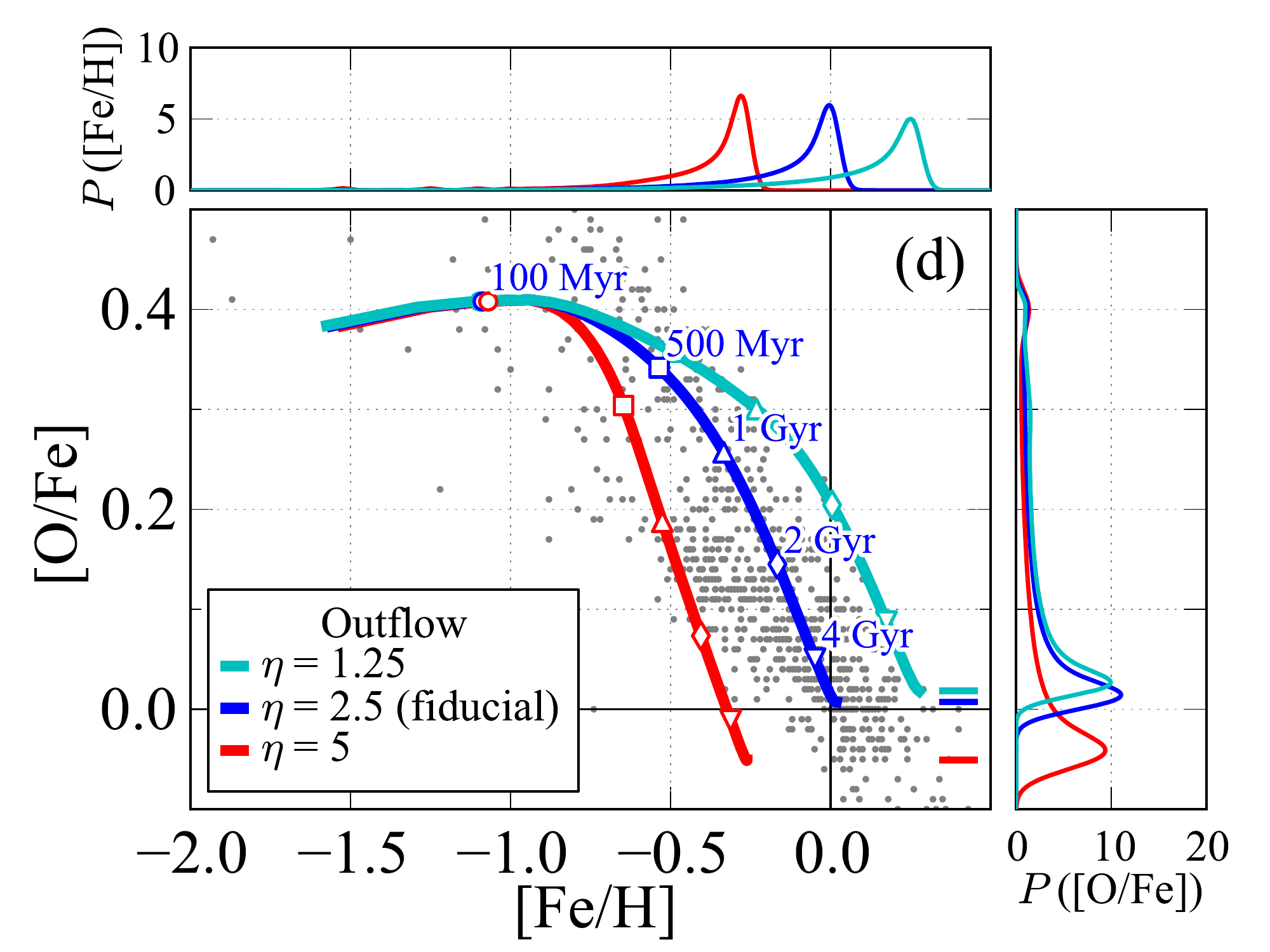}
\label{fig:ofe_mdf_outflow}}

\subfloat{
\includegraphics[width=9cm]
{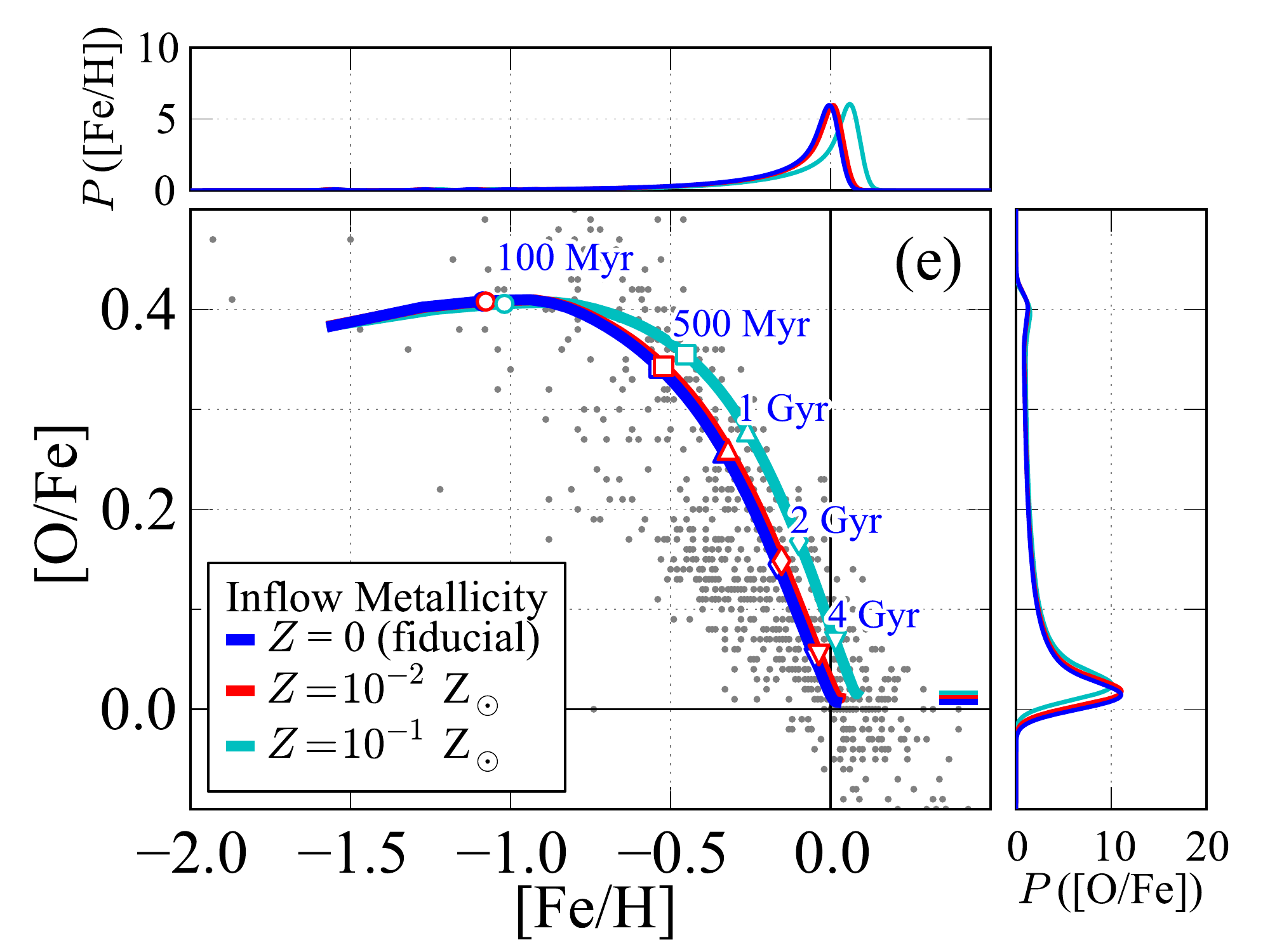}
\label{fig:ofe_mdf_inflow_metallicity}}
\subfloat{
\includegraphics[width=9cm]
{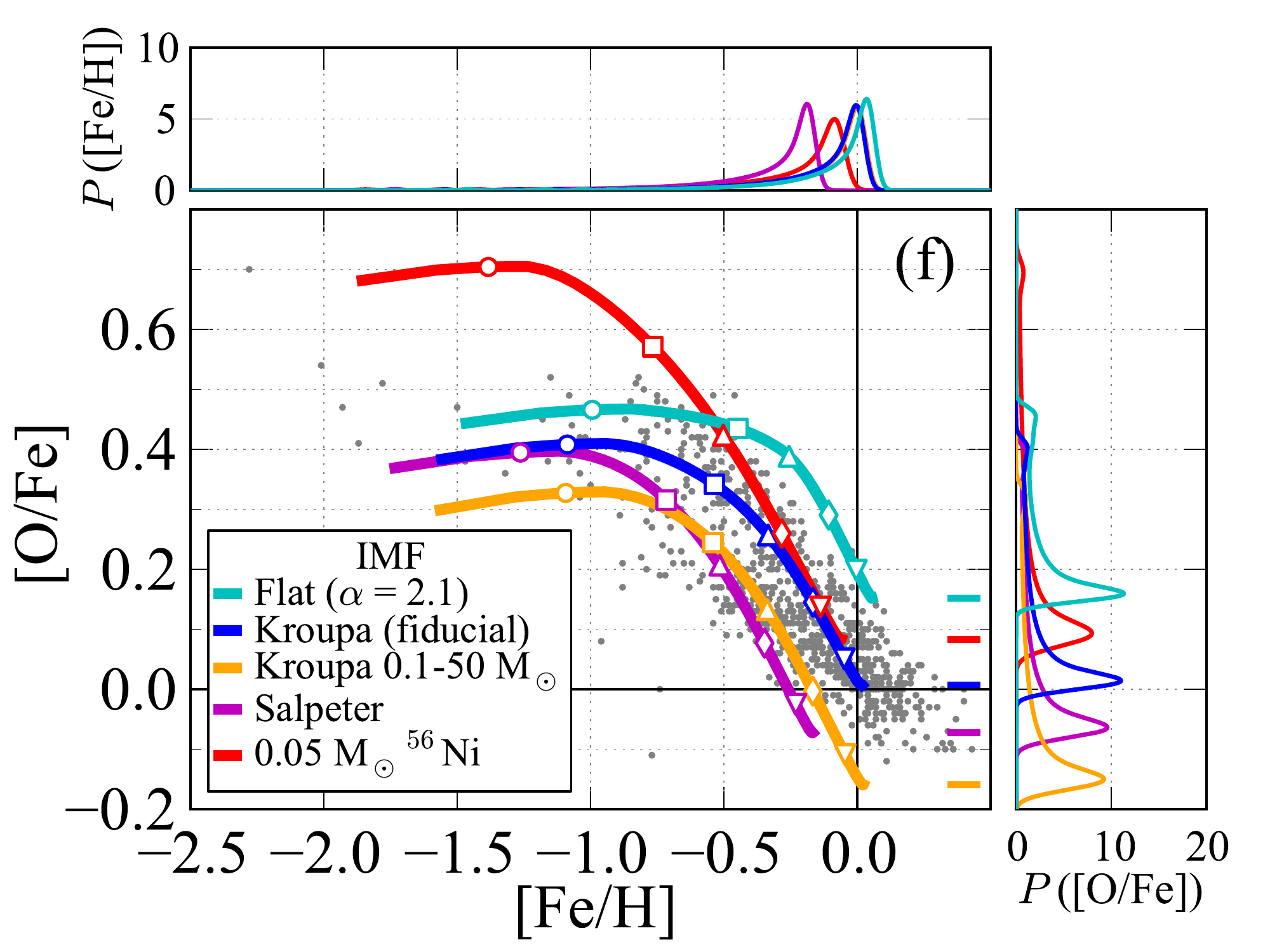}
\label{fig:ofe_mdf_imf}}

\caption{[O/Fe]--[Fe/H], MDFs, and [O/Fe]-DFs for variations in (a) the inflow
timescale, (b) the functional form of the inflow rate, (c) the star formation
efficiency, (d) the outflow mass-loading parameter $\eta$ (see Equation
\ref{eqn:eta}), (e) the inflow metallicity, and (f) the IMF/mass cut for CCSN.
Increasing the inflow timescale produces more peaked MDFs and [O/Fe]-DFs.
Increasing the star formation efficiency increases the [Fe/H] of the knee.
Increasing the outflow mass-loading parameter decreases the equilibrium [Fe/H]
and [O/Fe], steepening the trajectory of the trend in [O/Fe]--[Fe/H]. We note
that panel (f) has a different scale than the other panels.  Symbols and data
points are the same as in Figure \ref{fig:ofe_gas_flows}.
\label{fig:ofe_mdf_multi}}

\end{figure*}

In Figures \ref{fig:ofe_mdf_multi}--\ref{fig:ofe_warm_ism} the blue solid curves
always show the fiducial model described in Section \ref{sec:fiducial}. We
investigate changes of one model ingredient at a time, and one can infer from
our results what changes could compensate each other and what changes cannot
easily be compensated.

\subsection{Inflow Rate}
\label{sec:inflow_rate}

Closed box chemical evolution models generically suffer from the G-dwarf problem
\citep{vandenbergh1962}, namely that they produce MDFs for G-dwarfs (whose
lifetimes are comparable to the age of the Galactic disk) with too many
metal-poor stars relative to metal-rich stars.  Inflows allow galaxies to form
more stars later in their lifetimes when the cold ISM is metal-rich, increasing
the fraction of metal-rich stars. Continuing infall is, of course, the natural
expectation in analytic or numerical models of cosmological galaxy formation,
with star formation typically tracking gas accretion after a moderate delay
(e.g., \citealt{katz1996}).

To represent a range of inflow histories, we vary the inflow timescale
$\tau_1$ assuming an exponential inflow rate:
\begin{equation}
\dot{M}_\mathrm{in} = \frac{M_1}{\tau_1} e^{-t / \tau_1},
\label{eqn:exp_inflow}
\end{equation}
where $M_1$ sets the normalization of the inflow rate.  Figure
\ref{fig:ofe_mdf_tau_inflow} shows the effect of varying the inflow timescale
by a factor of two relative to the $\tau_1$~=~6~Gyr of the fiducial simulation.
All three simulations follow very similar tracks, but the equilibrium
abundances, [O/Fe]-DFs, and MDFs are distinct.  The $\tau_1$~=3~Gyr simulation
has a lower equilibrium [O/Fe] and a higher equilibrium [Fe/H].  It experiences
less late time accretion, star formation, and CCSN enrichment to counteract
SNIa Fe production, which continues to drive down [O/Fe] and increase [Fe/H].
Its [O/Fe]-DF and MDF are broader because it does not asymptote to an
equilibrium abundance as quickly as the fiducial simulation.  Conversely, the
$\tau_1$~=12~Gyr simulation reaches a higher [O/Fe] and lower [Fe/H] at the end
of the simulation and has more peaked [O/Fe]-DF and MDF. Overall, the inflow
timescale dictates how far along the track the simulation goes until it
achieves equilibrium abundance ratios at late times.  The faster a simulation
reaches the equilibrium abundance, the more peaked its [O/Fe]-DF and MDF will
be.

We experimented with several functional forms of the inflow history with
different ratios of early to late time accretion, including a double
exponential,
\begin{equation}
\dot{M}_\mathrm{in} = \frac{M_1}{\tau_1} e^{-t / \tau_1} + \frac{M_2}{\tau_2}
e^{-t / \tau_2},
\label{eqn:double_exp_inflow}
\end{equation}
and a linear-exponential product,
\begin{equation}
\dot{M}_\mathrm{in} = \left(\frac{M_1}{\tau_1}\right)
\left(\frac{t}{\tau_1}\right) e^{-t / \tau_1}.
\label{eqn:lin_exp_inflow}
\end{equation}
We also ran a constant SFR simulation, where the inflow rate was set to
maintain a constant gas mass at all times.  This simplified star formation
history (SFH) helps illustrate the differences between simulations with
different yields (see Figure \ref{fig:xfe}) and the relative contribution of
CCSN, SNIa, and AGB stars to enrichment (see Figures \ref{fig:ofe_river} and
\ref{fig:river_all}).

\begin{figure}
\centerline{\includegraphics[width=9cm]
{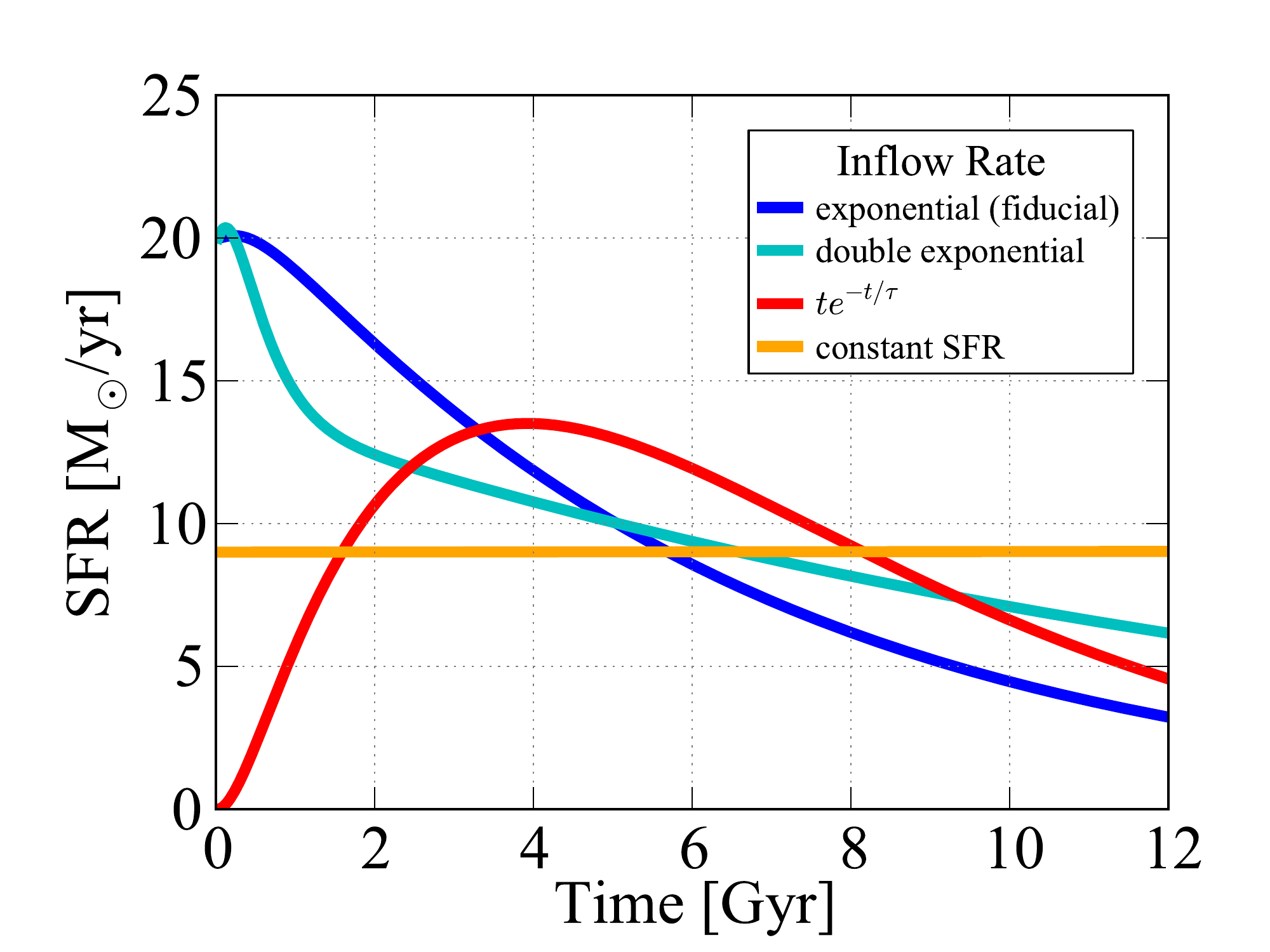}}
\caption{SFHs for exponential, double exponential, and linear-exponential
product inflow rates and a constant SFR.}
\label{fig:sfh}
\end{figure}

Figure \ref{fig:sfh} shows the SFR as a function of time for the fiducial,
double exponential, linear-exponential, and the constant SFR simulations.  The
double exponential simulation represents a scenario with early rapid accretion
($M_1$~=~$10^{10}$ \msun, $\tau_1$~=~300~Myr) followed by steady accretion on
long timescales ($M_2$~=~6$\times10^{11}$ \msun, $\tau_2$~=~14~Gyr), where the
timescales were adopted from \citet{schoenrich2009a}.  The linear-exponential
simulation is motivated by cosmological hydrodynamical simulations that find
inflow rates that increase at early times, reach a maximum at a characteristic
timescale, and then decline at late times \citep{simha2014}, so it starts with
no initial gas.  The timescale of the linear-exponential simulation
($\tau_1$~=~3.5~Gyr) was chosen so that the simulation reached solar
metallicity and oxygen abundance.  The normalizations of the double
exponential, linear-exponential, and the constant SFR simulations were chosen
to approximately match the same total amount of inflow and final stellar mass
as the fiducial simulation.

Figure \ref{fig:ofe_mdf_inflow_rate} shows the mean tracks in [O/Fe]--[Fe/H],
the [O/Fe]-DFs, and MDFs for these inflow histories.  The double exponential
simulation enriches slightly faster than the fiducial simulation for the first
2.5~Gyr, but its higher late time inflow rate results in a higher equilibrium
[O/Fe], a lower equilibrium [Fe/H], and more peaked [O/Fe]-DF and MDF.  The
linear-exponential simulation starts at a lower metallicity ([Fe/H]~=~$-$2.0 at
$t$~=~30~Myr) than the fiducial simulation because it does not start with any
gas.  Its lower early time SFR shifts the knee to lower metallicity and
produces fewer stars on the high [O/Fe] plateau.  It also broadens the
[O/Fe]-DF and MDF, especially towards the higher [O/Fe] and lower [Fe/H] side
of the peak.  However, its equilibrium [O/Fe] and [Fe/H] match those of the
fiducial simulation.  Since the constant SFR simulation effectively has an
infinite inflow timescale, it behaves similarly to the $\tau$~=~12~Gyr inflow
timescale simulation.  The constant SFR simulation follows nearly the same
track as the fiducial simulation, but its equilibrium [O/Fe] is higher,
equilibrium [Fe/H] is lower, and its [O/Fe]-DF and MDF are more peaked due to
the additional dilution and CCSN enrichment at late times from its higher late
inflow rate.

The total mass of the simulation does not affect the mean trend in
[O/Fe]--[Fe/H], as long as the initial mass and the mass inflow rate are scaled
together. However, larger masses decrease the effect of stochastic fluctuations
from incomplete sampling of the IMF or the SNIa DTD.  For a fixed inflow rate,
the initial gas mass of the model regulates the speed of enrichment but not the
final metallicity. In this case, a smaller initial gas mass extends the plateau
in [O/Fe]--[Fe/H] to lower metallicities.  A large initial gas mass produces a
galaxy that enriches quickly then asymptotes to a constant metallicity, whereas
a small initial gas mass results in a smoother, more continuous increase in
metallicity.

\subsection{Star Formation Rate and Efficiency}
\label{sec:sfe}

A common prescription for star formation rate is the Kennicutt-Schmidt Law
($\dot{\Sigma}_\star \propto \Sigma_{\rm gas}^N$; \citealt{schmidt1959};
\citealt{kennicutt1998}).  \citet{kennicutt1998} found a best fit power law
index of $N$~=~1.4 across a wide range in global galaxy SFRs, which implies a
higher SFE in denser regions.  More recent studies by \citet{leroy2008} and
\citet{bigiel2008} determined a value of $N$~=~1.0 for molecular-dominated gas
in spatially resolved observations of local galaxies, which implies a constant
SFE for gas that is sufficiently dense. \citet{scoville2016} also found a
nearly linear slope ($N$~=~0.9) for high redshift ($z$~=~1--6) star-forming
galaxies using the dust continuum to measure the total gas mass.  Likewise,
our model assumes a constant SFE (i.e., $N$~=~1.0):
\begin{equation}
\mathrm{SFR} = \frac{1}{t_\mathrm{gas}} M_\mathrm{gas}
    = \mathrm{SFE} \cdot M_\mathrm{gas},
\label{eqn:kslaw}
\end{equation}
where $t_\mathrm{gas}$ is the gas depletion timescale and SFE~$\equiv
t_\mathrm{gas}^{-1}$ is the star formation efficiency.  This parameterization
is the natural choice for models with a constant SFR or if the gas is
predominantly molecular, but for models with a declining gas supply and a
non-negligible fraction of atomic gas, a declining SFE is arguably a more
observationally motivated assumption.  That being said, we tested simulations
with a Kennicutt-Schmidt star formation law ($N$~=~1.4), and their evolution
was not significantly different from the fiducial simulation.  We note that
\citet{portinari1999} found that a Kennicutt-Schmidt law with either $N$~=~1.0
or 1.5 cannot simultaneously produce the radial metallicity gradient and
gas profile, though radial gas flows can alleviate this issue
\citep{portinari2000}.  As for the normalization of the SFE, \citet{leroy2008}
and \citet{bigiel2008} measured the SFE for molecular gas to be
5$\times$10$^{-10}$~yr$^{-1}$, which is a factor of two lower than the SFE
required in the fiducial simulation to match the knee in [O/Fe]--[Fe/H].
However, this comparison is dependent on the adopted minimum SNIa delay time
(see Figure \ref{fig:ofe_min_snia_time}), so it is inadvisable to draw strong
conclusions about the agreement between the overall normalization of the SFE
in observations and the model.

Figure \ref{fig:ofe_mdf_sfe} shows factor-of-three variations around the SFE of
the fiducial simulation (10$^{-9}$~yr$^{-1}$).  Increasing the SFE results in a
knee at a higher [Fe/H] because more CCSN produce more iron and more star
formation consumes more hydrogen (see also \citealt{matteucci1990}). Further
increases in the SFE result in diminishing increases in the [Fe/H] of the knee
because the simulation reaches a temporary equilibrium abundance. Increasing
the SFE also increases the relative number of stars on the high [O/Fe] plateau.
The final equilibrium [O/Fe] and [Fe/H] are mostly insensitive to SFE, though
SFE sets the speed of convergence, hence the width of the peaks of the
[O/Fe]-DF and MDF become narrower with increasing SFE.  To simplify the SFH, we
also performed constant SFR simulations with the same three SFEs and find
similar general characteristics for the SFE variations (modulo the effects of
using a constant SFR mentioned above).

\subsection{Outflows}
\label{sec:outflow}

The energy and momentum injection from massive stars can launch large-scale
galactic winds \citep[e.g.,][]{veilleux2005}.  Measurements of the mass
outflow rates of these winds, while challenging, point to values that are a
few times the SFR (e.g., \citealt{martin1999}, \citealt{heckman2000}).   We
parametrize the outflow rate ($\dot{M}_\mathrm{outflow}$) to be proportional
to the SFR,
\begin{equation}
\dot{M}_\mathrm{outflow} = \eta_\mathrm{wind} \mathrm{SFR},
\label{eqn:eta}
\end{equation}
where $\eta_\mathrm{wind}$ is the mass-loading factor of the wind.  We assume
that the outflowing material has the same abundance pattern as the cold ISM,
which would be expected if the mass in the outflow is dominated by entrained
cold gas as opposed to the hot gas ejected by SNe.  We performed simulations
that ejected a fraction of the stellar yields without any cold ISM leaving the
Galaxy, which required 75\% of the stellar yields to be ejected for the
simulation to have a equilibrium [O/Fe] and [Fe/H] at solar values. However, the
[O/Fe] knee of this simulation occurred at [Fe/H]~=~$-$1.5, which is
significantly lower than the observed metallicity of the knee.
Metallicity-enhanced winds can also spoil the otherwise good agreement of model
predictions with the observed ISM oxygen abundance and deuterium-to-hydrogen
ratio \citep{weinberg2016b}.

Figure \ref{fig:ofe_mdf_outflow} shows the mean track in [O/Fe]--[Fe/H], the
[O/Fe]-DF, and the MDF for factor of two variations in the mass-loading factor
around the fiducial value of $\eta_\mathrm{wind}$~=~2.5, which produces a track
that ends at solar metallicity and oxygen abundance.  Changing
$\eta_\mathrm{wind}$ strongly affects the shape of the track after the knee at
[Fe/H]~$\approx -0.9$, leading to very different equilibrium abundances,
especially [Fe/H]. Increasing the mass-loading factor removes more metal-rich
gas, which is replaced by inflowing pristine gas, thus decreasing the
equilibrium metallicity. The lower equilibrium metallicity results in a lower
equilibrium [O/Fe] because the CCSN oxygen yields slightly increase with
metallicity. The [O/Fe]-DF and MDF of the three simulations are offset because
of their different equilibrium abundances, but the shape of the peaks remain
similar.  A higher mass-loading factor also induces greater variance in the
abundances because it decreases the size of the gas reservoir and boosts the
impact of stochastic fluctuations, but this effect only becomes important at
much lower total masses than shown in Figure \ref{fig:ofe_mdf_outflow}.

\subsection{Inflow Metallicity and Metal Recycling}
\label{metal_recycling}

Gas that accretes onto galaxies is a mix of pristine or low metallicity IGM gas
($Z$~$\approx$~10$^{-3}$~\zsun; \citealt{songaila1996}) and enriched material
previously ejected from galaxies (e.g., \citealt{marasco2012};
\citealt{ford2014}).  The composition of accreting gas is difficult to determine
observationally, so the relative contribution of new vs.~recycled gas is
unknown.  We performed simulations where the overall metallicity of the
inflowing gas was
\begin{itemize}
    \item $Z$~=~0 (fiducial value used to simulate inflow of pristine gas),
    \item $Z$~=~10$^{-2}$~\zsun\ (to simulate inflow from an enriched IGM), and
    \item $Z$~=~10$^{-1}$~\zsun\ (to simulate a mix of pristine gas and recycled gas),
\end{itemize}
where \zsun~=~0.02 (this value is higher than the sum of the solar abundances of
individual elemental abundances of \citealt{lodders2003} that we have adopted,
though this difference does not affect our qualitative findings). The
composition of the inflow was the relative abundances of elements in the ISM at
the previous time step, which was then scaled to an overall metallicity. We also
experimented with $\alpha$-enhanced and scaled-solar abundance patterns but
chose not to use them because the $\alpha$-enhanced abundance pattern prevented
the gas from reaching the solar [$\alpha$/Fe] value and the scaled-solar
abundance pattern created very low metallicity stars with solar [$\alpha$/Fe]
abundances.

Figure \ref{fig:ofe_mdf_inflow_metallicity} shows the simulations with
enriched inflow.  The $Z$~=~10$^{-2}$~\zsun\ simulation is nearly identical to
the fiducial ($Z$~=~0) simulation, indicating that enriched inflow from the
IGM has little effect on chemical evolution (though it may be important at
extremely low metallicities; see \citealt{brook2014}).  The
$Z$~=~10$^{-1}$~\zsun\ simulation has a higher [Fe/H] and a higher [O/Fe]
(above the knee) at fixed time than the fiducial simulation, so the knee
occurs at a higher [Fe/H].  Its [O/Fe]-DF and MDF are peaked at higher values,
though the shape of these distribution functions is similar to the fiducial
simulation.  We conclude that enriched inflow only has a major effect on the
track in [O/Fe] vs.~[Fe/H] if the accreted gas has a metallicity of
$Z>10^{-1}$~\zsun.

In this experiment, we have only considered constant inflow metallicities that
stay fixed throughout the simulations.  In reality, the inflow metallicity
likely increases with time and would be lower than the ISM metallicity at early
times. This issue is significantly mitigated by the dominance of the large
reservoir of pristine gas relative to the early infalling gas, so the infall
metallicity has a small impact on the ISM metallicity provided that the infall
metallicity is low compared to the eventual equilibrium metallicity.  This
finding is confirmed in the analytic model of \citetalias{weinberg2016a} (see
their Section 5.2).  Cosmological simulations (e.g., \citealt{keres2009}) find
that the transition from pristine cold flows to enriched hot mode accretion
occurs around $z$~$\sim$~1, when the abundances are already approaching
equilibrium values. Thus, we expect that a time-dependent inflow metallicity
will have a negligible effect in moderate and high metallicity environments,
though it may be important in the low metallicity outer regions of galactic
disks.

\subsection{IMF}
\label{sec:imf}

The slope and mass range of the stellar initial mass function (IMF) sets the
relative number of high and low mass stars that form in a stellar population.
The main effect of the IMF is that it governs the relative mass locked up in low
mass stars and remnants versus the element production by high and intermediate
mass stars. Additionally, the high mass slope and cutoff can have a secondary
effect on elements whose CCSN yields are highly mass dependent, like oxygen.
\citet{cote2016a} found that the high mass slope of the IMF is a major source of
uncertainty in chemical evolution models.

Historically, the \citet{salpeter1955} IMF, with a slope of $\alpha$~=~2.35
and a mass range from 0.1--100~\msun, has provided a useful reference point
for IMFs.  However, the more recent \citet{kroupa2001} IMF, whose slope is
$\alpha$~=~(1.3, 2.3) from (0.1--0.5, 0.5--100~\msun) does a better job
describing the solar neighborhood, so we adopt it as the fiducial IMF.  We
have assumed that the IMF is constant in time, as suggested by the chemical
evolution study of \citet{romano2005}.

In Figure \ref{fig:ofe_mdf_imf}, we show the mean track in [O/Fe]--[Fe/H], the
[O/Fe]-DF, and the MDF for four IMFs:
\begin{itemize}
\item Kroupa IMF from 0.1--100~\msun\ (fiducial),
\item flat ($\alpha$~=~2.1) IMF from 0.1--100~\msun,
\item Kroupa IMF from 0.1--50~\msun, and
\item Salpeter IMF from 0.1--100~\msun.
\end{itemize}
Relative to the fiducial IMF, the flat IMF simulation always has a higher
[O/Fe].  It also has a slightly higher [Fe/H], especially at early times
($t<$~2~Gyr).  The Kroupa 0.1--50~\msun\ IMF simulation has a lower [O/Fe]
but a similar [Fe/H] compared to the fiducial simulation.  These changes are
due to the oxygen yield of CCSN increasing strongly with mass but iron yield
staying constant.

The Salpeter 0.1--100 \msun\ IMF simulation has a nearly identical [O/Fe]
value of the plateau because of the similar high mass slopes of the Kroupa and
Salpeter IMFs and the lack of SNIa iron production at these early time steps.
However, the larger number of low mass stars that are produced in the Salpeter
IMF simulation results in a lower [Fe/H] because more gas goes into long-lived
low mass stars as opposed to CCSN.  The [O/Fe]-DFs and MDFs for the IMF
variation simulations have similar shapes, though the locations of their peaks
are offset for the reasons given above.

Although Figure \ref{fig:ofe_mdf_imf} shows that chemical evolution tracks are
sensitive diagnostics of the IMF, these differences are partly degenerate with
SN yields (c.f., \citealt{molla2015}).  The mass cut for CCSN, which sets the
division between mass ejected in the explosion and mass that falls back on the
neutron star, has a large impact.  In particular, the iron yield is governed by
the mass cut because the material near the mass cut consists mostly of iron-peak
elements, whereas oxygen is produced further out in the star so its yield is
insensitive to the mass cut. Increasing the mass cut to produce 0.05~\msun\ of
ejected $^{56}$Ni instead of 0.1~\msun, increases the [O/Fe] level of the
plateau, decreases the [Fe/H] of the knee, and generally decreases [Fe/H] at all
times.  It also produces slightly broader [O/Fe]-DFs and MDFs.  We discuss mass
cut effects further in Section \ref{sec:ccsn_yields}.

We note that the [O/Fe]--[Fe/H] tracks for the various IMFs assume the outflow
mass loading parameter for the Kroupa IMF. The Kroupa IMF is more bottom
light/top heavy than the alternative IMFs that we consider here and the IMFs
others have considered elsewhere (e.g., \citealt{scalo1986},
\citealt{kroupa2003}).  Recalibrating the model for one of these IMFs would lead
to a smaller outflow mass loading parameter, but at least in the case of the
Salpeter IMF from 0.1--100 M$_\odot$, the model still requires outflows to
produce the knee at [Fe/H]~=~$-$1 and reach solar abundances.

\subsection{SNIa Delay Time Distribution}
\label{sec:snia_dtd}

The SNIa history cannot be predicted robustly due to uncertainty in the delay
time between the birth of a stellar population and the SNIa that it will
produce.  Our understanding of the SNIa delay time distribution (DTD) is
complicated by a few factors: (1) the evolutionary state of the progenitor
systems---double degenerate (WD + WD) or single degenerate (WD + red giant, WD
+ main sequence star, or WD + helium star)---remains unknown, (2) binary
evolution, especially the common envelope phase, is difficult to model, and
(3) dynamical effects in triple star systems could dramatically hasten the
merger of two WDs relative to naive expectations from binary evolution
\citep{thompson2011, katz2012}.

Because of these uncertainties, we consider variations in the minimum delay
time and multiple functional forms of the DTD.  The minimum delay time sets
the timescale for SNIa to become significant contributors of iron, which
produces the knee in [O/Fe]--[Fe/H].  The different functional forms of the
DTD govern the SNIa rate on longer timescales.

In the fiducial simulation, we adopt an exponential form of the SNIa DTD as in
\citet{schoenrich2009a}, where the mass in the WD reservoir produced by a
stellar population is depleted at a constant fractional rate after a minimum
delay time. The SNIa DTD (in units of SNIa yr$^{-1}$ \msun$^{-1}$) is
\begin{equation}
    f_\mathrm{SNIa}(t) =
    K \, \tau_\mathrm{SNIa}^{-1} \,
    e^{-(t - t_\mathrm{SNIa}^\mathrm{min}) / \tau_\mathrm{SNIa}},
\label{eqn:exp_dtd}
\end{equation}
where $t_\mathrm{SNIa}^\mathrm{min}$ is the minimum delay time for a SNIa and
$\tau_\mathrm{SNIa}$ is the timescale for SNIa, which are set to 150~Myr and
1.5~Gyr from the birth of a stellar population, respectively, in the fiducial
simulation (also adopted from \citealt{schoenrich2009a}, who calibrated these
values to reproduce oygen abundances).  The constant $K$ was chosen to match the
observed time-integrated (from 40~Myr to 10~Gyr) number of SNIa per stellar mass
formed  $N_\mathrm{Ia} / M_\star = 2.2 \times 10^{-3} \, \mathrm{M}_\odot^{-1}$
from \citet{maoz2012a}. \citet{maoz2012a} adopted the \citet{bell2003} IMF for
their stellar mass estimates, which is similar enough to the Kroupa IMF for our
purposes, but one would have to recalibrate the value of $K$ if using a
significantly different IMF.

\begin{figure}
\centerline{\includegraphics[width=9cm]
{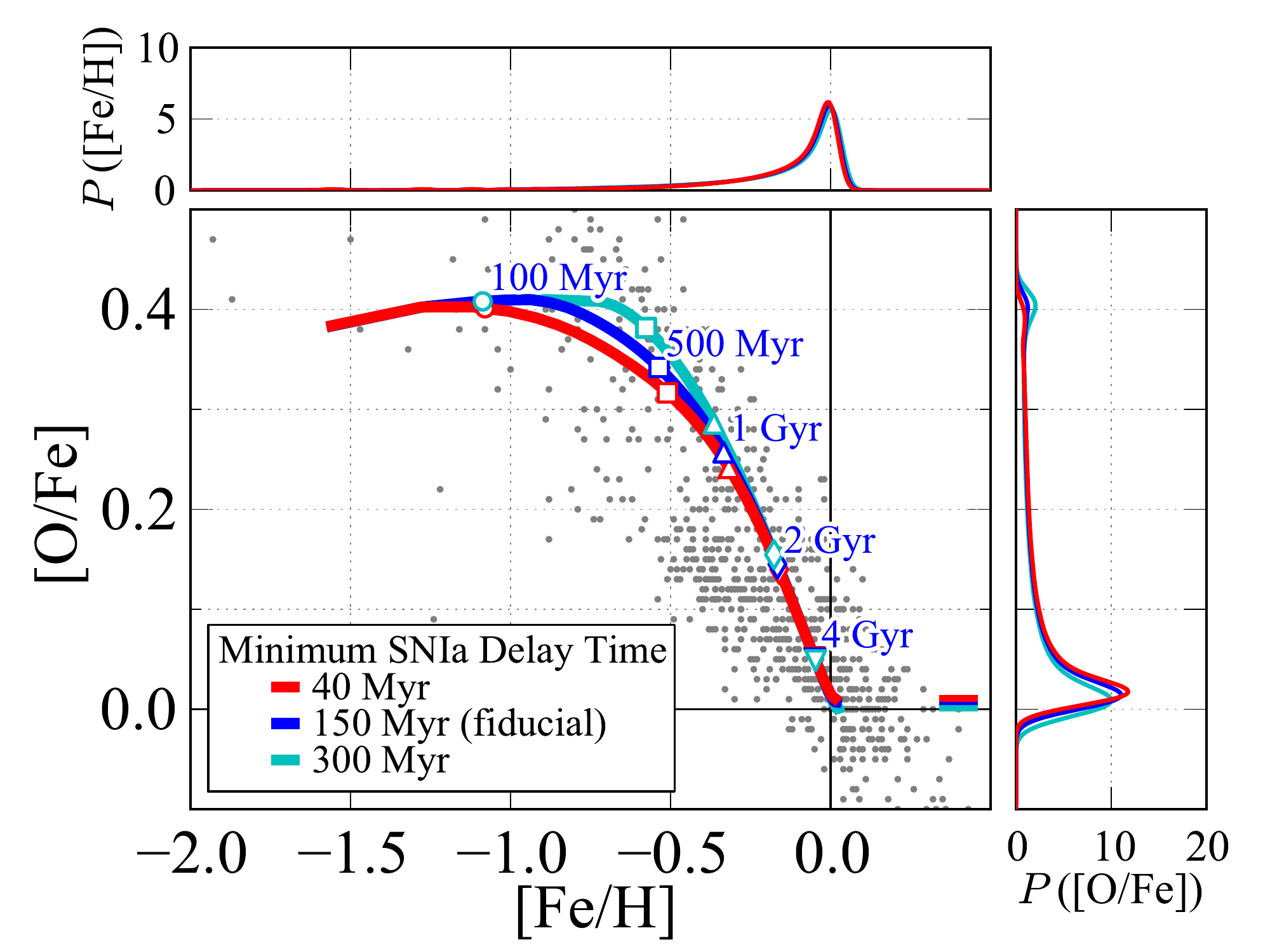}}
\caption{[O/Fe]--[Fe/H], MDFs, and [O/Fe]-DFs for variations in the minimum
SNIa delay time. Increasing the minimum SNIa delay time increases the [Fe/H] of
the knee and the number of stars on the high-$\alpha$ plateau.  Symbols and
data points are the same as in Figure \ref{fig:ofe_gas_flows}.}
\label{fig:ofe_min_snia_time}
\end{figure}

Figure \ref{fig:ofe_min_snia_time} shows the mean tracks in [O/Fe]--[Fe/H],
the [O/Fe]-DFs, and the MDFs for an exponential SNIa DTD with three different
minimum delay times: 40, 150 (fiducial value), and 300~Myr.  The fiducial
simulation turns over at [Fe/H]~$\sim -0.9$.  Increasing the minimum delay
time increases the [Fe/H] of the knee and makes the turnover sharper.  After
about 2~Gyr, the [O/Fe]--[Fe/H] tracks of the three simulations are very
similar.  The MDFs of the three simulations are almost identical, though the
relative number of stars on the high [O/Fe] plateau increases with the minimum
delay time, as can be seen in the high [O/Fe] peak of the [O/Fe]-DF.  The data
do not rule out any of these possible delay times.

\begin{figure*}
\centerline{\includegraphics[width=9cm]
{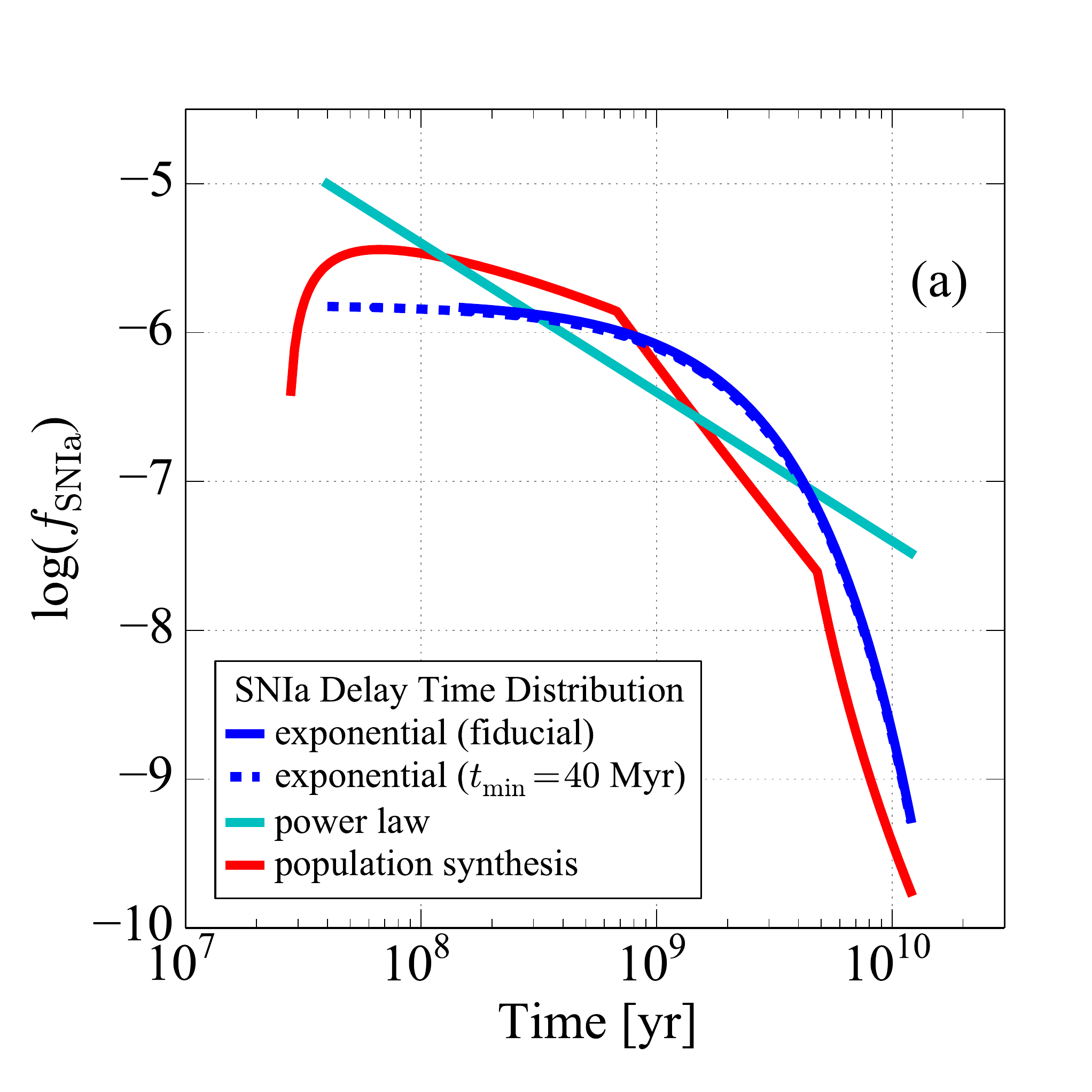}
\includegraphics[width=9cm]
{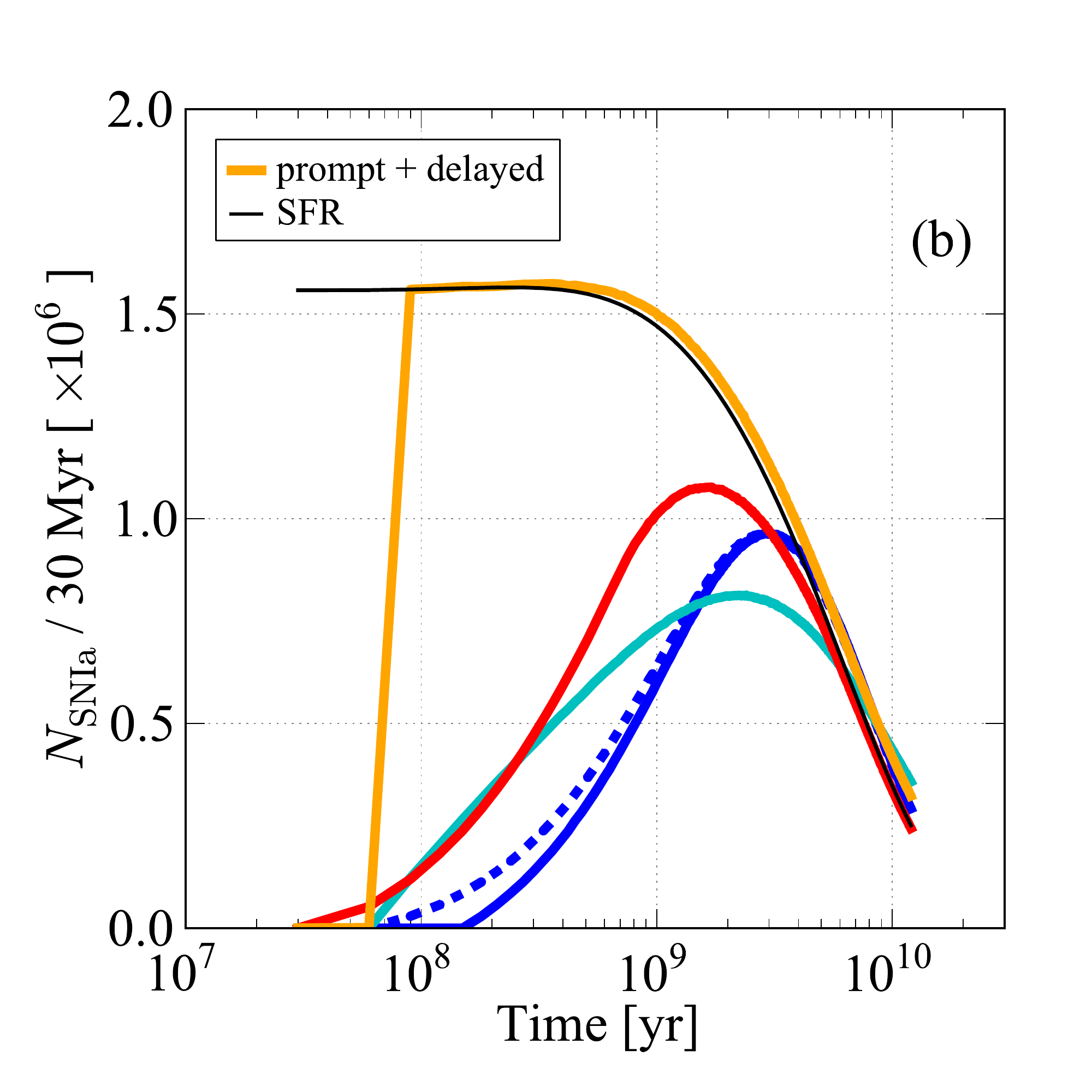}}

\caption{Panel (a): the exponential, power law, and population synthesis SNIa
DTDs (in units of $N_\mathrm{SNIa} \, \mathrm{yr}^{-1} \mathrm{M}_\odot^{-1}$)
for a single stellar population (born at $t=0$), with all distributions
normalized to produce 2.2 $\times 10^{-3}$ SNIa/\msun\ when integrated from 40
Myr to 10 Gyr.  Panel (b): number of SNIa per 30 Myr time step for simulations
with our fiducial SFH and different SNIa DTDs, now including the prompt +
delayed DTD as an orange line. The black curve shows the star formation rate
with arbitrary normalization.}
\label{fig:dtd}
\end{figure*}

We tested three additional functional forms of the SNIa DTD: a power law DTD,
a prompt + delayed DTD, and a theoretical population synthesis DTD, which are
compared in Figure \ref{fig:dtd}.  The power law DTD is motivated by recent
measurements of the SNIa rates and the SFHs of galaxies that have placed
interesting observational constraints on the SNIa DTD using a variety of
strategies \citep{totani2008, maoz2010a, maoz2010b, maoz2011, maoz2012b,
graur2011, maoz2012a}.  Encouragingly, the results from different techniques
agree that a $t^{-1}$ power law DTD provides a good fit to the data,
\begin{equation}
f_\mathrm{SNIa}(t) \propto \left(\frac{t}{\mathrm{yr}}\right)^{-1}.
\label{eqn:powerlaw_dtd}
\end{equation}
The evidence for a $t^{-1}$ power law DTD is strongest for $t$~=~1--10~Gyr
\citep{maoz2012a}, but the data are consistent with a $t^{-1}$ power law
continuing to shorter delay times with a minimum delay time of 40~Myr from the
birth of a stellar population, which is approximately the lifetime of the most
massive star that will produce a WD. We also consider an empirical prompt +
delayed DTD based on observations that the SNIa rate is correlated with SFR
tracers \citep{scannapieco2005, mannucci2006}.  In this scenario, the
``prompt'' SNIa are associated with young stellar populations, so their rate
scales with the SFR.  The ``delayed'' SNIa come from old stellar populations,
so their rate depends on stellar mass. The SNIa rate is
\begin{equation}
\frac{R_\mathrm{SNIa}(t)}{\mathrm{yr}^{-1}} = A
\left(\frac{M_\star}{M_\odot}\right) + B
\left(\frac{\mathrm{SFR}}{\mathrm{M}_\odot \, \mathrm{yr}^{-1}}\right),
\label{eqn:prompt_delayed_dtd}
\end{equation}
where $A$~=~4.4~$\times$~10$^{-14}$ and $B$~=~~2.6~$\times$~10$^{-3}$ are the
values from \citet{scannapieco2005} converted to the relevant units, and we have
implemented a 40 Myr minimum delay time for WD formation. \citet{maoz2012a}
point out that this scenario is similar to coarse time sampling of a $t^{-1}$
power law DTD, and the demarcation between the prompt and delayed components is
somewhat arbitrary. Finally, we implement the theoretical DTD from
\citet{greggio2005} (described in their Section 3.1) based on stellar population
synthesis models of single degenerate systems that reach the Chandrasekhar
limit.

Figure \ref{fig:dtd}a shows the DTDs for the exponential, power law, and
population synthesis DTDs for a stellar population born at $t=0$, normalized to
the observed time-integrated (from 40~Myr to 10~Gyr) SNIa rate of
$N_\mathrm{Ia} / M_\star = 2.2 \times 10^{-3} \mathrm{M}_\odot^{-1}$
(\citealt{maoz2012a}; though see \citealt{cote2016a} for a discussion of the
significant element-dependent impact caused by uncertainty in this
normalization). The power law DTD produces more SNIa at early times
($\lesssim$~200~Myr) and late times ($\gtrsim$~4~Gyr) than the exponential DTD
but fewer in between. The population synthesis DTD has a roughly similar shape
to the exponential DTD, except shifted to shorter delay times.  The breaks in
the population synthesis DTD are due to the requirement of a carbon--oxygen WD
and that the system must achieve the Chandrasekhar mass.

Figure \ref{fig:dtd}b shows the number of SNIa per 30~Myr time step for
simulations (with continuous star formation) with different DTDs but the same
SFH.  The numbers of SNIa per time step in the exponential, power law, and
population synthesis DTD simulations increase up to a peak at 1.5--3~Gyr
followed by a decline at late times.  The prompt + delayed scenario produces a
very high initial rate because it is tied to the SFR, which is high at early
times.  We adopted the normalization of \citet{scannapieco2005}, though these
numbers were determined for galaxies in equilibrium.  The large number of very
early SNIa has a dramatic effect in the [O/Fe]--[Fe/H] diagram. After 12 Gyr,
the prompt + delayed scenario produces 30\% more SNIa than the fiducial DTD due
to the normalization. The exponential DTD with $t_{\rm min}$ = 40~Myr and the
population synthesis scenario have nearly the same number of SNIa as the
fiducial DTD after 12 Gyr. The power law DTD has 5\% fewer SNIa after 12 Gyr
than the fiducial DTD even though they were normalized to the same value because
of the different relative number of short versus long delay SNIa. Another way to
understand this point is that there are more SNIa with delays of 4--10~Gyr for
the power law DTD than the fiducial DTD (see Figure \ref{fig:dtd}a). For
populations between 4--10~Gyr, more of their SNIa will have exploded by the end
of the simulation for the fiducial DTD than the power law DTD.

\begin{figure}
\centerline{
\includegraphics[width=9cm]
{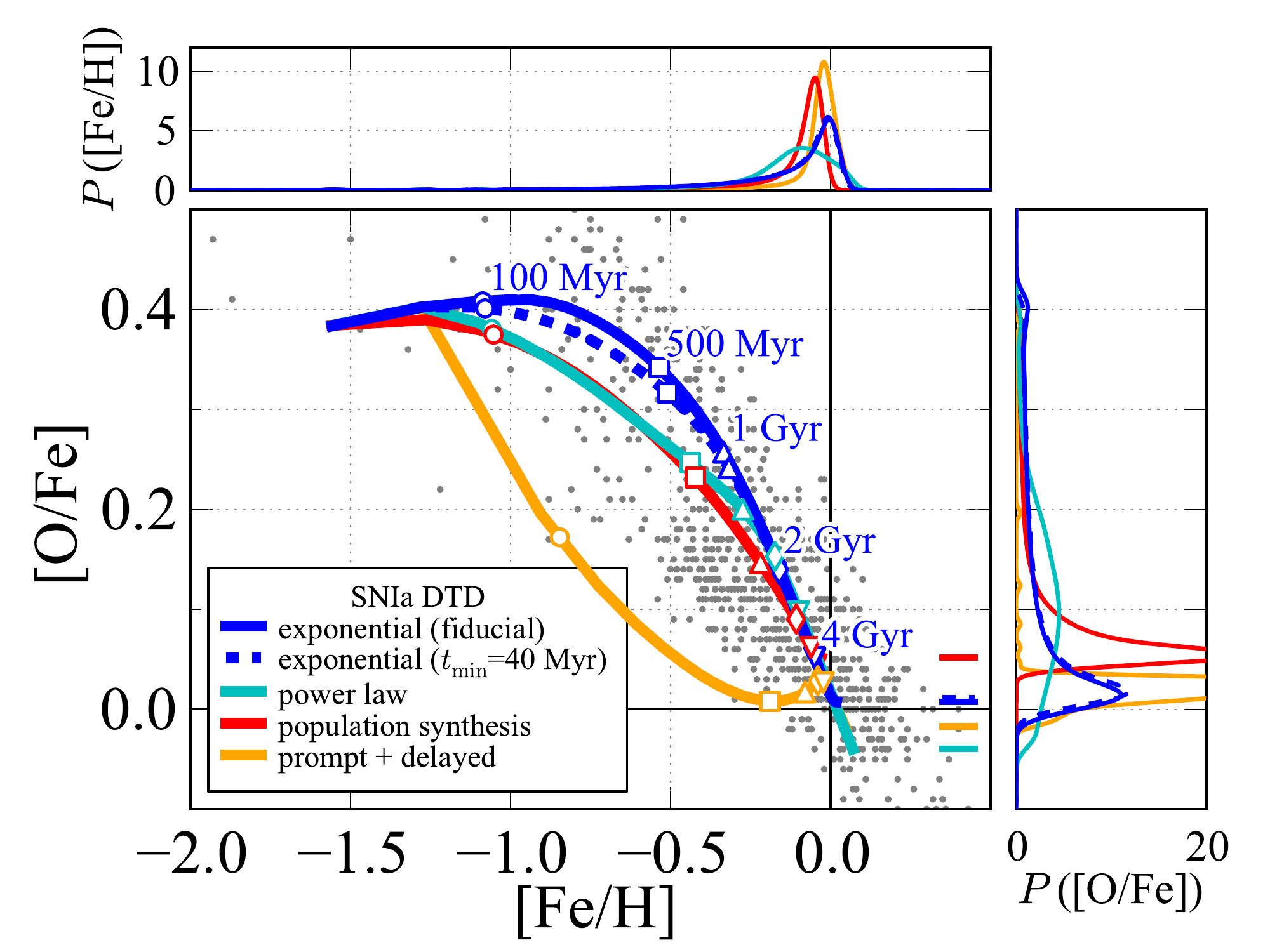}}

\caption{[O/Fe]--[Fe/H], MDFs, and [O/Fe]-DFs for various SNIa DTDs (see Figure
\ref{fig:dtd}).  The exponential, power law, and population synthesis DTD
simulations are broadly consistent with the data in [O/Fe]--[Fe/H].  The power
law DTD simulation extends to higher metallicity and lower [O/Fe] than the
fiducial simulation. This simulation has a broad MDF and even broader
[O/Fe]-DF, in stark contrast to the MDF and [O/Fe]-DF of the fiducial
simulation, which are more peaked and have a sharp cutoff.  The population
synthesis DTD simulation reaches an equilibrium at lower metallicity and higher
[O/Fe], with more peaked MDF and [O/Fe]-DFs than the fiducial simulation. The
prompt + delayed DTD simulation does not fit the [O/Fe]--[Fe/H] data because it
produces too much iron at early times, driving down [O/Fe].  This simulation
produces a sharp left edge to the [O/Fe]-DF, which is not seen in the
[O/Fe]-DFs of the other simulations or in observed [O/Fe]-DF.  Symbols and data
points are the same as in Figure \ref{fig:ofe_gas_flows}.}

\label{fig:ofe_snia_dtd}
\end{figure}

Figure \ref{fig:ofe_snia_dtd} shows the mean tracks in [O/Fe]--[Fe/H], the
[O/Fe]-DFs, and the MDFs of simulations with various SNIa DTDs.  We also
include an exponential DTD with \tmin~=~40~Myr (blue dashed line) to more
closely match the other DTDs.  The power law DTD simulation turns over at a
lower metallicity than the exponential DTD, but it reaches the same [Fe/H] and
[O/Fe] at about 2~Gyr.  It then continues along the same track at a slower pace
but ends at a higher equilibrium [Fe/H] and lower [O/Fe] than the exponential
DTD.  Its [O/Fe]-DF and MDF are significantly broader than those of the
fiducial simulation, especially the [O/Fe]-DF, and do not show a sharp cutoff.
The power law DTD does not reach an equilibrium metallicity or [O/Fe] because
of the large number of SNIa at late times (see also \citealt{kubryk2015a}). The
population synthesis DTD has a similar track to the power law DTD over the
first 500~Myr.  It reaches the same track as the exponential DTD but finishes
with a lower equilibrium [Fe/H] and a higher [O/Fe] (near the 4~Gyr symbol).
Its [O/Fe]-DF and MDF are more peaked than the fiducial simulation's.  The
exponential, power law, and population synthesis scenario DTD are all
consistent with the [O/Fe]--[Fe/H] data, especially if other parameters, such
as the inflow timescale or the SNIa normalization, are allowed to vary to
compensate for the different equilibrium [Fe/H] and [O/Fe]. On the other hand,
the prompt + delayed DTD creates a sudden drop in [O/Fe] before leveling off, a
feature that is not present in the data.  This behavior is caused by the large
number of prompt SNIa (see Figure \ref{fig:dtd}b). Its [O/Fe]-DF and MDF are
strongly peaked, and its [O/Fe]-DF has a sharp left edge, which is unique among
the models and inconsistent with the data. \citet{matteucci2009} found that
using an alternative parametrization of the prompt + delayed DTD from
\citet{mannucci2006} is also inconsistent with observations although
the drop off is less extreme and can be brought into agreement with data by
reducing the fraction of prompt SNIa from 50\% to 30\%. We conclude that the
``prompt + delayed'' description of the SNIa rate must be only an approximation
to a more smoothly declining DTD.

\subsection{Warm ISM}
\label{sec:warm_ism}

\begin{figure}
\centerline{
\includegraphics[width=9cm]
{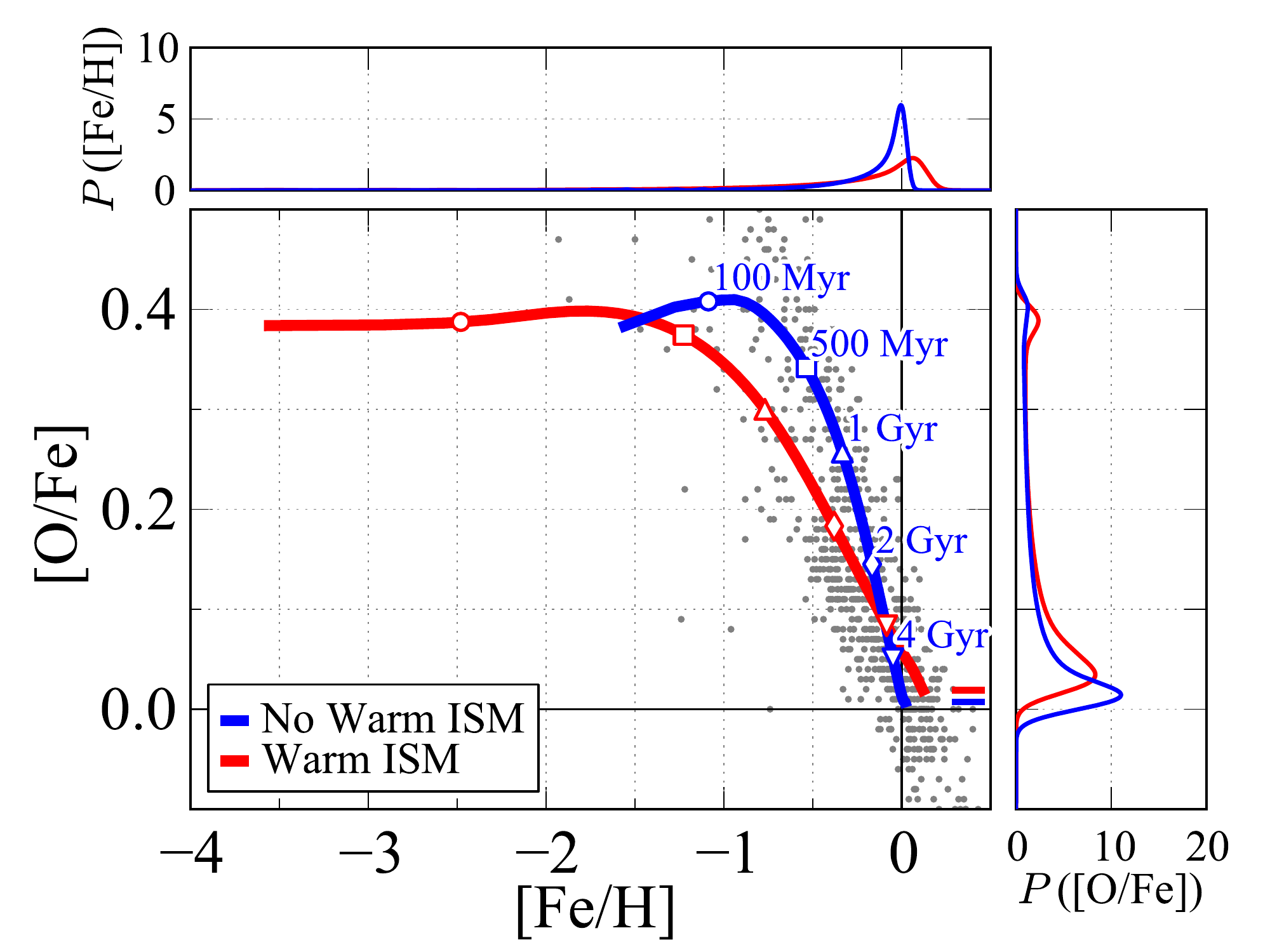}}
\caption{[O/Fe]--[Fe/H], MDFs, and [O/Fe]-DFs for a simulations with and
without (fiducial simulation) a warm phase of the ISM that does not form stars.
In the warm ISM simulation, 99\% of yields from all sources are injected into
the warm ISM, and this material returns to the cold ISM on a timescale of
1.2~Gyr, where it can be reincorporated into stars.  The other 1\% of the
yields go directly into the cold ISM.  The warm ISM simulation enriches more
slowly, produces much lower metallicity stars, and has a lower [Fe/H] of the
knee than the fiducial simulation.  Its main peaks in the MDF and [O/Fe]-DF are
broader and shifted to higher metallicity and [O/Fe].  It also shows a larger
high-$\alpha$ peak in the [O/Fe]-DF.  Symbols and data points are the same as
in Figure \ref{fig:ofe_gas_flows}.}

\label{fig:ofe_warm_ism}
\end{figure}

Much of the stellar ejecta from CCSN, SNIa, and AGB stars is not immediately
returned to the cold ISM (see, e.g., \citealt{walch2015}).  Instead, it is
injected into the warm ($T \gtrsim 10^4$~K) ISM that does not form stars. Over
time, gas in the warm ISM cools into the cold ISM and can get reincorporated
into future generations of stars. While we have not included a warm ISM in the
fiducial simulation, we explore its effects on chemical evolution in Figure
\ref{fig:ofe_warm_ism}.

For the warm ISM simulation, we follow the scheme from \citet{schoenrich2009a}
and inject 99\% of stellar yields from all sources (CCSN, SNIa, and AGB
stars) into the warm ISM with the remainder of the yields entering the cold
ISM directly.  In each time step, a fraction of the gas
(d$t$/$t_\mathrm{cool}$) cools out of the warm ISM and joins the cold ISM.  We
adopt a gas cooling timescale, $t_\mathrm{cool}$, of 1.2 Gyr.  The gas in the
warm ISM is not ejected in outflows.

Including a warm ISM delays the return of yields from all sources, in contrast
to changing the minimum time delay for SNIa (see Section \ref{sec:snia_dtd}),
which only affects one yield source.  The warm ISM simulation shows many signs
of slower enrichment.  It has much lower metallicity at early times, turns
over at a lower [Fe/H], produces more high-$\alpha$ stars, and has broader
main peaks to the MDF and [O/Fe]-DF than the fiducial simulation.  Its
equilibrium [O/Fe] and [Fe/H] are slightly higher than those of the fiducial
simulation because the warm ISM prevents the ejection of a large fraction of
the enrichment products, which is especially important for elements
synthesized early, such as oxygen in CCSN.

\subsection{Relation to Analytic Results}
\label{sec:relation_to_analytic}

\citetalias{weinberg2016a} derive analytic solutions for oxygen and iron
evolution by adopting two simplifying assumptions beyond those used here:
instantaneous recycling for material from AGB stars (but not for SNIa
enrichment), and supernova yields that are independent of metallicity.  They
assume an exponential or linear-exponential star formation history and an
exponential DTD for SNIa (or a sum of exponentials, which can be used to
approximate the $t^{-1}$ power-law).  Their analytic results capture many of
the key features of Figure~\ref{fig:ofe_mdf_multi}, including the dependence of
equilibrium abundances and abundance ratios on $\eta$ and SFE, the strong
impact of SFE on the location of the knee in evolutionary tracks, the influence
of the star formation history (which is tied to the inflow rate), and the small
impact of inflow metallicity. The analytic models also capture the dependence
on the minimum delay time and shape of the DTD seen in
Figures~\ref{fig:ofe_min_snia_time} and \ref{fig:ofe_snia_dtd}. These models do
not incorporate a warm ISM, though they could be adapted to do so by
introducing an exponential ``delay time distribution'' for return of CCSN
elements.


\section{Yields and Multi-element Abundances}
\label{sec:yields_multielement}

\subsection{Yields}
\label{sec:yields}

The adopted nucleosynthesis yields remain one of the largest sources of
uncertainty in chemical evolution models \citep{timmes1995, gibson1997,
portinari1998, romano2010}.  The yields of $\alpha$-elements from CCSN and
iron-peak elements from SNIa are relatively robust, but the yields of elements
that are sensitive to metallicity and elements produced in explosive burning
phases or by neutron-capture are much more uncertain.  Here we discuss our
choice of yields and how they affect the chemical evolution model. We note that
we use the term "yields" in the sense of solar masses of element $X$ returned,
as opposed to the classically defined sense of a mass fraction (e.g.,
\citealt{tinsley1980}).

\subsubsection{CCSN}
\label{sec:ccsn_yields}

CCSN produce the majority of elements heavier than helium, including
$\alpha$-elements (C, O, Mg, Si, S, Ca, and Ti), but there is not yet a
consensus on the yield of each element as a function of stellar mass and
metallicity. The yields of $\alpha$-elements produced in hydrostatic burning
phases (C, O, and Mg) increase nearly monotonically with stellar mass and show
a weak metallicity dependence (with the exception of C).  The yields of
$\alpha$-elements produced at least partially in explosive burning phases (Si,
S, Ca, and Ti) have a more complicated relation to stellar mass, which changes
with metallicity, though the IMF-integrated yields show a weak metallicity
dependence for these elements.  On the other hand, the yields of odd-$Z$
elements (odd number of protons), such as sodium and aluminum, increase
sharply with metallicity.  Higher initial abundances of carbon and oxygen lead
to larger neutron excesses and enhanced production of odd-$Z$ elements
\citep{truran1971a}.

We adopted the solar metallicity CCSN yields of \citet{limongi2006} and the
intermediate (sub-solar but non-zero) metallicity CCSN yields of
\citet{chieffi2004}\footnote{Both are available on the ORFEO database at
http://www.iasf-roma.inaf.it/orfeo/public\_html/~.}.  The \citet{limongi2006}
and \citet{chieffi2004} yields cover 11--120 \msun\ and 13--35 \msun,
respectively.  We created a coarse grid of net yields spanning 8--100~\msun\ at
the metallicities of the models ($Z$ = 10$^{-6}$, 10$^{-4}$, 10$^{-3}$,
6$\times$10$^{-3}$, and 2$\times$10$^{-2}$) by adopting the yields of the
closest mass model at fixed metallicity for masses that are higher or lower than
those covered by these models (but between 8--100~\msun).  Then we linearly
interpolated the net yields to create a uniform grid of net yields finely
sampled in mass and log metallicity.  Zero metallicity CCSN yields (especially
for nitrogen) are highly uncertain, because of the poorly constrained amount of
rotational mixing that occurs in zero metallicity stars, so we adopt the lowest
metallicity \citet{chieffi2004} yields ($Z$ = 10$^{-6}$) for stars with yet
lower metallicity.  We note that \citet{cote2016b} found that the choice of
including or excluding zero metallicity yields had a dramatic impact on
abundance trends at [Fe/H]~$\lesssim -2$, though the lowest non-zero metallicity
of their CCSN yield grid was higher than ours ($Z~=~1.53 \times 10^{-4}$).  For
super-solar metallicities, we used the solar metallicity \citet{limongi2006}
yields.  We linearly extrapolated the remnant mass to higher and lower masses at
fixed metallicity.  For more details on how we assigned yields outside of the
calculated grid, see Appendix \ref{sec:yield_grid_extension}.

\citet{chieffi2004} and \citet{limongi2006} report yields as a function of the
mass cut in CCSN, which is the dividing line between material that falls back
onto the neutron star and material that is ejected in the explosion.  Since
the mass cut is deep in the star, its location affects the yields of Fe-peak
elements but not those of elements created in the outer layers of the star
like oxygen.  As shown in Figure \ref{fig:ofe_mdf_imf}, a lower mass cut
decreases the [O/Fe] of the plateau, increases [Fe/H], and increases the
[Fe/H] of the knee.  This leads to a significant degeneracy between the mass
cut, the SFE (Figure \ref{fig:ofe_mdf_sfe}), and the SNIa DTD (Figure
\ref{fig:ofe_snia_dtd}). The appropriate mass cut is uncertain at the factor
of 2--3 level and might itself change as a function of stellar mass
\citep{limongi2003}.  We adopt a mass cut such that 0.1~\msun\ of $^{56}$Ni is
produced in all CCSN.  This mass cut reproduces the [O/Fe] abundances for
stars with [Fe/H]~$<-$1 (see Figure \ref{fig:ofe_mdf_imf}) and is the same
mass cut adopted by \citet{chieffi2004} and \citet{limongi2006}.  (However,
\citealt{schoenrich2009a} chose a mass cut of 0.05~\msun\ $^{56}$Ni to match
the [Ca/Fe] abundances of \citealt{lai2008}.) While the [O/Fe] plateau seems
like a powerful criterion for selecting the mass cut, there is observational
uncertainty in the measured value at the level of $\sim$0.2~dex.  For example,
the [O/Fe] plateau at low [Fe/H] shifted from $\sim$0.6 in \citet{ramirez2007}
to $\sim$0.45 in \citet{ramirez2013} using much of the same data due to the
adoption of the improved calibration of the temperature scale from
\citet{casagrande2010}.

The yields for neutron-capture elements are not well-predicted by theoretical
yield calculations.  We adopt the ``empirical'' $r$-process yields for Eu and Ba
from \citet{cescutti2006}, which were determined by fitting a chemical evolution
model to Eu and Ba abundances with different yields.  The \citet{cescutti2006}
yields do not depend on metallicity.  We report the evolution of Eu abundances
even though the application of ``empirical'' yields is partially circular (we
use a different chemical evolution model). We do not show the evolution of Ba
abundances because Ba $s$-process nucleosynthesis is currently challenging and
model-dependent but would be interesting to investigate in future work.  The
$s$-process contributes significantly to Ba abundances but not Eu abundances
\citep{burbidge1957}.

\subsubsection{SNIa}
\label{sec:snia_yields}

SNIa synthesize at least half of the iron and iron-peak elements in the Galaxy.
They are rarer than CCSN by a factor of $\sim$4--5 (\citealt{mannucci2005,
li2011, maoz2011}; c.f., \citealt{cote2016a} for implications of the uncertainty
in the total number of SNIa), but they make $\sim$0.6--0.7 \msun\ of $^{56}$Fe
whereas CCSN produce $\sim$0.05--0.1 \msun\ of $^{56}$Fe. Depending on the
adopted yields, SNIa also can create $\alpha$-elements like oxygen, calcium, and
titanium, but in quantities that are insignificant relative to the CCSN
contribution to the galactic budget of these elements (see Figures
\ref{fig:ofe_river} and \ref{fig:river_all}).

We adopted the SNIa yields of the W70 model from \citet{iwamoto1999}.  Unlike
CCSN yields, chemical evolution is relatively insensitive to the specific
choice of SNIa yield model because most models produce similar amounts of iron
and iron-peak elements.  This is also borne out by observations of SNIa light
curves, which are powered by the radioactive decay of $^{56}$Ni (eventually
into $^{56}$Fe) and constrain the amount of iron produced to be $\sim$0.6
\msun\ on average.  The SNIa models do differ in the quantities of other
elements synthesized, but SNIa are not the dominant contributors of these
elements.  We assume that SNIa yields are independent of the mass and
metallicity of the progenitor stars, which is probably a good assumption given
the arguments above.

\subsubsection{AGB Stars}
\label{sec:agb_yields}

AGB stars dominate the production of nitrogen and $s$-process elements; are
important producers of carbon; and return large amounts of hydrogen and helium
to the Galaxy (see Figure \ref{fig:river_all}).  During hydrogen-burning, they
convert carbon and oxygen into nitrogen, which is the bottleneck of the CNO
cycle.  They also synthesize carbon and oxygen, but most of it remains in the
core that forms the white dwarf, and the net yield for these elements can even
be negative at certain stellar masses and metallicities.  AGB stars generate
$s$-process elements in the thermal pulses at the end of their lives, but
modeling these pulsations is complex, so predicting $s$-process yields is
challenging.  Thus, we do not try to model $s$-process elements, though they
are important chemical tracers of enrichment from low mass stars.

We used the AGB yields from \citet{karakas2010} that span 1--6.5 \msun\ and
$10^{-4}$--1.0 \zsun.  We linearly interpolated between the grid points in mass
and log metallicity, and linearly extrapolated the yields up to 8 \msun. CCSN
consume hydrogen but have positive or zero yields for other elements.  However,
AGB stars can either destroy or produce carbon and oxygen, depending on
metallicity and mass, so interpolating and extrapolating the yields of these
elements can be uncertain.

\subsection{Multi-element Abundances}
\label{sec:multielement}

\begin{figure*}
\centerline{\includegraphics[width=19cm]
{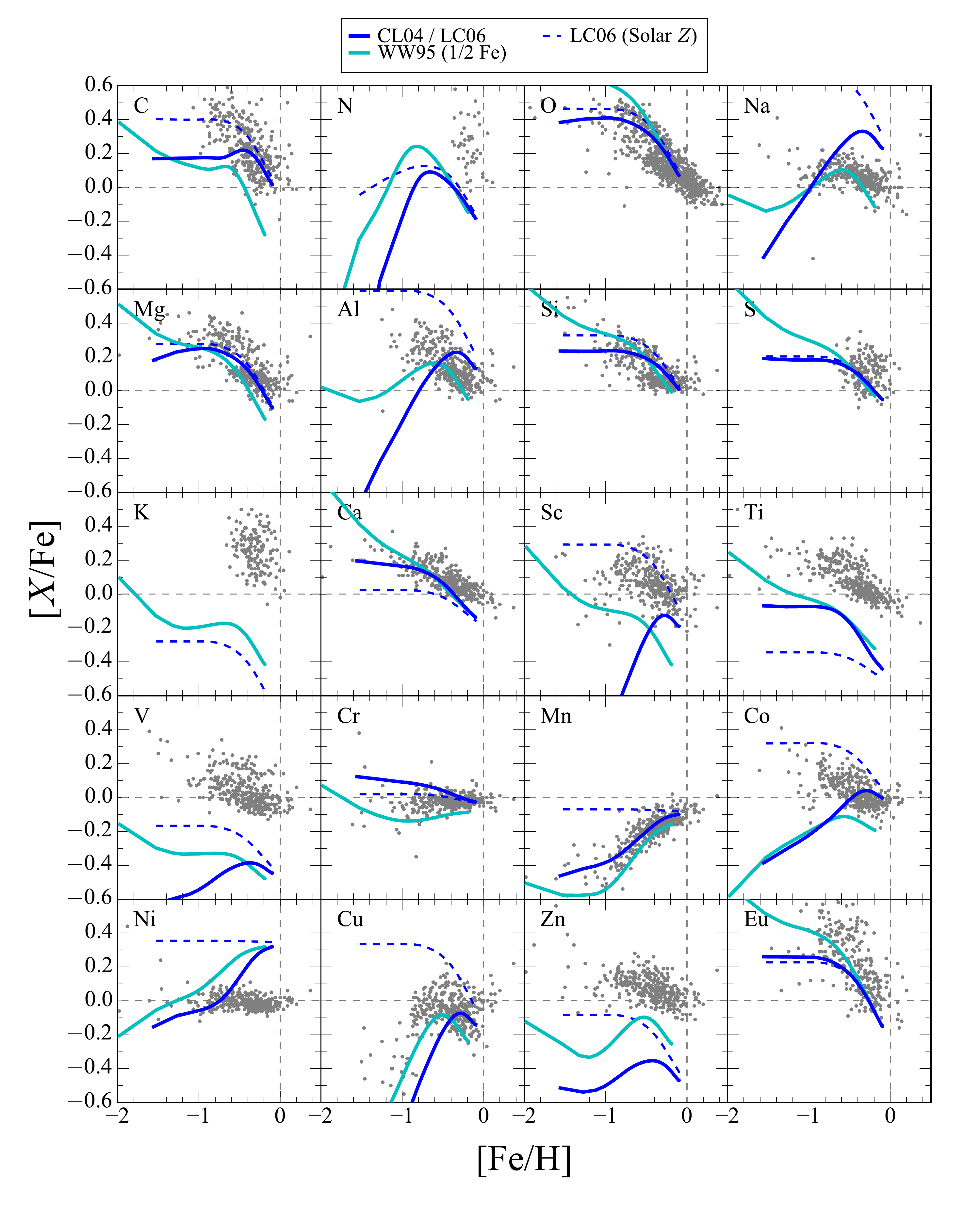}}
\caption{[$X$/Fe]--[Fe/H] for \citetalias{chieffi2004} and
\citetalias{limongi2006} yields (solid blue), \citetalias{woosley1995} yields
(cyan), and \citetalias{limongi2006} yields computed at solar metallicity
(dashed blue). A constant SFR is assumed in all cases. The gray points show
data from \citet{ramirez2013} (oxygen only) and from \citet{reddy2003,
reddy2006} (all other elements).}
\label{fig:xfe}
\end{figure*}

In Figure \ref{fig:xfe}, we compare the abundances from three simulations with
different CCSN yields: \begin{itemize} \item \citet{chieffi2004} and
\citet{limongi2006} yields (the same as the fiducial simulation; hereafter
\citetalias{chieffi2004}), \item \citet[hereafter WW95]{woosley1995} yields, and
\item \citet[hereafter LC06]{limongi2006} solar metallicity only yields.
\end{itemize} We reduced the \citetalias{woosley1995} yields of the iron-peak
elements (Cr, Mn, Fe, Co, Ni, Cu, and Zn) by a factor of two as recommended by
\citet{timmes1995} based on the normalization of [$X$/Fe]--[Fe/H] tracks. When
we used the original \citetalias{woosley1995} yields, we found the expected
$\sim$0.3~dex offset in [$X$/Fe] at low metallicity for non-iron-peak elements,
though these yields were able to increase the final [Fe/H] by 0.1~dex to match
the \citetalias{chieffi2004} simulation.  \citet{timmes1995} note that this
reduction is consistent with the uncertainty in the explosion energy that
affects the mass cut and hence the yields of iron-peak elements. The
\citetalias{woosley1995} iron yields have a metallicity dependence at low
metallicity, which causes the rise in [O/Fe] as [Fe/H] decreases below $-1$. Our
\citetalias{chieffi2004} models, on the other hand, have metallicity independent
iron yields by construction, so they produce nearly flat plateaus at low
metallicity for elements such as O, Mg, Si, and S. Since the
metallicity-dependence of the yields is a significant physical uncertainty, we
also show the \citetalias{limongi2006} solar metallicity yields applied at all
metallicities.  The [$X$/Fe] trends for this simulation are often higher than
the \citetalias{chieffi2004} simulation, indicating that the solar metallicity
CCSN yields are higher than the sub-solar metallicity CCSN yields (with the
notable exceptions of Ca and Ti).  Ni and Mn have similar SNIa and solar
metallicity CCSN yields, which produces their flat abundance trends.   Elements
with significant secondary production (e.g., N) show large discrepancies between
the \citetalias{chieffi2004} and \citetalias{limongi2006} trends.  The Eu yields
are metallicity independent by construction, so the differences between the
\citetalias{chieffi2004} and the solar metallicity CCSN yields show the slight
metallicity dependence of the \citetalias{chieffi2004} Fe yields.

\begin{figure*}
\centerline{\includegraphics[width=19cm]
{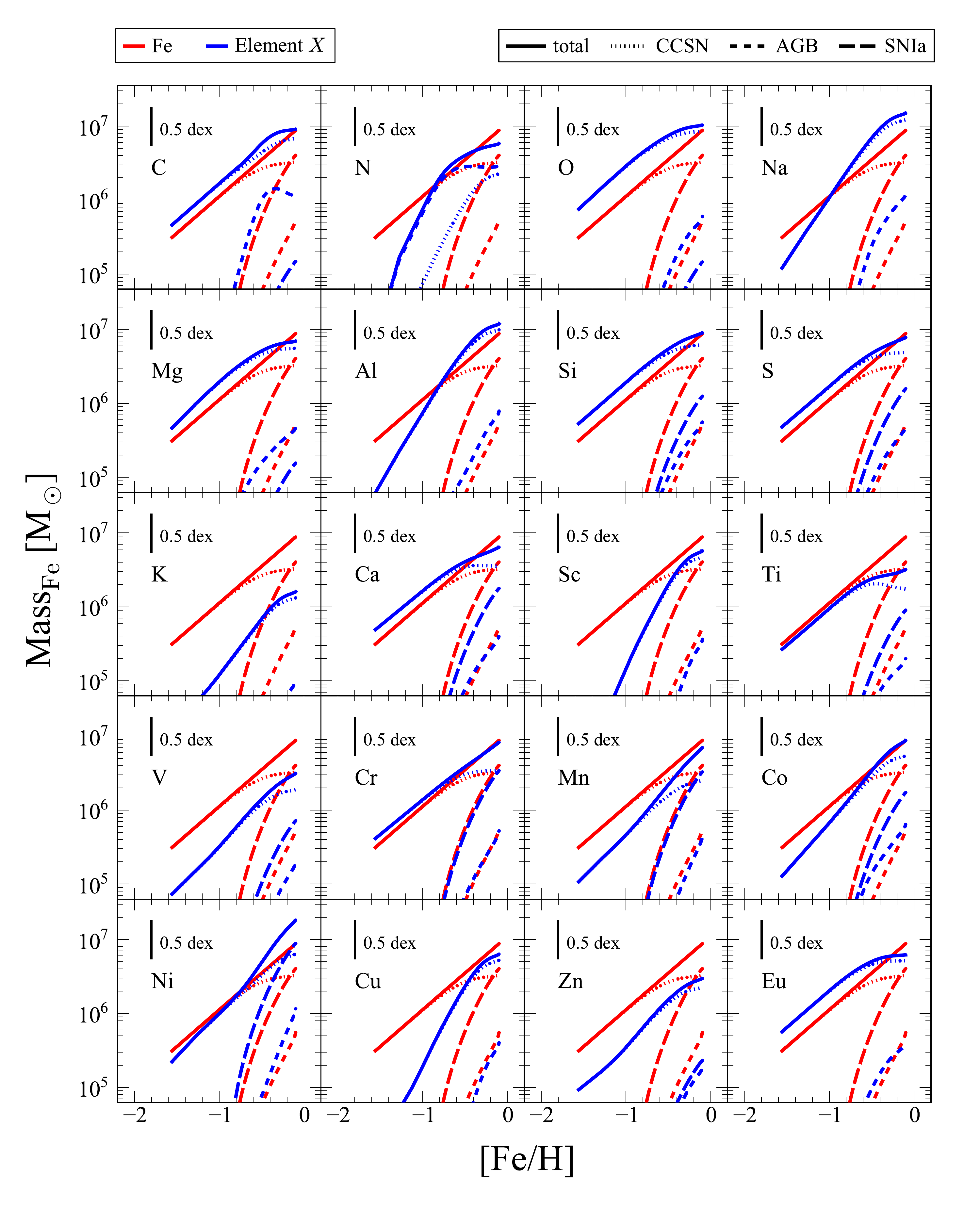}}

\caption{The mass of various elements in the gas-phase as a function
of [Fe/H] for the constant SFR simulation with \citetalias{chieffi2004} yields.
As in Figure~\ref{fig:ofe_river}, solid curves show total mass and dotted,
long-dashed, and short-dashed curves show CCSN, SNIa, and AGB contributions,
respectively.  Red curves show results for iron and are the same in all panels.
The total ISM mass is $9\times 9^{10} M_\odot$ at all times.}

\label{fig:river_all}
\end{figure*}

These simulations have a constant SFR but otherwise adopt the parameters of
the fiducial simulation, including the SNIa and AGB yields described above.
Relative to the fiducial simulation, the constant SFR simulation has a lower
equilibrium abundance and higher equilibrium [$X$/Fe] values (except Mn and
Ni).  Below we compare the predicted abundances of twenty elements for the two
yield sets and discuss the agreement or disagreement with data from
\citet{reddy2003, reddy2006} and \citet{ramirez2013} (oxygen only). Figure
\ref{fig:river_all} shows the mass enrichment histories for these elements in
the same format as Figure \ref{fig:ofe_river}.  Elements with
metallicity-dependent yields have a steeper blue solid curve in Figure
\ref{fig:river_all} than the red solid curve (iron mass).  We discuss results
element-by-element with reference to both figures. Our physical descriptions of
the dominant enrichment sources and their metallicity dependence are based on
our adopted yield models, which may or may not reflect reality.

\paragraph{Carbon}

Carbon enrichment, like oxygen, is dominated by massive stars.  With our adopted
yields, the contribution of AGB stars is sub-dominant but non-negligible above
[Fe/H]~$\approx -$0.6 (see Figure \ref{fig:river_all}). The [C/Fe] data follow
an $\alpha$-element trend except that the [C/Fe] turnover appears displaced
towards somewhat higher [Fe/H]. The predicted [C/Fe] actually increases at
[Fe/H]~=~$-$0.7, where carbon production by AGB stars peaks, which may explain
this offset.  At higher [Fe/H], AGB stars become net consumers of carbon.  The
CCSN C yields are mildly metallicity-dependent above $Z \sim 1/3 \, Z_\odot$;
however, they increase with $Z$ for the \citetalias{chieffi2004} yields but
decrease for the \citetalias{woosley1995} yields.  The \citetalias{chieffi2004}
simulation matches the observational data well for [Fe/H]~$>-$0.4, but it
predicts too low an abundance for the plateau (c.f., the observed [C/Fe] trend
from \citealt{bensby2006}, which the  \citetalias{chieffi2004} simulation agrees
with fairly well). The \citetalias{woosley1995} simulation produces a similarly
shaped trend to the data that is systematically too low by $\sim$0.3~dex.
Interestingly, the \citetalias{limongi2006} solar metallicity yields reproduce
the data very well.

\paragraph{Nitrogen}

Most nitrogen production occurs in the CNO cycle, where carbon and oxygen get
converted into nitrogen.  Figure \ref{fig:river_all} shows that AGB stars
dominate the production of nitrogen, especially for [Fe/H]~$<-$1, so it is not
surprising that the \citetalias{chieffi2004} and \citetalias{woosley1995}
simulations result in similarly shaped tracks in [N/Fe]--[Fe/H].  The [N/Fe] of
the \citetalias{chieffi2004} and \citetalias{woosley1995} simulations increases
quickly from low metallicity to [Fe/H]~$\sim -0.6$ and $-$0.8, respectively,
which reflects the secondary nature of nitrogen (i.e., its yield increases with
the initial carbon and oxygen abundances).  This metallicity dependent yield is
also evident in the steepness of the blue dashed curve in Figure
\ref{fig:river_all}.  It then declines as the nitrogen production from AGB stars
is offset by losses to star formation and outflows while iron production from
SNIa continues to increase (Figure \ref{fig:river_all}).  Both simulations
underpredict [N/Fe] by $\sim$0.5~dex over the limited range of the data $-0.4
\lesssim \mathrm{[Fe/H]} \lesssim 0.1$.

Massive stars, (i.e., ones that will explode as CCSN) could be an important
source of nitrogen at zero and low metallicities. However, the uncertainty in
the nitrogen yield of zero metallicity massive stars is believed to be at least
an order of magnitude \citep{heger2010}.  These stars are born with low
abundances of carbon and oxygen (relative to hydrogen), but rotational mixing
could transport freshly synthesized carbon and oxygen from helium-burning
layers to hydrogen-burning layers, making nitrogen production extremely
sensitive to the strength of mixing.  Nitrogen abundances are also difficult to
measure due to the weakness of the N{\sc i} lines and their strong dependence
on the measured effective temperature \citep{reddy2003}.  The [N/Fe]--[Fe/H]
trend from the \citet{reddy2003} data implies a super-solar [N/Fe] at solar
metallicity, which suggests that either the Sun has a lower than typical [N/Fe]
or, more likely, that the normalization of the \citet{reddy2003} data is too
high. Nonetheless, if observational estimates are approximately correct, then
either the AGB or CCSN yields (or both) must be substantially in error.

\paragraph{Oxygen}

Oxygen is produced almost exclusively by CCSN (see Figure \ref{fig:river_all}).
Its yield has weak dependence on metallicity (see \citetalias{limongi2006}
curve) but a strong dependence on stellar mass, as can be seen from the effects
of changing the slope or cutoff of the high mass end of the IMF in Figure
\ref{fig:ofe_mdf_imf}.  The parameters of the fiducial simulation were chosen to
match the data in [O/Fe]--[Fe/H] and reach equilibrium at solar oxygen abundance
and metallicity, so the \citetalias{chieffi2004} simulation fits the data by
construction. (While our simulation in Figures~\ref{fig:xfe}
and~\ref{fig:river_all} assumes a constant SFR instead of exponentially
declining infall, this has little impact on the shape of tracks as shown
previously in Figure~\ref{fig:ofe_mdf_inflow_rate}.) This model was not intended
to fit the \citet{ramirez2013} data above solar metallicity, as these stars were
likely born interior to the solar annulus in the metal-rich inner disk and
migrated to the solar neighborhood. The \citetalias{woosley1995} simulation fits
the data well for [Fe/H]~$>-$0.4, though it overpredicts [O/Fe] at lower
metallicities ([O/Fe] declines from 0.85 to 0.6 from [Fe/H]~=~$-$2.0 to $-$1.0).
Our model using the \citetalias{woosley1995} yields produces a 0.2--0.3~dex
higher [O/Fe] plateau than \citet{timmes1995} found with the same yields. While
there are many differences between the two models, this discrepancy is almost
certainly due to the fact that their model does not include yields for stars
with M~$>$~40~\msun unlike our model (see Appendix \ref{sec:yield_grid_extension}
for more details).  The differences between the \citetalias{woosley1995} and
\citetalias{chieffi2004} simulations at low metallicity are partially due to the
metallicity dependence of the \citetalias{woosley1995} iron yields, whereas the
\citetalias{chieffi2004} iron yields are metallicity-independent by design since
we chose a fixed mass cut. The range of plausible mass cuts and iron yields for
CCSN spans the range of uncertainty in the oxygen abundance measurements, which
are challenging, even for the Sun \citep{caffau2011}.  In particular, abundances
determined from the strong O{\sc i} triplet at 777~nm are sensitive to the
adopted temperature scale and require corrections from the typical assumption of
local thermodynamic equilibrium (LTE), adjustments of 0.1--0.5~dex depending on
gravity and temperature \citep{ramirez2013}.

\paragraph{Sodium}

Nearly all sodium is synthesized in CCSN.  Its CCSN yield is strongly
metallicity-dependent because it is an odd-$Z$ element, hence the steepness of
the blue dotted curve in Figure \ref{fig:river_all} and the rising trend of
[Na/Fe] up to [Fe/H]~$<-$0.4 for the \citetalias{chieffi2004} simulation. At
higher metallicities, [Na/Fe] decreases due to increased iron enrichment from
SNIa.  The \citetalias{chieffi2004} simulation produces a dramatic rise in
[Na/Fe] that is not seen in the observed [Na/Fe] trend, which is flat at
[Na/Fe]~$\approx$~0.1 for [Fe/H]~$<-$0.4 with a slow decline toward solar
[Na/Fe] at solar [Fe/H].  However, low metallicity stars
([Fe/H]~$\lesssim$~$-$1) can have sub-solar [Na/Fe] (e.g, \citealt{nissen1997,
nissen2010}), suggesting that the rising trend may qualitatively reflect
observations at metallicities below that of the \citet{reddy2003, reddy2006}
data.  The \citetalias{woosley1995} simulation initially declines from
[Fe/H]~=~$-$2 to $-$1.6 due to its metallicity-dependent iron yields.  It then
rises from [Fe/H]~=~$-$1.6 to $-$0.6 but more gradually than the
\citetalias{chieffi2004} simulation due to a weaker metallicity dependence for
the sodium yields.  At [Fe/H]~=~$-$0.6, it turns over because of SNIa iron
enrichment.  The normalization of the \citetalias{woosley1995} trend is a
better match to the data than the \citetalias{chieffi2004} simulation, and it
does a reasonable job of fitting the data albeit with a sharper turnover.  The
\citetalias{limongi2006} trend always using the solar metallicity Na yield
predicts an extremely high [Na/Fe] that is mostly off the top of the plot.
There are some observational uncertainties in measuring sodium abundances,
including non-LTE effects (see \citealt{smiljanic2012} and references therein).

\paragraph{Magnesium}

Like oxygen, magnesium enrichment is dominated by CCSN, and the predicted yield
is only mildly metallicity-dependent.  The [Mg/Fe] data follow the typical
$\alpha$-element trend and are well reproduced by the \citetalias{chieffi2004}
simulation, though it slightly underpredicts [Mg/Fe], by $\sim$0.05--0.1~dex.
The \citetalias{woosley1995} simulation also underpredicts [Mg/Fe], by
$\sim$0.15--0.2~dex.  Several authors (\citealt{timmes1995}, \citet{thomas1998},
and \citealt{francois2004}) have found that magnesium is underproduced by the
\citetalias{woosley1995} yields. \citealt{francois2004} noted that magnesium
yields depend on the treatment of convection.

\paragraph{Aluminum}

Aluminum is an odd-$Z$ element produced almost entirely by CCSN.  The
\citetalias{chieffi2004} and \citetalias{woosley1995} simulations produce
aluminum trends that are similar to those of sodium because the physics of
production is similar.  The \citetalias{chieffi2004} simulation does a poor job
of fitting the observational data for [Al/Fe], which instead resembles the
$\alpha$-element trend.  The \citetalias{chieffi2004} simulation overpredicts
[Al/Fe] above [Fe/H]~=~$-$0.4 by $\sim$0.15~dex but dramatically underpredicts
it at low metallicity.  The weaker metallicity dependence of the
\citetalias{woosley1995} yields relative to the \citetalias{chieffi2004} yields
results in a flatter [Al/Fe] trend that provides a better fit to the data.  It
underpredicts the normalization of the observations by $\sim$0.1~dex, but its
shape is consistent with the data for [Fe/H]~$>-$1. Though the normalization of
the \citetalias{limongi2006} trend is too high, its shape matches the data well,
adding further evidence that Al has a much weaker metallicity-dependence than
the \citetalias{chieffi2004} yields predict.  Al abundance estimates can suffer
non-LTE effects (see \citealt{andrievsky2008}), especially at extremely low
metallicities, which would affect any assessment of the low metallicity Al
abundance trend and low metallicity CCSN yields.

\paragraph{Silicon}

Silicon is an $\alpha$-element made mostly in CCSN like oxygen and magnesium
but with a larger, though still small, contribution from SNIa.  In CCSN, it is
produced in both hydrostatic and explosive burning phases \citep{woosley2002},
and its yield is independent of metallicity. Both the \citetalias{chieffi2004}
and the \citetalias{woosley1995} simulations do an excellent job of
reproducing the observed [Si/Fe] data. \citet{francois2004} found good
agreement with data when they adopted the solar metallicity
\citetalias{woosley1995} silicon yields.

\paragraph{Sulfur}

Sulfur is also an $\alpha$-element, and its production is very similar to that
of silicon with a slightly larger contribution from SNIa.  Like silicon,
sulfur is synthesized in hydrostatic and explosive burning phases
\citep{woosley2002} of CCSN, and it has a metallicity-independent yield.  The
\citetalias{chieffi2004} and \citetalias{woosley1995} simulations agree with
the mean of the [S/Fe] measurements, though these data span almost 0.5~dex in
[S/Fe] and only 0.8~dex in [Fe/H].

\paragraph{Potassium}

Potassium is an odd-$Z$ element produced almost exclusively in CCSN.  Its
yield is predicted to be metallicity-dependent, as can be seen from the
super-linear slope of the blue dotted line in Figure \ref{fig:river_all},
though its metallicity dependence is weaker than sodium or aluminum.  The
\citetalias{chieffi2004} simulation significantly underpredicts the amount of
potassium needed to achieve the solar value of [K/Fe] and the observed [K/Fe]
data in Figure \ref{fig:xfe}.  The \citetalias{chieffi2004} curve falls off
the bottom of the plot, but the highest value it reaches is [K/Fe]~=~$-$0.65.
The \citetalias{woosley1995} simulation also falls far below the data, though
by 0.6~dex.  The super-solar [K/Fe] values suggest that potassium may behave
like an $\alpha$-element at lower metallicities, but more data are needed to
confirm such a trend.

\paragraph{Calcium}

Calcium is an $\alpha$-element made mainly in CCSN, but with a significant
SNIa contribution.  The \citetalias{chieffi2004} simulation matches the level
of the plateau in the [Ca/Fe] data, but its track turns over more sharply than
the data and ends at a sub-solar [Ca/Fe] value.  The \citetalias{chieffi2004}
CCSN yields decrease with metallicity, as shown by the
\citetalias{limongi2006} track lying at lower [Ca/Fe].  The
\citetalias{woosley1995} simulation produces higher [Ca/Fe] values at low
metallicity, but it overlaps with the \citetalias{chieffi2004} track for
[Fe/H]~$>-$0.8. \citet{mulchaey2014} argued that calcium-rich gap transients,
a subclass of supernovae whose nebular spectra are dominated by calcium lines,
are an important contributor of calcium in the intracluster medium, so it is
possible that they might be a meaningful source of calcium in galaxies, too.

\paragraph{Scandium}

CCSN are the dominant source of scandium.  Like other odd-$Z$ elements, the
\citetalias{chieffi2004} scandium yields have a very strong metallicity
dependence that results in a rising trend with [Fe/H], which turns over at
[Fe/H]~$\approx$~$-$0.3.  However, the observed scandium abundances more
closely follow the typical $\alpha$-element trend as in the
metallicity-independent \citetalias{limongi2006} track.  The
\citetalias{woosley1995} simulation predicts an $\alpha$-element trend for
scandium, in stark contrast to the \citetalias{chieffi2004} simulation.  Both
the \citetalias{chieffi2004} and \citetalias{woosley1995} simulations
underpredict the overall normalization of [Sc/Fe].  The offset for the
\citetalias{woosley1995} simulation is $\sim$0.3 --0.4~dex.  A generic offset
cannot be determined for the \citetalias{chieffi2004} simulation because of its
dramatically different shape compared to the data, though its peak is below the
main trend by $\sim$0.15~dex.

\paragraph{Titanium}

Titanium is both an $\alpha$-element and a low iron-peak element.  It is
mostly produced in CCSN, though SNIa contribute about 1/3 of the solar
titanium abundance.  Both the \citetalias{chieffi2004} and the
\citetalias{woosley1995} simulations produce $\alpha$-element trends with
similar normalizations.  The data also follow an $\alpha$-element trend.
However, both simulations underpredict the titanium abundance data by
$\sim$0.3--0.4~dex at all metallicities, and this underprediction is much more
severe for the solar metallicity \citetalias{limongi2006} calculation.  The
underproduction of titanium is a generic problem for SN yields, and its cause
is not well understood.

\paragraph{Vanadium}

Vanadium is an odd-$Z$ element produced predominantly in CCSN but with a
significant SNIa contribution, so its predicted abundance trend behaves like a
combination of odd-$Z$ and Fe-peak elements.  The \citetalias{chieffi2004}
vanadium yields have a weak metallicity dependence, and the
\citetalias{chieffi2004} simulation predicts a slightly rising [V/Fe] trend
with [Fe/H] that peaks at [V/Fe]~$\approx$~$-$0.4.  However, the [V/Fe] data
show a weak $\alpha$-element trend that declines from [V/Fe]~$\sim$0.2 to
$-$0.1, which is roughly similar in shape to the \citetalias{limongi2006} curve
but 0.3~dex higher in normalization.  The \citetalias{woosley1995} simulation
produces an $\alpha$-element trend like the data, but it underpredicts the
normalization by $\sim$0.4--0.5~dex.

\paragraph{Chromium}

Chromium is an iron-peak element made equally in CCSN and SNIa.  The
\citetalias{chieffi2004} yields produce slightly super-solar [Cr/Fe], which
declines to slightly sub-solar [Cr/Fe] due to the sub-solar [Cr/Fe] SNIa
yield ratio.  The measured [Cr/Fe] data are centered on solar [Cr/Fe] with
little scatter down to [Fe/H]~=~$-$1.  The \citetalias{woosley1995} simulation
declines by $\sim$0.2~dex from [Fe/H]~=~$-$2 to $-$1 and then rises slightly.
It underpredicts the data by about 0.1~dex.

\paragraph{Manganese}

SNIa and CCSN contribute to the enrichment of manganese, an iron-peak element,
at similar levels.  The \citetalias{chieffi2004} manganese yields increase
with metallicity, resulting in a rising trend with [Fe/H]. The data show a
unique observational trend of a plateau at [Mn/Fe]~$\approx$~$-$0.4 for
[Fe/H]~$<-$1 that rises to solar [Mn/Fe] at solar metallicity.  The
\citetalias{chieffi2004} simulation reproduces the data nearly perfectly, and
the \citetalias{woosley1995} simulation reproduces the shape of the observed
trend with a minor offset to lower [Mn/Fe].  The CCSN manganese yields are
sensitive to the adopted mass cut, in a way that is not offset by similar
changes to the iron yields. While the agreement between models and data in this
panel is good, \citet{bergemann2008} argue that the {\it observed} trend is
caused by erroneously assuming local thermodynamic equilibrium, and that the
trend of rising [Mn/Fe] towards higher metallicity goes away if non-LTE effects
are taken into account.

\paragraph{Cobalt}

Cobalt is an iron-peak element whose CCSN contribution is three times its SNIa
contribution.  The \citetalias{chieffi2004} cobalt yields are
metallicity-dependent, which produces the rising [Co/Fe] trend with increasing
[Fe/H].  In contrast to the \citetalias{chieffi2004} simulation, however, the
observed [Co/Fe] data follow an $\alpha$-element trend, like that predicted by
the solar-metallicity \citetalias{limongi2006} simulation but with a $\sim
0.2$~dex offset.  The \citetalias{woosley1995} simulation predicts a nearly
identical trend to the \citetalias{chieffi2004} simulation for [Fe/H]~$<-$0.8,
but it turns over at [Fe/H]~=~$-$0.6, whereas the \citetalias{chieffi2004}
simulation continues to rise until [Fe/H]~=~$-$0.3.

\paragraph{Nickel}

Nickel is an iron-peak element that has slightly more contribution from SNIa
than CCSN.  CCSN nickel yields are metallicity-dependent, so the
\citetalias{chieffi2004} simulation predicts a monotonically rising [Ni/Fe]
trend.  However, the observations show a flat trend at solar [Ni/Fe] with
little scatter.  The \citetalias{limongi2006} simulation produces a flat trend
because the \citetalias{limongi2006} Ni/Fe ratio is similar to that of the SNIa
yields, though the normalization is 0.35~dex above the data. The
\citetalias{woosley1995} prediction is similar to that of the
\citetalias{chieffi2004} simulation but offset towards higher [Ni/Fe] by
$\sim$0.1--0.2~dex.

\paragraph{Copper}

Copper is an iron-peak element that is exclusively produced in CCSN.  Both
the  \citetalias{chieffi2004} and the \citetalias{woosley1995} copper yields
have a strong metallicity dependence, and they predict rising [Cu/Fe] trends
that turn over when SNIa begin to contribute significant amounts of iron.
The [Cu/Fe] measurements roughly agree with this prediction, but the
mean trend is offset to higher [Cu/Fe] by $\sim$0.1~dex.  The data have
significant scatter, especially at low metallicity.  The
\citetalias{woosley1995} simulation fits the low metallicity data better than
the \citetalias{chieffi2004} simulation because the former is offset towards
lower [Fe/H].

\paragraph{Zinc}

Zinc is an iron-peak element, whose yields are dominated by CCSN.  Both
simulations produce an undulating trend that first declines, then rises due to
a weak metallicity dependence of the zinc yields, and finally declines because
of SNIa iron enrichment.  The observed [Zn/Fe] data show a weak
$\alpha$-element trend with large scatter below [Fe/H]~=~$-$1.  The shape of
the simulated trends may be consistent with the shape of the observed trend.
However, the overall normalizations of the \citetalias{woosley1995} and
\citetalias{chieffi2004} predictions are too low by $\sim$0.25~dex and
$\sim$0.5~dex, respectively, indicating that the CCSN yields are grossly
underproducing zinc.

\paragraph{Europium}

Europium is produced almost exclusively in the $r$-process.  The astrophysical
site of the $r$-process is unknown, but the two most likely candidates, CCSN
and neutron star mergers, are associated with massive stars.  The [Eu/Fe] data
follows an $\alpha$-element trend, which suggests that europium yields
are independent of metallicity.  Neither the \citetalias{chieffi2004} nor the
\citetalias{woosley1995} yields included europium, so we adopted the
\citet{cescutti2006} Eu yields (their model 1), which are
metallicity-independent by construction.  The differences between the two
simulations result from differences in the CCSN iron yields, and the small
offset between the \citetalias{chieffi2004} and the \citetalias{limongi2006}
curves shows the slight metallicity dependence of the \citetalias{chieffi2004}
iron yields. The \citet{cescutti2006} europium yields were determined
empirically by fitting a chemical evolution model to observed [Eu/Fe] data from
[Fe/H]~=~$-$3.4 to +0.15.  Given the empirical nature of the europium yields,
it is not surprising that the simulations and the data agree, but it is
reassuring that we successfully reproduce a different observational [Eu/Fe]
data set than the calibrating one with a different chemical evolution model.

For any individual element, the one-zone model predictions have freedom
associated with the choice of SFE, outflow rate, and so forth, as discussed in
Section \ref{sec:variations}.  Once these parameters are chosen to produce a
given [O/Fe] track, however, the predictions for other elements are sensitive
mainly to yields.  Thus, major discrepancies in Figure \ref{fig:xfe} probably
indicate either incorrect data or incorrect yields.  The most common forms of
discrepancy are either an overall normalization offset or predictions of rising
trends for elements with metallicity-dependent yields where observed trends are
similar to those of $\alpha$-elements. Adding more measurements at
[Fe/H]~$<$~$-$1 would greatly increase the constraining power of the
observational data.


\section{Forming the High-$\alpha$ and Low-$\alpha$ Sequences}
\label{sec:scatter}

The distribution of stars in [$\alpha$/Fe]--[Fe/H] separates into
high-$\alpha$ and low-$\alpha$ sequences \citep[e.g.,][]{bensby2003,
bensby2005, bensby2014, reddy2003, reddy2006, fuhrmann2011}.  The bimodality
in [$\alpha$/Fe] is enhanced by kinematic selection, but samples without
kinematic selection still clearly show bimodality \citep{fuhrmann1998,
fuhrmann2004, fuhrmann2008, fuhrmann2011, adibekyan2012}.  This bimodality is
also evident throughout the disk in the APOGEE red clump \citep{nidever2014}
and main \citep{anders2014, hayden2014, hayden2015} samples.

The high-$\alpha$ sequence starts at super-solar [$\alpha$/Fe] at low
metallicity, turns over at [Fe/H]~$\sim$~$-0.7$, then declines to solar
[$\alpha$/Fe] at super-solar metallicity.  Recent APOGEE results from
\citet{nidever2014} and \citet{hayden2015} found that the location of the
high-$\alpha$ sequence is ubiquitous throughout the disk from $R$~=~3--15~kpc
and up to heights of 2~kpc.  The low-$\alpha$ sequence lies at nearly solar
[$\alpha$/Fe] and slowly declines with metallicity.  In the solar
neighborhood, it runs from [Fe/H]~=~$-0.5$ to $+0.3$, with a roughly Gaussian
MDF peaked at solar metallicity.  Unlike the high-$\alpha$ sequence, it shifts
from higher metallicities in the inner disk to lower metallicities in the
outer disk \citep{nidever2014, hayden2015}.  Furthermore, \citet{hayden2015}
found that the thin disk MDF shifts in skewness from negative (sharp right
edge) to positive (sharp left edge) from the inner to the outer disk.

One-zone chemical evolution models with simple SFHs can reproduce the
high-$\alpha$ sequence (e.g., \citealt{nidever2014} used \flexce\ to reproduce
the high-$\alpha$ sequence of the APOGEE red clump stars).  They also readily
reproduce the merged high-$\alpha$ and low-$\alpha$ sequences in the inner disk
and the negative skewness of the inner disk MDF.  However, they cannot
simultaneously create the high-$\alpha$ and low-$\alpha$ sequences in the solar
annulus and the roughly Gaussian shape of the solar annulus MDF.  We
investigate two classes of modifications to resolve these differences, one
inspired by the two infall model of \citet{chiappini1997, chiappini2001}, and
one inspired by the multi-zone model with stellar migration of
\citet{schoenrich2009a, schoenrich2009b}.  The two infall model relies on a
hiatus in the SFH, which produces a gap between the high-$\alpha$ and
low-$\alpha$ sequences.  In the \citet{schoenrich2009a, schoenrich2009b}
stellar migration model, the high-$\alpha$ sequence results from similar
high-$\alpha$ sequences in all radial zones, whereas the low-$\alpha$ sequence
is a superposition of the equilibrium abundances from multiple radial zones.
\cite{hayden2015} argue that the change of shape of the observed MDF between
the inner and outer disk, including the Gaussian shape in the solar
neighborhood, are a consequence of radial mixing.


\begin{figure*}
\centerline{\includegraphics[width=9cm]
{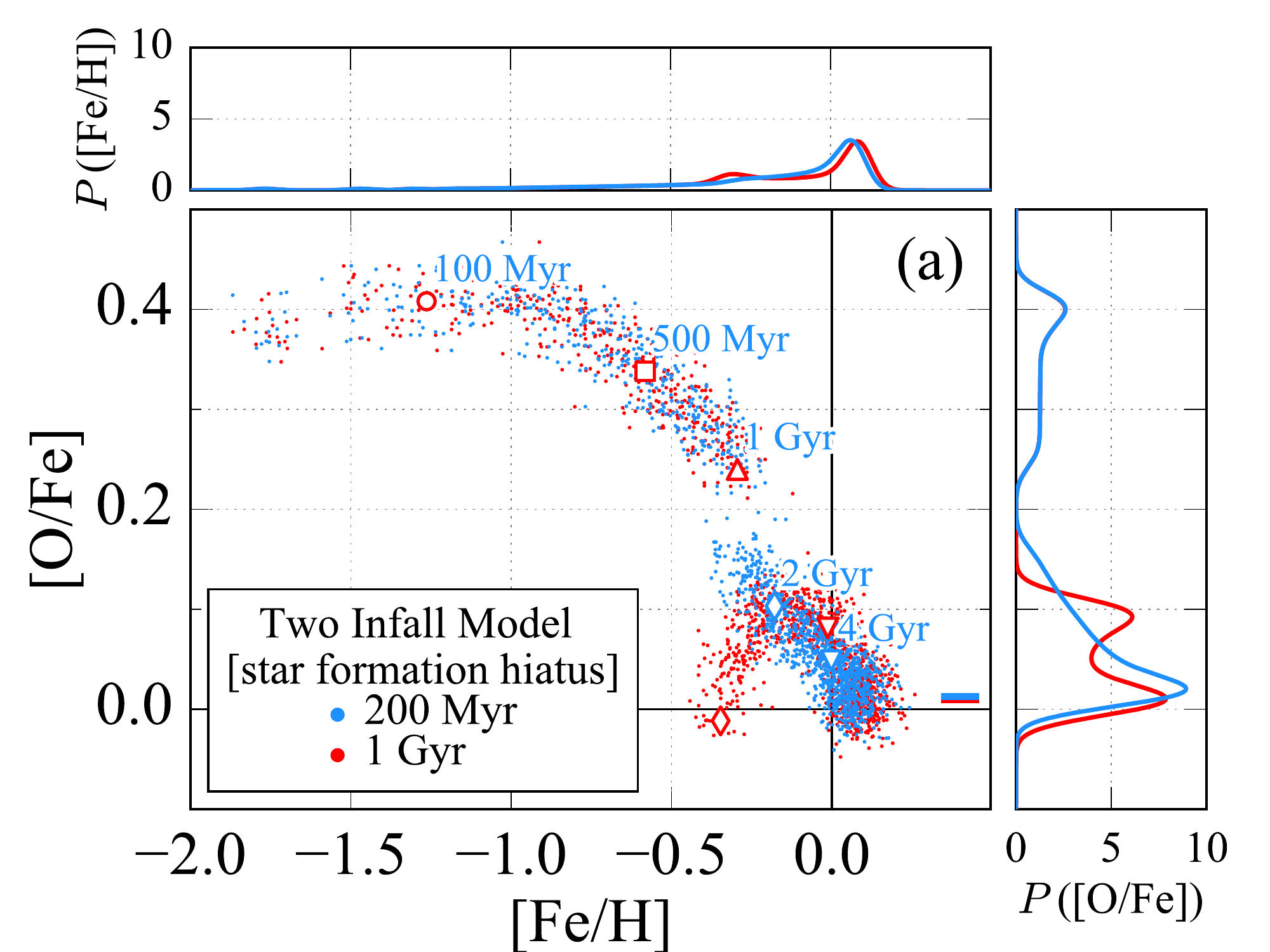}
\includegraphics[width=9cm]
{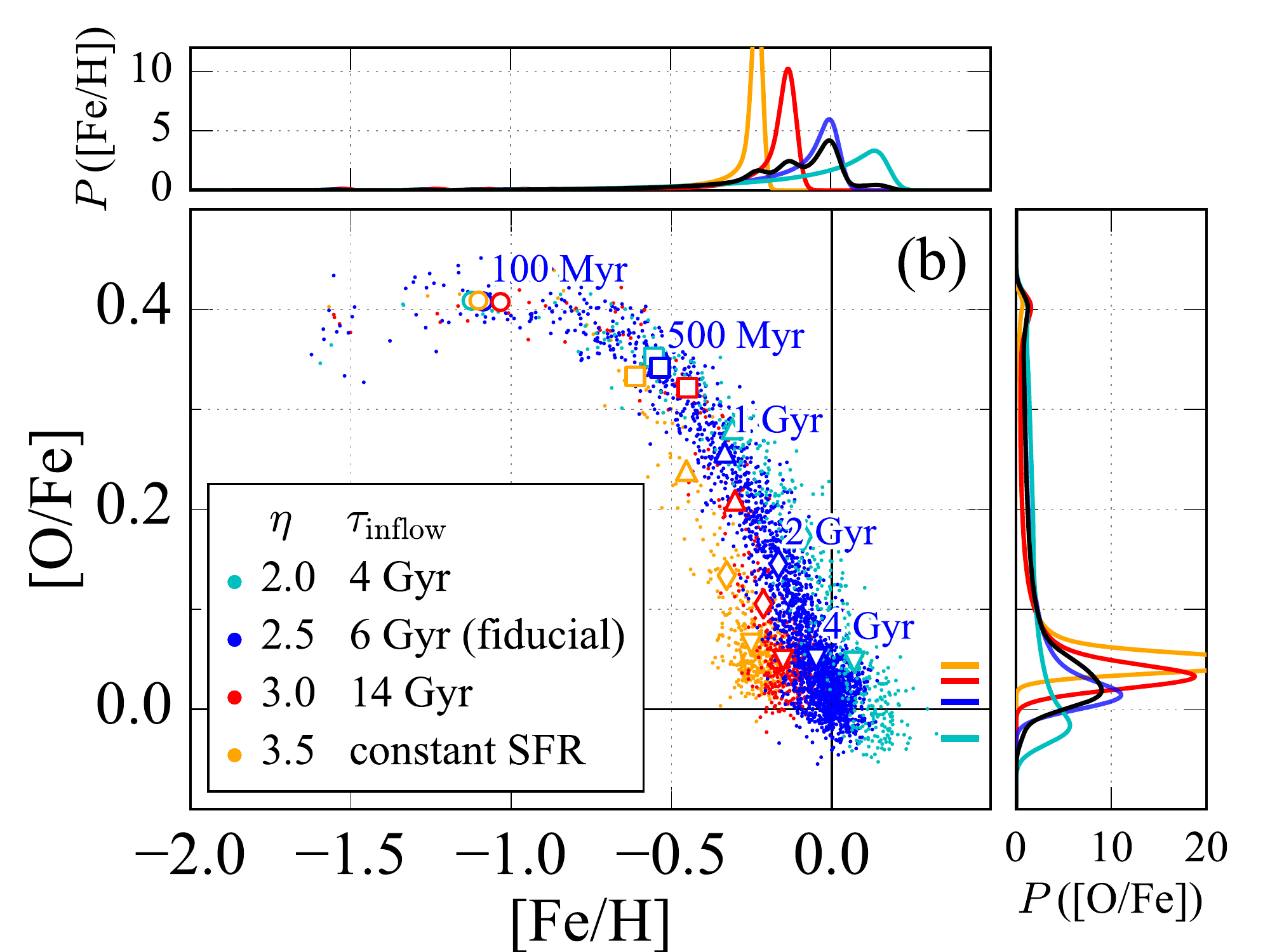}}

\caption{Two scenarios for creating the high-$\alpha$ and low-$\alpha$
sequences.  The points are randomly selected stars with noise added
($\sigma([\mathrm{Fe/H}]) = 0.05$ and $\sigma([\mathrm{O/Fe}]) = 0.02$). Panel
(a) shows a two infall model inspired by \citet{chiappini1997, chiappini2001},
with hiatuses in star formation of 200~Myr (blue points) and 1~Gyr (red
points). The two infall model produces a true gap in [O/Fe].  Panel (b) shows a
superposition of simulations with variations in the outflow mass-loading
parameter ($\eta$) and inflow timescale designed to reflect enrichment
histories at different radii, inspired by the \citet{schoenrich2009a,
schoenrich2009b} model in which stars formed at different galactocentric radii
migrate to the solar neighborhood. The superposition scenario produces a
low-$\alpha$ sequence that is not an evolutionary track but rather a ridge line
formed from the equilibrium abundances of the simulations. The black line shows
the composite MDF and [O/Fe]-DF for all simulations.  The open symbols are the
same as in Figure \ref{fig:ofe_gas_flows}.}

\label{fig:ofe_mdf_bimodality}
\end{figure*}

\subsection{Two Infall Model}
\label{sec:two_infall}

Connecting the high-$\alpha$ sequence to the low-$\alpha$ sequence remains
challenging for chemical evolution models. Motivated by early observational
evidence of a gap in [Fe/O]--[O/H] data from \citet{gratton2000},
\citet{chiappini2001} found that a track could produce a gap in the abundance
trend if the disk formed in two major infall episodes separated by a cessation
of star formation.  The gap in the SFH is caused by a slow infall rate for the
second episode coupled with a threshold gas density for star formation
\citep{kennicutt1989}.  The effect of this threshold is magnified in the
\citet{chiappini2001} model because it increases from 4 to 7 \msun~pc$^{-2}$
between the first and second infall episodes. Figure
\ref{fig:ofe_mdf_bimodality}a shows a pair of simulations of the two infall
model for hiatuses in star formation of 200~Myr (blue points) and 1~Gyr (red
points).  The points show stars randomly chosen from the simulations with
Gaussian noise of $\sigma$~=~0.05 in [Fe/H] and $\sigma$~=~0.02 in [O/Fe] added.
For the first Gyr, we set the inflow timescale to be 1~Gyr and set the SFE to be
6~$\times 10^{-10}$~$\mathrm{Gyr}^{-1}$. Next, we completely suppressed star
formation starting at $t$~=~1.0~Gyr for 200~Myr or 1~Gyr. Then, we set the
inflow timescale to be 6~Gyr and reduced the SFE by a factor of two (to
3~$\times 10^{-10}$~$\mathrm{Gyr}^{-1}$) with a maximum inflow
($\tau_\mathrm{max}$) at 1~Gyr ($\propto e^{-(t - \tau_\mathrm{max})/\tau_1}$).

The inflow timescale for the first epoch and the ratio of the SFE between the
epochs (a factor of 2) were adopted from the original formulation of the two
infall model in \citet{chiappini1997}, but absolute values of SFE are not easily
comparable because they adopted a different star formation law.  However, the
exact values of these parameters are much less significant than the choice of
the length of the hiatus in star formation. Our choices of hiatuses of 200~Myr
and 1~Gyr are intended to demonstrate the sensitivity of the abundances to this
parameter.

The two infall model produces a gap in [O/Fe]--[Fe/H] because of the hiatus in
star formation.  During the hiatus, SNIa iron enrichment decreases [O/Fe], but
its effect of increasing [Fe/H] is offset almost perfectly by dilution from
inflow.  The magnitude of the gap depends sensitively on the length of the
hiatus in star formation. For a 200 Myr hiatus, [O/Fe] drops by 0.07 dex, and
the subsequent evolutionary track is roughly a continuation of the pre-hiatus
track but with a shallower slope. For a 1 Gyr hiatus, [O/Fe] drops 0.23 dex to
[O/Fe]$~\approx 0.0$. However, the sudden return of star formation boosts the
rate of CCSN relative to SNIa, which quickly drives [O/Fe] back up to $\sim
+0.1$, after which the model evolves to its solar abundance equilibrium. In
both models, the pause in star formation leads to a separate peak in the MDF at
[Fe/H]$~\approx -0.3$. The large gap, inverted-U shape of the low-$\alpha$
sequence, and the bimodal low-$\alpha$ sequence of the 1~Gyr hiatus simulation
are not seen in the observational data, implying that the length of the hiatus
must fall in a narrow range around 200~Myr to produce the right size gap and
the morphology of the low-$\alpha$ sequence. However, we have not found any
combination of parameters that reproduces the morphology of the observed
stellar distribution, with two distinct sequences showing high and low
[$\alpha$/Fe] over a substantial range in [Fe/H].

\subsection{Mixing Stellar Populations}
\label{sec:superposition}

\citet{schoenrich2009a, schoenrich2009b} incorporated the effects of radial
mixing of gas and stars into a chemical evolution model.  They included the
effects of changing the epicyclic amplitude (``blurring'') and guiding center
radius (radial migration or ``churning'').  Their model produces the
high-$\alpha$ and low-$\alpha$ sequences from a superposition of stars born at
a range of radii, without a hiatus in star formation.  The high-$\alpha$
sequence is formed by overlapping tracks from different radii. Once SNIa become
significant contributors of iron, the tracks move quickly across the valley in
[$\alpha$/Fe]--[Fe/H].  Like all models with continuous SFHs, their model
produces some but not many intermediate [$\alpha$/Fe] stars. The low-$\alpha$
sequence is a superposition of the stars near the equilibrium abundances of a
range of radii, not an evolutionary track. Since individual tracks rapidly
asymptote to their equilibrium abundance, the low-$\alpha$ sequence contains
many more stars than the valley at intermediate [$\alpha$/Fe].

Figure \ref{fig:ofe_mdf_bimodality}b shows a superposition of simulations that
was designed to reproduce the high-$\alpha$ and low-$\alpha$ sequences by
varying the outflow mass-loading parameter and inflow timescale to mimic the
enrichment histories of several different galactocentric radii, whose stars were
then mixed together.  Increasing $\eta$ decreases the equilibrium [Fe/H] (see
Figure \ref{fig:ofe_mdf_outflow}) and results in a range of equilibrium
metallicities that form a ridge line at low-$\alpha$. Increasing the inflow
timescale maintains the same trajectory of a simulation but moves its
equilibrium abundance higher up the track to higher [O/Fe] and lower [Fe/H],
which produces the negative slope of the low-$\alpha$ sequence.  To reproduce
the slope, we adopted a constant SFR for the highest-$\eta$ simulation that
effectively acts as an infinite inflow timescale.  As the inflow timescale and
$\eta$ increase along our model sequence, the final stellar mass of the models
decreases (by a factor of three from the lowest-$\eta$ to the highest-$\eta$
simulations) because the simulations accrete less gas and retain that gas less
efficiently. Our choices of outflow mass-loading parameter and inflow timescale
are intended to be physically plausible values that qualitatively reproduce a
two sequence abundance trend.

The points in Figure \ref{fig:ofe_mdf_bimodality}b are stars randomly drawn
from the simulations with Gaussian noise of $\sigma = 0.05$ in [Fe/H] and
$\sigma = 0.02$ in [$\alpha$/Fe] added to show the 2-D distribution of stars in
[O/Fe]--[Fe/H].  The composite [O/Fe]-DF (black line) shows a single peak at
low-$\alpha$, in contrast to the two infall model with a 1~Gyr hiatus.  To
account very roughly for the fact that stars born in situ are more common than
migrated stars, we downselected stars from the simulations by a factor of 0.2,
1.~(fiducial), 0.2, and 0.1 (from lowest to highest $\eta$).  This suite of
simulations is intended to illustrate how the superposition scenario works, and
a more complete multi-zone model with an accurate treatment of stellar
migration and radial gas flows (such as \citealt{schoenrich2009a,
schoenrich2009b}) is needed to make more quantitative comparisons with data.

While these parameter variations were motivated by the results shown in Figures
\ref{fig:ofe_mdf_tau_inflow} and \ref{fig:ofe_mdf_outflow}, they reflect
differences in star formation history as a function of galactocentric radius.
An increasing inflow timescale with radius is a generic aspect of inside-out
galaxy formation \citep{larson1976}.  The outflow mass-loading parameter also
might increase with radius because the outer disk may be less effective at
retaining gas and metals due to a weaker vertical potential and a lower gas
density to counteract energy injection by SNe.  Furthermore, gas and metals
flow inward through the disk \citep{stark1984}, enriching the inner disk at the
expense of the outer disk, which mimics the effect of the outflow mass-loading
parameter increasing with radius. While the superposition model of
Figure~\ref{fig:ofe_mdf_bimodality}b does have high-$\alpha$ and low-$\alpha$
stars at the same [Fe/H], the bimodality of the distribution is arguably weaker
than observed (though one must be cautious of observational selections that may
amplify this bimodality).  The low-$\alpha$ ridge line could be extended to
lower [Fe/H] by adding models with still higher $\eta$, representing migration
from the far outer disk.

To summarize, a single simple chemical evolution model with constant parameters
cannot reproduce both the high-$\alpha$ and low-$\alpha$ sequences.  The two
infall model produces a gap between the high-$\alpha$ and low-$\alpha$
sequences due to a gap in the SFH. The length of this hiatus in star formation
must be narrowly confined to about 200 Myr to avoid overshooting the
low-$\alpha$ sequence.  Alternatively, the high-$\alpha$ and low-$\alpha$
sequences can be formed from a mix of stellar populations, born at different
galactocentric radii with distinct enrichment histories, that migrated to the
solar neighborhood. 

\citetalias{weinberg2016a} show that one can produce a bimodal
[$\alpha$/Fe]-[Fe/H] distribution in a one-zone model by sharply increasing
$\eta$ after the model has evolved to an initial equilibrium, so that the
stellar population then evolves backward to \textit{lower} [Fe/H] at roughly
constant [$\alpha$/Fe]. We have not examined such a scenario here, but evolution
of $\eta$, radial mixing, and unsteady star formation may all play a role in
shaping the observed distribution of stars in [$\alpha$/Fe]--[Fe/H] space.


\section{Principal Component Abundance Analysis}
\label{sec:pcaa}

The new generation of massive, multi-element stellar abundance surveys will
require a shift in data--model comparisons from qualitative assessments in two
dimensions (e.g., showing model tracks and data in [O/Fe]--[Fe/H]) to
quantitative methods that simultaneously utilize information in many
dimensions.  The most direct way to interpret the results of these methods in
the context of galaxy evolution scenarios is to apply the same methods to
chemical evolution models.  One such method that has been applied to existing
multi-element data sets is Principal Component Abundance Analysis
\citep[PCAA;][]{ting2012, andrews2012}, and below we apply it to various
simulations.  Our goal here is to understand in general terms how PCAA
responds to the physics of chemical evolution models and to illustrate its
ability to separate populations, but we defer detailed comparisons to
observations to future work.

PCAA is standard principal component analysis (PCA; e.g.,
\citealt{jolliffe1986}) applied to the relative elemental abundances [$X$/Fe]
of stars, and here we apply PCAA across a wide range in [Fe/H].  PCA shifts the
origin to the mean of the data set and rotates the axes of the distribution to
align the first principal component PC1 with the direction of maximum
variation, the second principal component PC2 with the direction of the most
remaining variation, and so on, while maintaining a linear and orthogonal set
of basis vectors.  PCA is a form of dimensionality reduction, so the data can
often be closely approximated by the first few PCs (for a full description of
PCAA see Section 2 of \citealt{andrews2012}).  As in \citet{andrews2012}, we
apply PCAA to the logarithmic relative abundances [$X$/Fe] but do not include
[Fe/H] as one of the variables, since we wish to focus on the trends beyond the
general increase of all elements over time.

\subsection{PCAA of Fiducial Simulation and Parameter Variations}

\begin{figure*}
\centerline{\includegraphics[width=18cm]
{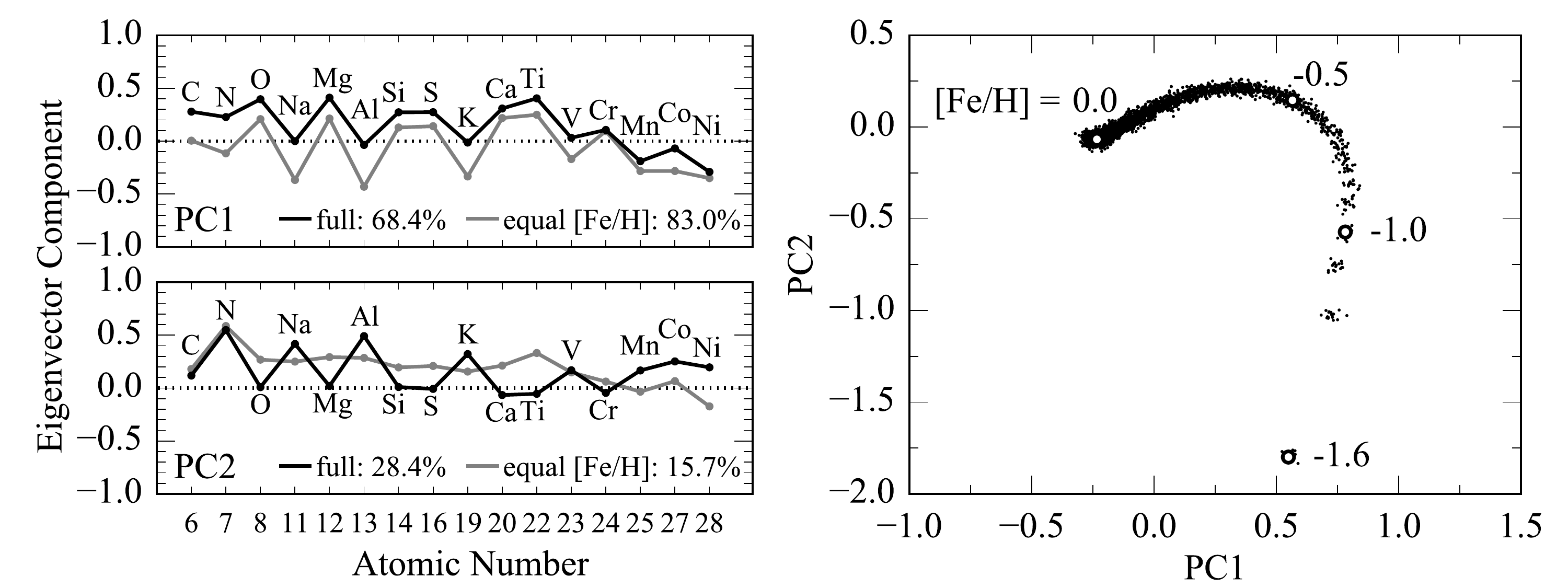}}

\caption{The left panels show the PC1 and PC2 eigenvector components of each
elemental abundance relative to Fe with larger absolute values corresponding to
stronger contributions.  Elements with the same sign are correlated, and
elements with opposite signs are anticorrelated.  An eigenvector component of 0
means that a PC is independent of that elemental abundance. The quadratic sum
of the eigenvector components is unity.  The percentage of the variation
attributed to each PC is given in the legends. Grey points and curves show PCs
of the same simulation when the stellar population is sampled to have equal
numbers of stars in each 0.1~dex bin of [Fe/H]. The right panel shows the
distribution of randomly selected stars from the full sample of the fiducial
simulation in PC1--PC2 space, with Gaussian noise of $\sigma$~=~0.02 added in
both PCs for display purposes. Metallicity increases along the general trend
starting from the bottom and moving upward then leftward.}

\label{fig:pcaa_fiducial}
\end{figure*}

The left panels of Figure \ref{fig:pcaa_fiducial} shows the eigenvector
components for the first two PCs of the multi-element abundance space for the
fiducial simulation.  The points represent the contribution of the elemental
abundances relative to Fe for each PC.  A large eigenvector component (positive
or negative) indicates that an elemental abundance contributes heavily to a PC,
whereas an eigenvector component close to 0 implies that the PC is nearly
independent of that elemental abundance.  Two elements have correlated
contributions to a PC if their eigenvector components have the same sign and
anticorrelated contributions if their eigenvector components have opposite
signs.  The eigenvector components are normalized such that they quadratically
sum to unity.  In Figure \ref{fig:pcaa_fiducial}, we show PC1 and PC2 for the
full sample (black) and for a population of stars with an equal number sampled
in 0.1~dex bins of [Fe/H] from $-1.0$ to +0.1 (gray).  The equal metallicity
sampling was designed to roughly represent selection biases in existing
multi-element surveys such as \citet{reddy2003, reddy2006} and
\citet{bensby2003, bensby2005, bensby2014}.

PC1 of the full sample is dominated by a correlation amongst the abundances of
$\alpha$-elements (C, O, Mg, Si, S, Ca, and Ti) and N.  These abundances show
a slight anticorrelation with the abundances of Mn and Ni, two Fe-peak
elements with metallicity-dependent CCSN yields.  PC1 is mostly independent of
the abundances of the odd-$Z$ elements (Na, Al, and K), and the Fe-peak
elements, V, Cr, and Co.  Thus, PC1 mainly captures the relative contribution
of CCSN and SNIa, the primary factor in multi-element abundance evolution.

PC2 of the full sample is characterized by a strong correlation amongst the
abundances of N, the odd-$Z$ elements, and the Fe-peak elements V, Mn, Co, and
Ni.  PC2 is independent of $\alpha$-element abundances.  The main contributing
elemental abundances to PC2 all have rising trends at low metallicity because of
metallicity-dependent CCSN or AGB star (N) yields.

When we apply PCAA to the equal metallicity sampling, however, PC1 and PC2 are
swapped. PC1 of the equal metallicity sampling shows an anticorrelation between
elements with metallicity-dependent yields (odd-$Z$ elements, Mn, and Ni) and
$\alpha$-elements.  Instead of tracking the decline from the knee due to SNIa,
it tracks the progression along the high-$\alpha$ plateau at low [Fe/H], like
PC2 of the full sample.  PC2 of the equally metallicity sampling is a
correlation amongst all non-Fe-peak elements, which shows the dilution of the
abundances of these elements by Fe production from SNIa.  This PC is similar to
PC1 of the full sample, except that PC1 of the equally metallicity sampling has
already accounted for the odd-even effect.

The right panel of Figure \ref{fig:pcaa_fiducial} shows the distribution in
PC1--PC2 space of randomly selected stars (analogous to the full sample) with
Gaussian noise of $\sigma$~=~0.02 in both PC1 and PC2 added to display
overlapping points.  Each cluster of points is a stellar generation, and these
clusters overlap as the simulation approaches the equilibrium abundances.  The
abundances evolve in PC1--PC2 space from low to high PC2 values in the first
few hundred Myr and then move towards lower PC1 values over the remainder of
the simulation.  The initial increase in PC2 values at early times is driven by
the rapid increase in the abundances of elements with metallicity-dependent
yields.  The PC1 values decrease as SNIa Fe enrichment (starting around
[Fe/H]~=~$-$0.8) decreases the relative abundances of elements with
metallicity-independent CCSN yields (C and $\alpha$-elements) and N, whose AGB
star yields are less metallicity-dependent than at lower metallicities.

The lack of contribution, positive or negative, to PC1 from the abundances of
the odd-$Z$ elements Na, Al, K, and V is a consequence of the offset between
the knee in the abundances of $\alpha$-elements (plus C and N) at
[Fe/H]~$\approx$~$-$1 and the turnover in the abundances of odd-$Z$ elements at
[Fe/H]~$\approx$~$-$0.4.  The abundances of odd-$Z$ elements continue to
increase after the knee in $\alpha$-element abundances.  The abundances of both
$\alpha$- and odd-$Z$ elements decline after the turnover in odd-$Z$ element
abundances.  These two trends cancel out because the PCs are linear by
construction.  A non-linear dimensionality reduction technique would be better
suited to this case and possibly in general given the non-linearity of
nucleosynthesis processes, but we defer any such explorations to future work.

We applied PCAA to each of the parameter variation simulations in Figures
\ref{fig:ofe_mdf_multi}, \ref{fig:ofe_min_snia_time}, \ref{fig:ofe_snia_dtd},
and \ref{fig:ofe_warm_ism}.  The PC1s of the full sample in these simulations
are generally similar to PC1 of the fiducial simulation and all have
significant contributions from $\alpha$-elements.  The contribution to PC1
from elements with metallicity-dependent CCSN yields can be correlated,
uncorrelated, or anti-correlated with the $\alpha$-elements depending on
\begin{enumerate}
\item the offset in metallicity of the $\alpha$-element knee and the turnover
of the  metallicity-dependent elemental abundance trends and
\item the magnitude of the decline in the metallicity-dependent elemental
abundance trends after the turnover.
\end{enumerate}
A smaller offset between the knee and the turnover produces a correlation
between $\alpha$- and metallicity-dependent elements, such as in the high SFE
(higher metallicity knee) and the high outflow mass-loading parameter (lower
metallicity turnover) simulations. These two simulations also feature a steep
decline in the metallicity-dependent elemental abundances above the turnover,
which contributes to the correlation.  On the other hand, simulations that
have nearly monotonically rising metallicity-dependent elemental abundance
trends---like the low SFE, population synthesis SNIa DTD, and warm ISM
simulations---show a strong anti-correlation between $\alpha$- and
metallicity-dependent elements.

The difference between full sampling and equal metallicity sampling is a
caution that PCs depend on not just the enrichment physics but the sampling of
stars.  This is unsurprising, as it is capturing the directions of greatest
variance, and the sampling will affect which variations have the biggest
quantitative effect.  When comparing models to data one therefore needs to
match the sample properties.  However, with fixed sampling the structure of
the PCs does depend on the chemical evolution physics.  For subsequent
applications of PCAA, we will use the full sample unless specified otherwise.

\subsection{PCAA of Yield and IMF Variations}

\subsubsection{\citetalias{chieffi2004} vs.~\citetalias{woosley1995} CCSN
Yields}

\begin{figure}
\centerline{\includegraphics[width=9cm]
{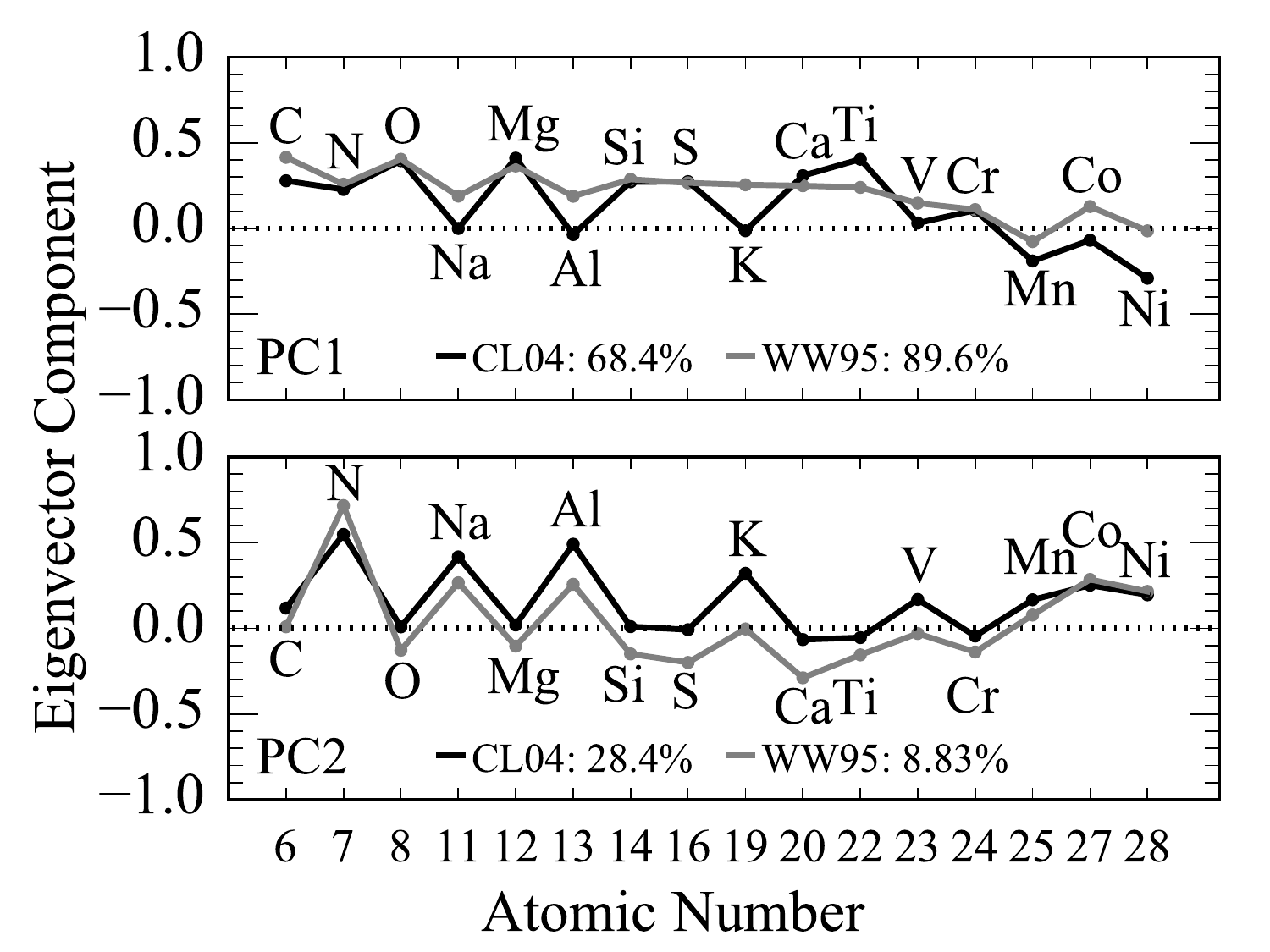}}

\caption{PC1 and PC2 eigenvector components of the \citetalias{chieffi2004}
yields (black) and \citetalias{woosley1995} yields (gray) simulations.  The
panels show the contribution of each elemental abundance relative to Fe to PC1
and PC2, with larger absolute values corresponding to stronger contributions.
PC1 describes a larger percentage of the variation (see legends) in the
\citetalias{woosley1995} simulation (89.6\%) than the \citetalias{chieffi2004}
simulation (68.4\%).}

\label{fig:pcaa_ww95}
\end{figure}

Figure \ref{fig:pcaa_ww95} shows the effect of varying the CCSN yields on the
PC1 and PC2 eigenvector components.  The black and gray lines represent the
\citetalias{chieffi2004} yield (same as Figure \ref{fig:pcaa_fiducial}) and
the \citetalias{woosley1995} yield simulations, respectively.  Both
simulations were run with the fiducial parameters.  PC1 of the
\citetalias{woosley1995} simulation shows a correlation between the abundances
of $\alpha$- and odd-$Z$ elements because of the weaker odd--even effect in
the yields.  The \citetalias{woosley1995} PC1 describes more of the variation
than the \citetalias{chieffi2004} PC1 (89.6\% vs.~68.4\%).  Like the
\citetalias{chieffi2004} PC2, the \citetalias{woosley1995} PC2 captures the
correlated abundances of N, Na, Al, Co, and Ni, though it is nearly
independent of the abundances of K, V, and Mn and shows a weak anticorrelation
with the heavier $\alpha$-element abundances.  This comparison demonstrates
that PC structure is a potential probe of SN yield physics provided that the
CCSN contribution can be separated from other effects.

\subsubsection{IMF Variations}

\begin{figure*}
\centerline{\includegraphics[width=18cm]
{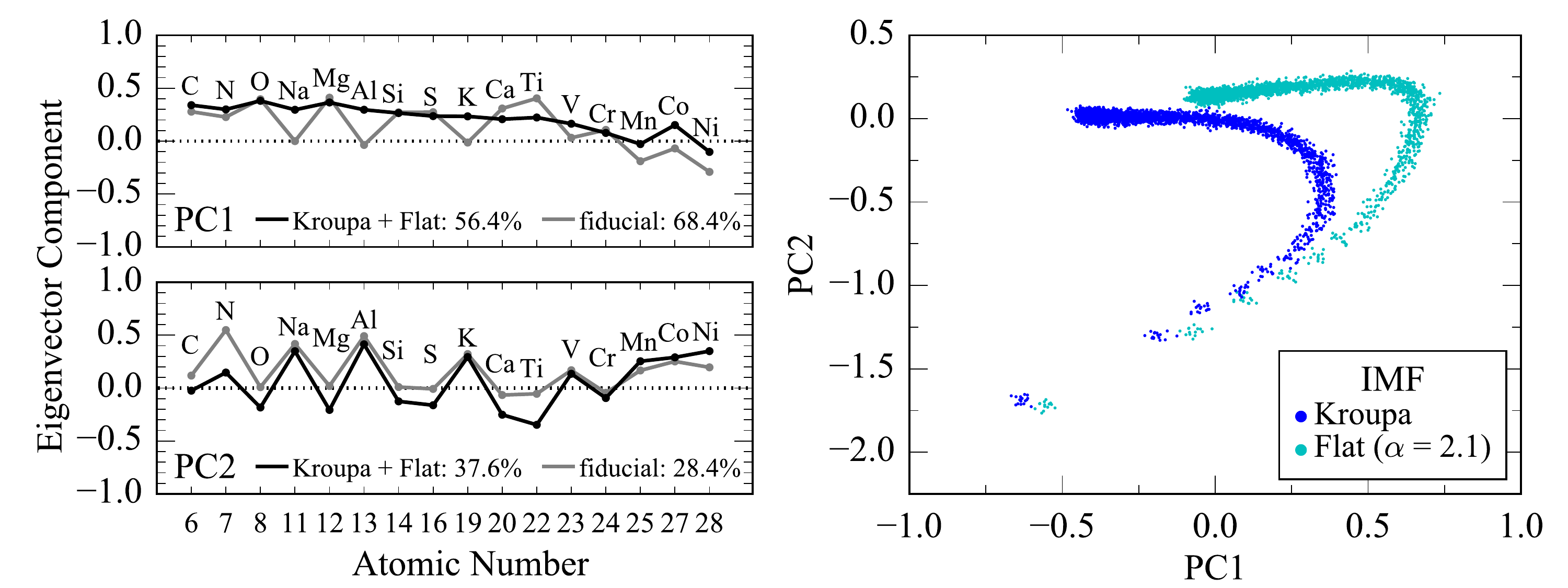}}

\caption{Connected black points in the left panels show the PC1 and PC2
eigenvector components for PCAA of a 1:1 mixture of populations formed with a
Kroupa and flat ($\alpha$~=~2.1) IMF simulations. Gray points show results for
a pure Kroupa IMF for comparison (same as the fiducial black points in
Fig.~\ref{fig:pcaa_fiducial}). PC1 of the mixture simulation has a correlation
amongst $\alpha$- and odd-$Z$ elements, unlike PC1 of the Kroupa IMF
simulation.  The right panel shows PC1--PC2 space for PCAA of the mixture
model.  Stars from the Kroupa IMF and flat IMF populations separate cleanly in
this space.  The points are randomly selected stars with Gaussian noise of
$\sigma$~=~0.02 added in both PCs for display purposes.}

\label{fig:pcaa_imf}
\end{figure*}

Figure \ref{fig:pcaa_imf} shows PC1, PC2, and PC1--PC2 space for PCAA of a 1:1
\textit{mixture} of the Kroupa and flat ($\alpha$~=~2.1) IMF simulations.  We
applied PCAA to this mixture of simulations as a conceptual model of the
mixture of two stellar populations formed with different IMFs that might result
from the accretion of satellites or radial mixing in the disk.  PC1 and PC2
mostly resemble PC1 and PC2 of the fiducial simulation with some key
differences, such as a correlation amongst odd-$Z$ and $\alpha$-elements in
PC1.  More importantly, the two sub-populations are cleanly separated in
PC1--PC2 space.  While their tracks in [O/Fe]--[Fe/H] are also distinct (see
Figure \ref{fig:ofe_mdf_imf}), PCAA coherently adds the differences in all
elements to amplify small offsets between populations, which could be used to
identify more subtle IMF variations between populations with otherwise similar
formation histories.

\subsection{PCAA of the Superposition Scenario}

\begin{figure}
\centerline{\includegraphics[width=9cm]
{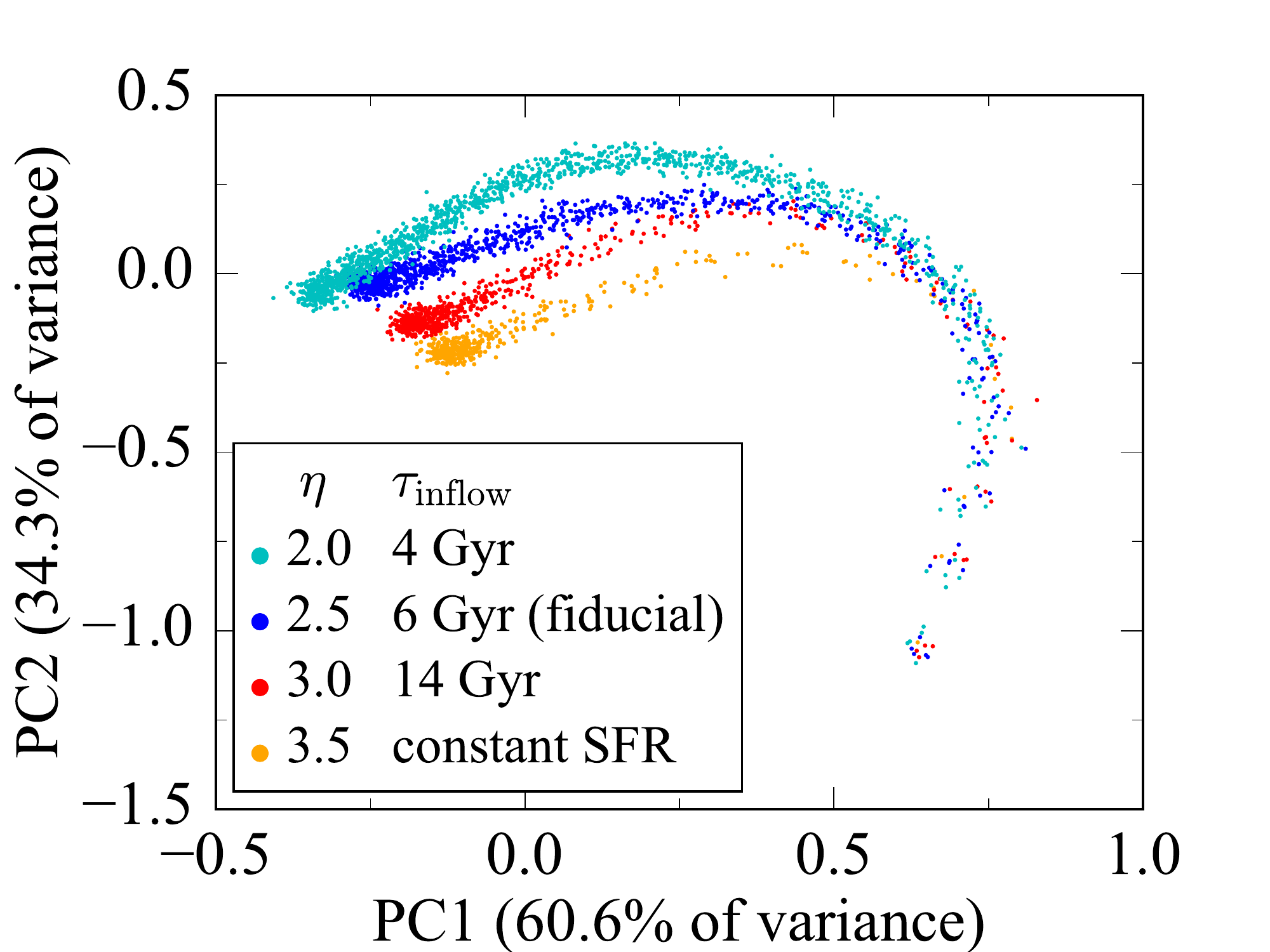}}

\caption{PC1--PC2 space for the superposition scenario simulation. PC1 and PC2
are nearly identical to those of the fiducial simulation (see Figure
\ref{fig:pcaa_fiducial}), but the distribution of stars within the space has
changed.  The points show randomly selected stars with Gaussian noise of
$\sigma$~=~0.02 added in both PCs for display purposes. Cyan, blue, red, and
orange points represent the same four populations shown previously in the right
panel of Fig.~\ref{fig:ofe_mdf_bimodality}.}

\label{fig:pcaa_superposition}
\end{figure}

In Figure \ref{fig:pcaa_superposition}, we show PC1--PC2 space for the
superposition scenario discussed in Section \ref{sec:superposition}.  PC1 and
PC2 (and PC3) of the superposition scenario are nearly identical to those of the
fiducial simulation (left panels of Figure \ref{fig:pcaa_fiducial}) for either
the full sample or equal metallicity sampling.  The superposition scenario
differs not in the structure of the PCs but in the distribution of stars in the
space.  The four simulations follow the same path in PC1--PC2 space at early
times (high PC1 and low PC2) but separate at late times (low PC1 and high PC2).
This separation is cleaner than in [O/Fe]--[Fe/H] (Figure
\ref{fig:ofe_mdf_bimodality}b) and is particularly noticeable in PC2 because of
its large contributions from elements with metallicity-dependent yields.
Consequently, the fractions of the variation attributed to PC1 and PC2 are
slightly smaller and larger, respectively, than those of the fiducial
simulation.  This scenario shows that PCAA can separate stellar populations with
distinct histories, using empirical correlations to take advantage of all the
information in the data, in contrast to focusing on a single element or group of
elements to define populations.

\subsection{PCAA of Reddy--Ram{\'{\i}}rez Data}

\begin{figure}
\centerline{\includegraphics[width=9cm]
{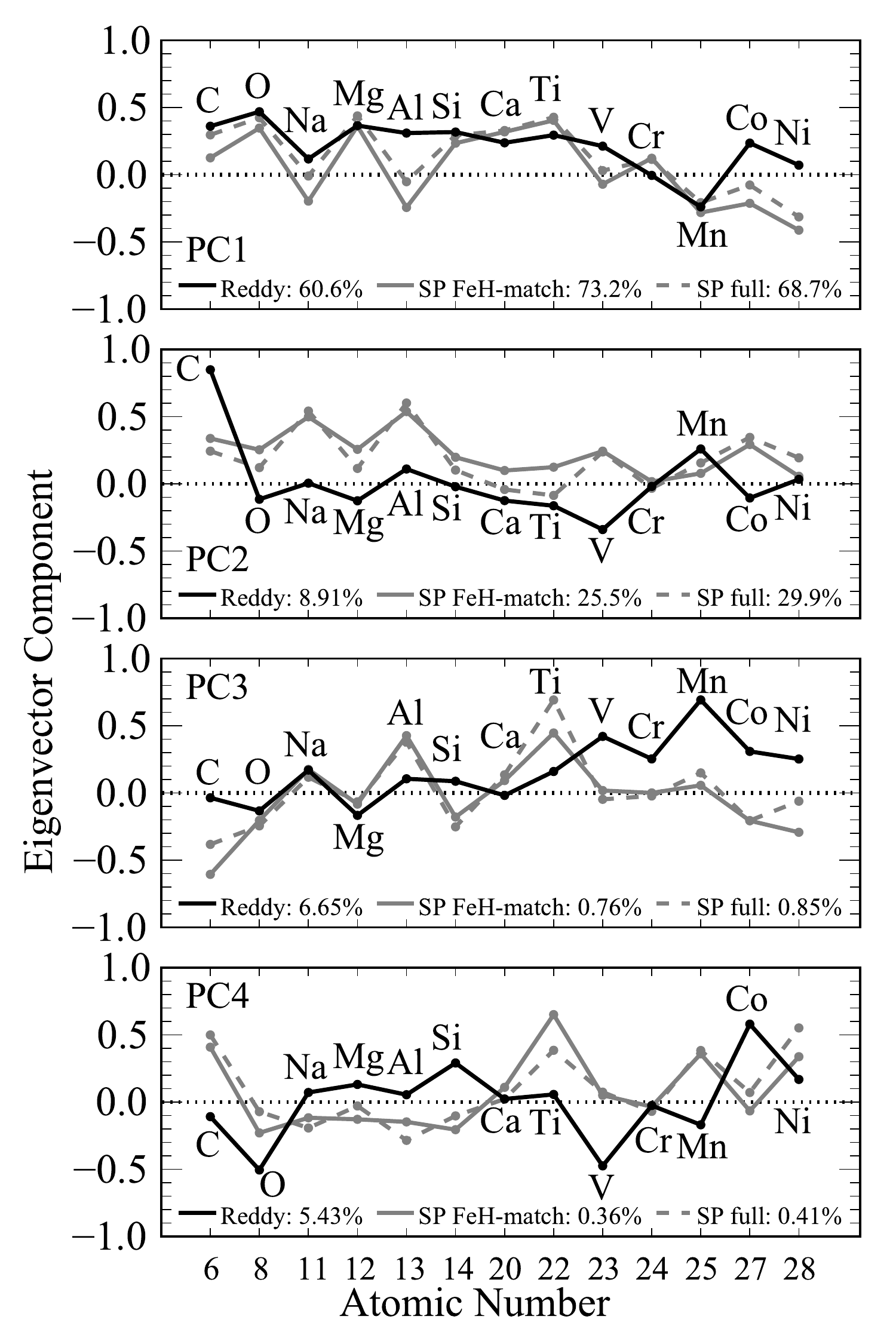}}

\caption{PC1--4 eigenvector components of the \citet{reddy2003, reddy2006} and
\citet{ramirez2013} data (black solid line), the superposition scenario
simulations sampled to match the Reddy--Ram{\'{\i}}rez MDF (gray solid line),
and the full superposition scenario sample (gray dashed line).  The abundances
of N, S, and K have been excluded due the paucity of data, and the PCs for the
superposition scenario have been recalculated with a consistent set of
elements. The panels show the contribution of each elemental abundance
relative to Fe for the PCs, with larger absolute values corresponding to
stronger contributions.}

\label{fig:pc1234_reddy}
\end{figure}

\begin{figure}
\centerline{\includegraphics[width=9cm]
{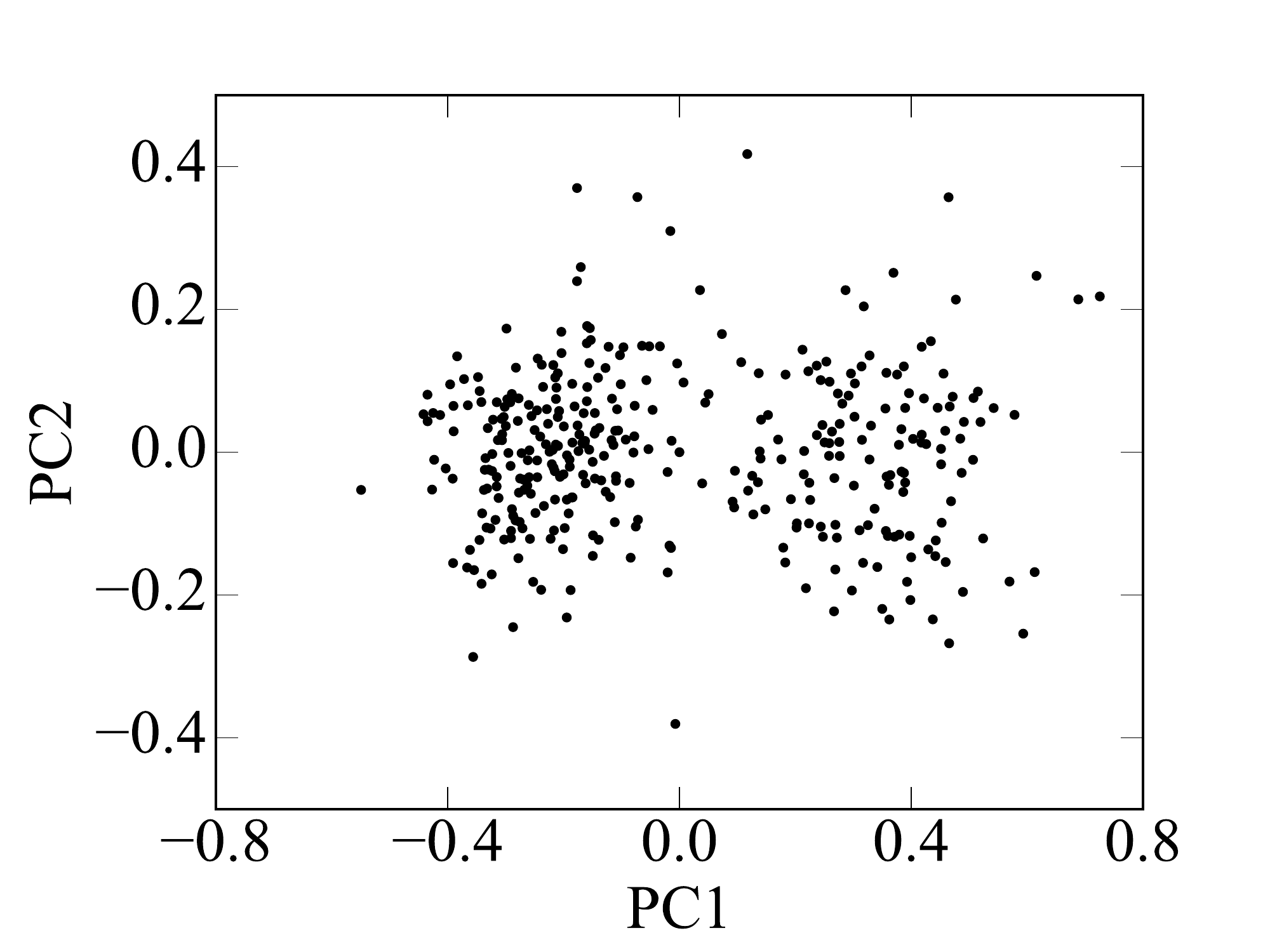}}

\caption{PC1--PC2 space of the \citet{reddy2003, reddy2006} and
\citet{ramirez2013} data for the elements in Figure \ref{fig:pc1234_reddy}.
Metallicity increases from right to left with decreasing PC1 (roughly
decreasing [$\alpha$/Fe]). PC2 is heavily dominated by C.  The high-$\alpha$
and low-$\alpha$ sequences show up as a clear bimodality in PC1.}

\label{fig:pc12_reddy}
\end{figure}

\begin{figure}
\centerline{\includegraphics[width=9cm]
{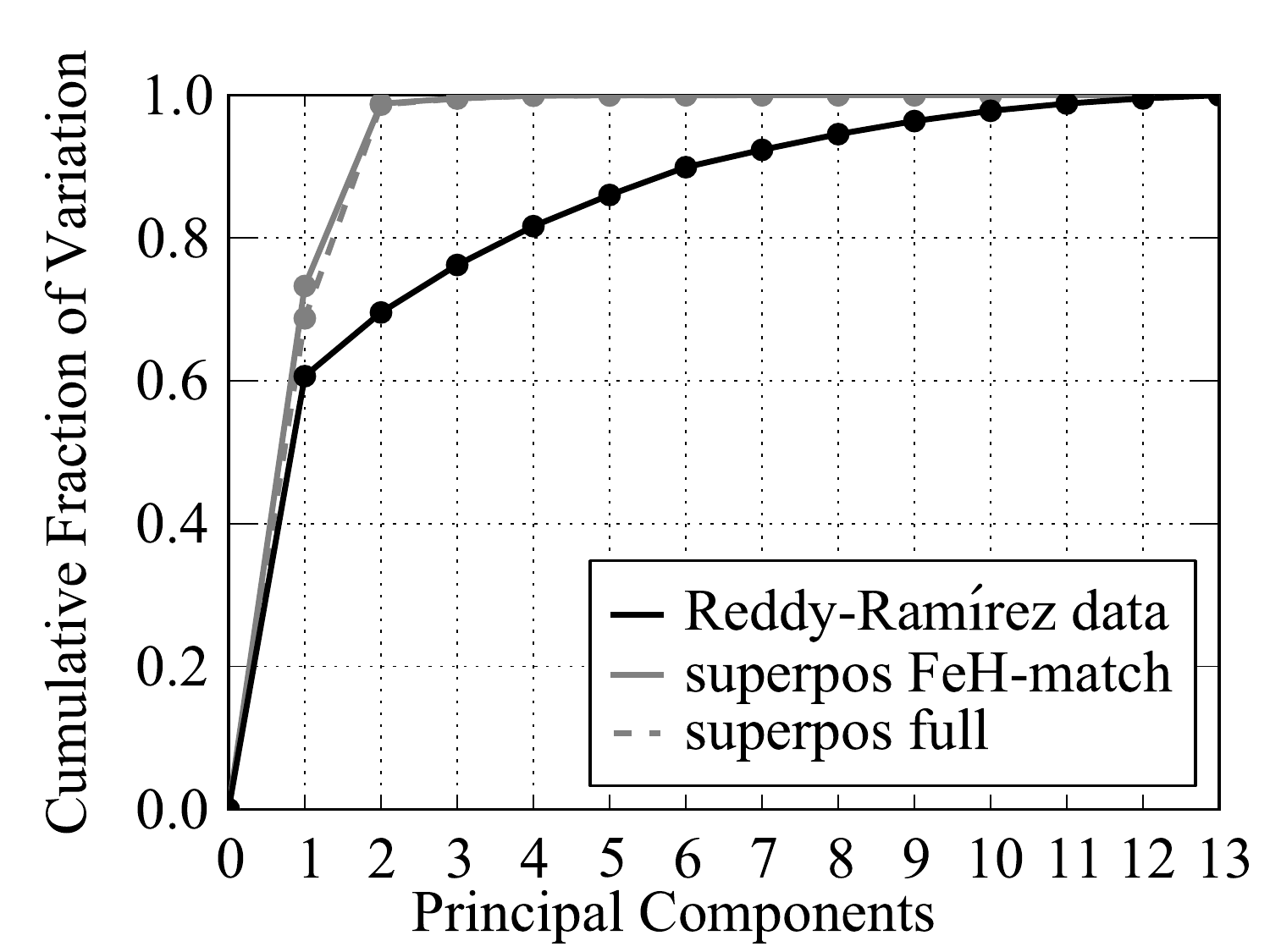}}

\caption{Cumulative fraction of the variation described by the PCs for the
\citet{reddy2003, reddy2006} and \citet{ramirez2013} data (black solid line),
the superposition scenario simulations sampled to match the
Reddy--Ram{\'{\i}}rez MDF (gray solid line), and the full superposition
scenario sample (gray dashed line).}

\label{fig:cumfracvar_reddy}
\end{figure}

In Figure \ref{fig:pc1234_reddy}, we show PC1, PC2, PC3, and PC4 for the
observed abundances of \citet{reddy2003, reddy2006} and \citet{ramirez2013}
shown in Figure \ref{fig:xfe} (which we will refer to as the
Reddy--Ram{\'{\i}}rez data).  We adopted the superposition scenario as the best
model comparison for the Reddy--Ram{\'{\i}}rez data because it more closely
reproduces the observed scatter in abundances than one-zone simulations.
Additionally, the gray solid line shows a downselection of stars from the
superposition scenario to match the Reddy--Ram{\'{\i}}rez iron MDF.  We note
that the abundances of N, S, and K were not included in the analyses discussed
in this section for either the superposition scenario or the observed
abundances due to limited data.

PC1 of the Reddy--Ram{\'{\i}}rez data broadly matches the $\alpha$-element
abundance nature of PC1 of the metallicity-matched and full samples of the
superposition scenario. However, the abundances of the odd-$Z$ elements Na,
Al, and V and the Fe-peak elements Co and Ni are correlated with the
abundances of $\alpha$-elements. This is evidence that the metallicity
dependence of the \citetalias{chieffi2004} yields is too strong (the
metallicity-dependence for these elements is weaker for the
\citetalias{woosley1995} yields) or else that delayed gas mixing can
significantly dampen its effects.  The abundance of Al contributes more
strongly to PC1 of the data than the abundance of Na, whereas the abundances of
these elements generically contribute to PC1 and PC2 of the simulations in
lockstep for both the \citetalias{chieffi2004} and \citetalias{woosley1995}
yields.  This suggests that the metallicity dependence of Na and Al CCSN yields
is more distinct than the \citetalias{chieffi2004} and \citetalias{woosley1995}
yields predict, though it is also possible that non-LTE effects are biasing the
observed abundances of Na and Al.  Both the \citetalias{chieffi2004} and
\citetalias{woosley1995} yields overpredict the metallicity dependence of Co
and Ni.

PC2 of the data is dominated by the abundance of C, in contrast to PC2 of the
metallicity-matched and full samples of the superposition scenario, which are
driven by the abundances of elements with metallicity-dependent CCSN yields.
The overwhelming importance of the C abundance to PC2 suggests that PC2 could
reflect variation due to AGB star nucleosynthesis, either in the standard
sense of a well-mixed gas reservoir or in binary mass transfer systems, or
observational scatter in the C abundances.  We note that PC2 of the data
accounts for a much lower fraction of the variation (9\% vs.~26\%) than PC2 of
the metallicity-matched superposition scenario, so any interpretation of its
potential physical origins should be treated with caution.

PC3 describes a comparable fraction of the variation as PC2 (7\%).  It is
dominated by Mn and the other Fe-peak elements, tentatively hinting at
variations in the SN yields of these elements.  Alternatively, and less
interestingly, PC3 might result from variance caused by random observational
errors in the Fe abundance, shifting the observed Fe coherently relative to
these other elements that are tied closely to the true Fe abundance.

PC4 also accounts for a non-negligible fraction of the variation (5\%).
However, it shows Co anti-correlated with O and V, whose physical
interpretation is unclear and is likely not warranted. We conjecture that this
PC reflects mainly observational noise and limited sample size.

Figure \ref{fig:pc12_reddy} shows PC1--PC2 space for the Reddy--Ram{\'{\i}}rez
data.  Metallicity generally increases from right to left.  The data exhibit a
bimodal distribution in PC1, reflecting the bimodality between the
high-$\alpha$ and low-$\alpha$ sequences.  However, the kinematic selection of
thick disk stars in this sample has enhanced this bimodality to create two
distinct populations. In contrast to the superposition scenario (see
Fig.~\ref{fig:pcaa_superposition}), PC2 shows no trend with PC1 and substantial
scatter. However, the nature of PC2 is very different in these two cases.

Figure \ref{fig:cumfracvar_reddy} shows the cumulative fraction of variation
attributed to the first $N$ PCs for the Reddy--Ram{\'{\i}}rez data (black solid
line), the superposition scenario sampled to match the Reddy--Ram{\'{\i}}rez
MDF (gray solid line), and the full superposition scenario sample (gray dashed
line).  The first two PCs of the superposition scenario explain almost 99\% of
the variation, while eleven PCs are required to reach the same threshold for
the data.  The much larger number of PCs required to explain the data could be
a consequence of more complex physics or more stochastic enrichment affecting
the data, or it could be a consequence of observational errors producing
variance, or both.  To fully explore the contribution of observational errors,
one should create synthetic spectra from the model distribution of stars, add
noise, and analyze them like one does the data.  This large undertaking is
beyond the scope of this paper. An alternative to fraction of variance is to
ask what fraction of stars are well fit in a $\chi^2$ sense by 1 PC, 2 PCs, ...
$N$ PCs, or to plot the distribution of $\chi^2$ values for stars after fitting
with 1 PC, 2 PCs, etc. These quantifications should be more robust to
observational noise, at least if the noise itself is accurately determined.

To summarize the PCAA data--model comparisons:
\begin{itemize}

\item The models and data both have the abundance of CCSN elements relative to
Fe-peak elements as the strongest PC.

\item The models with our fiducial CCSN yields (\citetalias{chieffi2004})
predict a strong signature of metallicity-dependent yields for Na and Al (also
for N and K but these elements have limited data).  This trend is weaker in the
data, particularly for Al. It would be very interesting to have N and K
observations as well.

\item The models predict that C behaves similarly to other $\alpha$-elements,
but the data shows additional variation in C essentially on its own as PC2.
This could result from enrichment physics, binary physics, or observational
errors.

\item The data have a correlated contribution of Fe-peak elements relative to Fe,
though it is not clear whether this is physically significant or an
observational effect.

\item The data require many more PCs to explain 99\% of the variance.
Observational noise certainly contributes, but more detailed study is needed
to decide whether it is the dominant factor or whether the larger number of
PCs reflects more complex enrichment physics.

\end{itemize}


\section{Conclusions}
\label{sec:conclusions}

The new generation of massive multi-element stellar abundance surveys, such as
APOGEE, GALAH, and Gaia-ESO, will dramatically increase our knowledge of the
enrichment history of the Milky Way. Deriving insights in the Milky Way's
formation history requires the ability to interpret elemental abundances with
galactic chemical evolution models. To this end, we have developed a flexible
chemical evolution model called \flexce.  In particular, it has been or will be
useful for:
\begin{itemize}

\item demonstrating the sensitivity of chemical evolution models to parameter
variations (Section \ref{sec:variations}),

\item showing the effect of CCSN yields on chemical evolution models (Section
\ref{sec:yields_multielement}),

\item modeling the ubiquitous high-$\alpha$ sequence as probed by APOGEE
\citep{nidever2014},

\item reproducing the two dimensional distribution in [O/Fe]--[Fe/H] by mixing
models with a range of inflow and outflow histories (Section
\ref{sec:scatter}),

\item providing a comparison data set to test the advanced statistical
techniques, such as PCAA, that will be crucial for fully exploiting the
multi-dimensional nature of current and upcoming surveys (Section
\ref{sec:pcaa}), and

\item post-processing cosmological simulations to predict element
  distributions for flexible assumptions about nucleosynthetic yields.

\end{itemize}

Our most important results appear in Figure~\ref{fig:ofe_mdf_multi}, which
presents [O/Fe]--[Fe/H] tracks for a variety of model assumptions, and
Figure~\ref{fig:xfe}, which compares observed [$X$/Fe]--[Fe/H] distributions for
20 different elements to predicted tracks for three different assumptions about
SN yields.  Figure~\ref{fig:ofe_mdf_multi} shows that for a specified set of
yields, the shape of [O/Fe]--[Fe/H] tracks is determined mainly by the outflow
mass loading factor $\eta$, which sets the equilibrium iron abundance reached at
late times, and by the star formation efficiency, which sets the location of the
knee in these tracks.  The inflow history, inflow metallicity, and delay time
distribution of SNIa play subdominant roles in governing tracks in
[O/Fe]--[Fe/H], but delaying enrichment by cycling SN ejecta through a warm ISM
phase can significantly shift the knee and alter the shape of the MDF. The
multi-element comparison in Figure~\ref{fig:xfe} shows several major
discrepancies with the data, especially for elements predicted to have strongly
metallicity-dependent yields. These discrepancies may reflect inaccurate
supernova physics, systematic errors in the abundance measurements, or both.
Homogeneous analyses of large data samples from APOGEE, Gaia-ESO, and GALAH will
allow more stringent comparisons applied over a wide range of Galactic
locations.

We have made \flexce\footnote{\flexceurl} a publicly available python
package to facilitate easy but realistic data--model comparisons.


\acknowledgments

We thank Jon Bird, Marc Pinsonneault, Todd Thompson, Andy Gould, Carles Badenes,
and Brian O'Shea for stimulating conversations, and the anonymous referee for a
thorough, constructive review and helpful suggestions. This work was supported
by NSF grant AST-1211853.

\appendix

\section{Mechanics of Model}
\label{sec:mechanics}
While many aspects of our formalism are similar to those of analytic chemical
evolution models, our numerical approach allows for additional flexibility,
including our treatment of stochastic star formation and SNIa explosions. In
this section, we describe the mechanics of the fiducial model in more detail. 

\subsection{Gas Reservoir}
We start by initializing the model with the initial gas mass with 
primordial mass
fractions. At each time step, we update the mass of each isotope, $i$, as
follows,
\begin{equation}
\begin{split}
    M_i(t_j) = M_i(t_{j-1}) + M_{i, \, \mathrm{inflow}}(t) - M_{i, \, \mathrm{SF}}(t) -  M_{i, \mathrm{outflow}}(t) + \\ CCSN_i(t) + AGB_i(t) + SNIa_i(t),
\label{eqn:gas_reservoir}
\end{split}
\end{equation}
where $M_{i, \, \mathrm{inflow}}$ is the inflowing mass, $M_{i, \, \mathrm{SF}}$
is the mass removed by star formation, $M_{i, \mathrm{outflow}}$ is the
outflowing mass, and $CCSN_i$, $AGB_i$, and $SNIa_i$ are the absolute yields
from each of these sources.  We describe each of these terms in more detail
below.

\subsection{Inflow}
The inflow mass of a given isotope is determined by multiplying the inflow
rate function ($\dot{M}_\mathrm{inflow}(t)$), the mass fraction of the isotope
in the inflowing gas ($f_i^\mathrm{inflow}$), and the length of the time step
$\Delta t$:
\begin{equation}
    M_{i, \, \mathrm{inflow}}(t) = \dot{M}_\mathrm{inflow}(t) \cdot f_i^\mathrm{inflow} \cdot \Delta t ~.
\end{equation}
In the fiducial model the inflow mass fractions are primordial.
The inflow rate can be an exponential (Equation \ref{eqn:exp_inflow}), a double
exponential (Equation \ref{eqn:double_exp_inflow}), a linear-exponential product
(Equation \ref{eqn:lin_exp_inflow}), or set to produce a constant SFR by
maintaining a constant total gas reservoir mass by exactly offsetting the other
terms in Equation \ref{eqn:gas_reservoir}.

\subsection{Star Formation}
One of the more unique features of our model is that it does not rely on
population averages but stochastically forms every individual star. In each time
step, the number of stars born in each mass bin (of width 0.1~\msun\ from
0.1--8.0~\msun\ and 1.0~\msun\ from 8.0--100~\msun) is determined by the
following steps.
\begin{enumerate}
    \item Calculate the expectation value of the SFR ($SFR^\mathrm{EV}$) from the Kennicutt--Schmidt law,
    \begin{equation}
        SFR^\mathrm{EV}(t) = \frac{1}{t_\mathrm{gas}} Area \left( \frac{M_{\mathrm{gas}}}{Area}\right)^N,
    \label{eqn:sf_general}
    \end{equation}
    where $N$ is the Kennicutt--Schmidt law index. In the fiducial model, we adopt $N=1$, so the $Area$ terms cancel and Equation \ref{eqn:sf_general} reduces to Equation \ref{eqn:kslaw}.
    \item Multiply $SFR^\mathrm{EV}$ by the length of the time step to get the expectation value of the mass of newly forming stars ($SF^\mathrm{EV}$),
    \begin{equation}
        SF^\mathrm{EV}(t) = SFR^\mathrm{EV}(t) \cdot \Delta t.
    \end{equation}
    \item Distribute $SF^\mathrm{EV}$ into mass bins according to the IMF, so the expectation value of the mass of newly formed stars in bin $m_j$ ($SF_ {m_j}^\mathrm{EV}$) is
    \begin{equation}
        SF_{m_j}^\mathrm{EV}(t) = SF^\mathrm{EV}(t) \left(a \int_{m_\mathrm{bin \, low}}^{m_\mathrm{bin \, up}} m^{-\alpha+1} \mathrm{d}m \right),
    \end{equation}
    where $m_\mathrm{bin \, low}$ and $m_\mathrm{bin \, up}$ are the lower and upper mass limits of mass bin $m_j$, $a$ is the appropriate normalization constant for the chosen IMF, and $\alpha$ is the IMF power law slope over the
	mass bin.
    \item Convert $SF_{m_j}^\mathrm{EV}$ into the expected number of stars in mass bin $m_j$ ($N_{\mathrm{SF}, {m_j}}^\mathrm{EV}$) by dividing by the average mass of the bin:
    \begin{equation}
        N_{\mathrm{SF}, {m_j}}^\mathrm{EV}(t) = SF_{m_j}^\mathrm{EV}(t) \frac{\int_{m_\mathrm{bin \, low}}^{m_\mathrm{bin \, up}} m^{-\alpha} \mathrm{d}m}{ \int_{m_\mathrm{bin \, low}}^{m_\mathrm{bin \, up}} m^{-\alpha+1} \mathrm{d}m}.
    \end{equation}
    \item Stochastically sample the expected number of stars in mass bin $m_j$ according to a Poisson distribution to get the actual number of stars formed in each mass bin ($N_{\mathrm{SF}, {m_j}}$):
    \begin{equation}
        N_{\mathrm{SF}, {m_j}}(t) = Poisson(N_{\mathrm{SF}, {m_j}}^\mathrm{EV}(t)).
    \end{equation}
    For the fiducial model, the effect of random sampling of the IMF is small because of its large mass, so the maximum difference between the expected value of the SFR and the actual SFR was only 0.06\%.
    \item Compute the actual total mass in newly formed stars by multiplying the number of stars formed in each mass bin by the average mass in the bin and then summing over all mass bins:
    \begin{equation}
        SF(t) = \sum SF_{m_j}(t) = \sum N_{\mathrm{SF}, {m_j}}(t) \frac{ \int_{m_\mathrm{bin \, low}}^{m_\mathrm{bin \, up}} m^{-\alpha+1} \mathrm{d}m}{\int_{m_\mathrm{bin \, low}}^{m_\mathrm{bin \, up}} m^{-\alpha} \mathrm{d}m}.
    \end{equation}
    \item Finally, multiply the total mass in newly formed stars by the mass fraction of isotope $i$ in the ISM at time $t$ ($f_i(t)$) to get the mass of isotope $i$ that goes into newly formed stars,
    \begin{equation}
        SF_i(t) = SF(t) \cdot f_i(t).
    \end{equation}
\end{enumerate}

\subsection{Outflow}
In the fiducial model, the outflow rate is a constant factor times the SFR. The composition of the outflowing gas is the same as the ISM composition, so the mass of isotope $i$ ejected from the galaxy in outflows is simply:
\begin{equation}
    outflow_i(t) = \eta \cdot SF_i(t),
\end{equation}
where $\eta$ is the outflow mass-loading parameter.

\subsection{CCSN and AGB Stars}
For each time step, we determine the mass range of stars from each previous time
step that will evolve in the current time step by inverting the stellar lifetime
function. If only some of the stars in a mass bin will evolve because the age
range of the mass bin straddles the edge of the time step, then the model calculates
the fraction of stars in that mass bin that will evolve. This step is
particularly helpful in smoothing out the gas return from very long-lived stars.
We then look up the appropriate CCSN or AGB star yields for each mass bin and
the metallicity of the ISM at birth to compute the yields. We also add the newly
formed remnant mass to the existing remnant population.

\subsection{SNIa}
To calculate the mass return from SNIa, we use the SNIa DTD to compute the
expectation value of the number of SNIa that will explode in a given time step
and then stochastically sample from a Poisson distribution to determine the
actual number of SNIa. The calculation of the expected number of SNIa varies
amongst the four different DTDs that we considered (exponential, power law,
population synthesis, and prompt + delayed), so we will describe the unique
aspects of the procedures for each DTD separately.

\subsubsection{Exponential SNIa DTD}
For the exponential DTD, we track a reservoir of WDs that will explode as SNIa
with mass $M_\mathrm{WD}^\mathrm{SNIa}$ using the following steps:
\begin{enumerate}
    \item calculate the mass of WDs that will be formed from stars of the stellar population of the current time step with initial masses between 3.2--8.0 \msun\ ($M_{\mathrm{WD},t}^{3.2-8.0}$),
    \item multiply $M_\mathrm{WD,t}^{3.2-8.0}$ by the fraction that will explode as SNIa to get the mass of WDs that will explode as SNIa ($M_{\mathrm{WD}, t}^\mathrm{SNIa}$), then
    \item add $M_{\mathrm{WD},t}^\mathrm{SNIa}$ to the reservoir of WDs that will explode as SNIa (i.e., $M_\mathrm{WD}^\mathrm{SNIa}$) after the minimum SNIa delay time.
\end{enumerate}
The model then calculates the expectation value of the number of SNIa to explode
in the current time step:
\begin{enumerate}
    \item compute the fraction of the WD reservoir to explode ($f_\mathrm{WD}^\mathrm{SNIa}$),
    \begin{equation}
        f_\mathrm{WD}^\mathrm{SNIa} = \Delta t \cdot \tau_\mathrm{SNIa},
    \end{equation}
    \item multiply the mass of the reservoir of WDs that will explode as SNIa by the fraction to explode and divide by ejected mass of each SNIa ($m_\mathrm{SNIa}^\mathrm{ej}$) to get the expectation value of the number of SNIa ($N_\mathrm{SNIa}^\mathrm{EV}$),
    \begin{equation}
        N_\mathrm{SNIa}^\mathrm{EV} = \frac{M_\mathrm{WD}^\mathrm{SNIa} \cdot f_\mathrm{WD}^\mathrm{SNIa}}{m_\mathrm{SNIa}^\mathrm{ej}}.
    \end{equation}
\end{enumerate}

\subsubsection{Power Law and Population Synthesis SNIa DTDs}

For the power law and population synthesis DTDs, we first compute the SNIa rate
per solar mass of star formation, given by Equation \ref{eqn:powerlaw_dtd} for
the power law DTD and described in detail in Section 3.1 of \citet{greggio2005}
for the population synthesis DTD.  Next, we normalize the SNIa rate to match
observational constraints of the total number of SNIa from 40 Myr to 10 Gyr of
an individual stellar population (see Section \ref{sec:snia_dtd} for details).
For previous stellar generations, we calculate the product of the SNIa rate and
the stellar mass formed in that time step ($SF$), which gives us the expected
number of SNIa ($N_\mathrm{SNIa}^\mathrm{EV}(t_1)$) from stars born in that time
step:
\begin{equation}
    N_\mathrm{SNIa}^\mathrm{EV}(t_1) = R_\mathrm{SNIa}(t_1) \cdot SF(t_1).
\end{equation}
Then, we sum over the contributions of the previous time steps older than the
minimum delay time to get the total expected number of SNIa
($N_\mathrm{SNIa}^\mathrm{EV}$):
\begin{equation}
    N_\mathrm{SNIa}^\mathrm{EV} = \sum_{t_1=0}^{t_1=t-t_\mathrm{min}} N_\mathrm{SNIa}^\mathrm{EV}.
\end{equation}

\subsubsection{Prompt + Delayed DTD}
The expected number of SNIa for the prompt + delayed DTD is simply the SNIa rate
as expressed in Equation \ref{eqn:prompt_delayed_dtd} multiplied by the length
of the time step (for all time steps after the minimum delay time):
\begin{equation}
    N_\mathrm{SNIa}^\mathrm{EV}(t) = R_\mathrm{SNIa}(t) \cdot \Delta t.
\end{equation}

\subsubsection{Stochastic Realization of SNIa}
Once we have the expectation value of the number of SNIa to explode, we compute the actual number of SNIa by drawing from a Poisson distribution:
\begin{equation}
    N_\mathrm{SNIa}(t) = Poisson(N_\mathrm{SNIa}^\mathrm{EV}(t)).
\end{equation}
The mass of isotope $i$ returned to the ISM from SNIa is simply the number of
SNIa multiplied by the absolute SNIa yield of isotope $i$ ($Y_\mathrm{SNIa, i}$)
in solar masses,
\begin{equation}
    SNIa_i(t) = N_\mathrm{SNIa}(t) * Y_{\mathrm{SNIa}, i}.
\end{equation}
Finally, we subtract the total mass returned from SNIa from the total mass in
remnants (for all DTDs) and from reservoir of WDs that will explode as SNIa (i.e.,
$M_\mathrm{WD}^\mathrm{SNIa}$) (for the exponential DTD only).

\section{CCSN and AGB Star Yield Calculation}
\label{sec:yield_grid_extension}

Stellar yields are one of the largest sources of uncertainty in chemical
evolution models \citep{timmes1995, gibson1997, portinari1998, romano2010}. In
addition to the uncertainties in the nucleosynthesis calculations themselves,
the yield grids typically do not span the full range in stellar mass and
metallicity required by chemical evolution models. Consequently, chemical
evolution models must include assumptions for the treatment of the regions that
are not covered by the nucleosynthesis calculations. These regions tend to be
zero and very low metallicities, super-solar metallicities, stellar masses near
the AGB star--CCSN boundary (8 \msun), and high stellar masses ($M
\gtrsim$40~\msun).

For CCSN, we adopt the \citetalias{limongi2006} net yields for 11--120 \msun\ at
solar metallicity ($Z$~=~2$\times10^{-2}$) and the \citetalias{chieffi2004} net
yields for 13--35 \msun\ at sub-solar metallicities down to $Z$~=~10$^{-6}$, so
we need to assign yields for the following regions of stellar mass--metallicity
space:
\begin{itemize}
\item 8--13 \msun\ and 35--100 \msun\ at sub-solar metallicities (down to $Z$~=~10$^{-6}$),
\item 8--11 \msun\ stars at solar metallicity, and
\item 8--100 \msun\ stars at metallicities below $Z$~=~10$^{-6}$ and above solar.
\end{itemize}

For AGB stars, we use the \citet{karakas2010} yields, which cover 1.0--6.0
\msun\ from $Z$~=~10$^{-4}$--8$\times10^{-3}$ and 1.0--6.5 \msun\ at solar
metallicity ($Z$~=~2$\times10^{-2}$).  Again, we are faced with the problem of
assigning yields to stars on both ends of the mass and metallicities ranges of
the yield grid, specifically:
\begin{itemize}
    \item 0.1--1.0 \msun\ and 6.0--8.0 \msun\ for metallicities between $Z$~=~10$^{-4}$--8$\times10^{-3}$,
    \item 0.1--1.0 \msun\ and 6.5--8.0 \msun\ for solar metallicity ($Z$~=~2$\times10^{-2}$), and
    \item 0.1--8.0 \msun\ for metallicities below $Z$~=~10$^{-4}$ or above solar.
\end{itemize}

From a modeling perspective, there are several reasonable, though unsatisfying,
assumptions that could be made to assign yields to stars of all masses and
metallicities:
\begin{enumerate}
    \item ignore yields from stars produced outside of the yield grid,
    \item use the yields from the nearest grid point, or
    \item extrapolate the yields to higher and/or lower masses and/or metallicities.
\end{enumerate}
Many existing GCE models have ignored yields from stars outside of the
calculated yield grid. However, given the significant gaps in the yield models
between 6--13 \msun\ and 35--100 \msun, this strategy misses a significant
amount of enrichment.

Our strategy for extending the yield grids consists of three steps:
\begin{enumerate}
    \item Create a uniform but coarsely sampled grid in mass and metallicity by extending the grid to the full mass range but only at the metallicities of the yield models. For CCSN, we use the yields of the closest mass grid point at fixed metallicity, and we conserve total mass by extrapolating the mass ejected and the remnant mass.  For AGB stars, we extrapolated the yields to higher and lower masses at fixed metallicity.
    \item Linearly interpolate the yields at fixed metallicity onto a finely sampled grid in mass at the metallicities of the yield calculations.
    \item Linearly interpolate the yields at fixed mass onto a finely sampled grid of log metallicity.
\end{enumerate}

The transition between the \citetalias{chieffi2004} and the
\citetalias{limongi2006} yields for stars with mass between 11--13~\msun\ and
35--100~\msun\ is smoothed out by extending the \citetalias{chieffi2004} grid
down to 11~\msun\ and up to 100~\msun\ (Step 2) and then linearly interpolating
between those yields and the \citetalias{limongi2006} yields (Step 3). For
metallicities outside of the yield models' ranges, below $Z$~=~10$^{-6}$ for
CCSN or $Z$~=~10$^{-4}$ for AGB stars and above solar metallicity for both
enrichment sources, we adopt the yields of the closest metallicity grid point at
fixed mass. We linearly interpolate the yields for all elements, even those
known to have a metallicity dependence, due to the difficulty in constraining
higher order interpolation schemes.

We do not extrapolate the yields for CCSN to higher and lower masses because it
leads to unphysical results, especially at the high mass end due to the long
mass baseline.  However, stars above 35~\msun\ can contribute a lot of enriched
material, so ignoring the return from these stars will underestimate the metal
return to the ISM.  For example, the 120~\msun\ model of the
\citetalias{limongi2006} solar metallicity yields produces roughly 2--30 times
the yields of the 35~\msun\ model for all elements. As a compromise, for
sub-solar metallicity CCSN with $M>$~35~\msun, we returned the net yields
appropriate for a 35~\msun\ CCSN at the same metallicity (and extrapolate the
total mass ejected and remnant mass). The yield grid for AGB stars, on the other
hand, covers most of the needed range, so we chose to linearly extrapolate the
yields to cover the mass range up to the AGB star--CCSN boundary and down to the
lower mass limit of the IMF.

\begin{figure*}
\centerline{\includegraphics[width=19cm]
{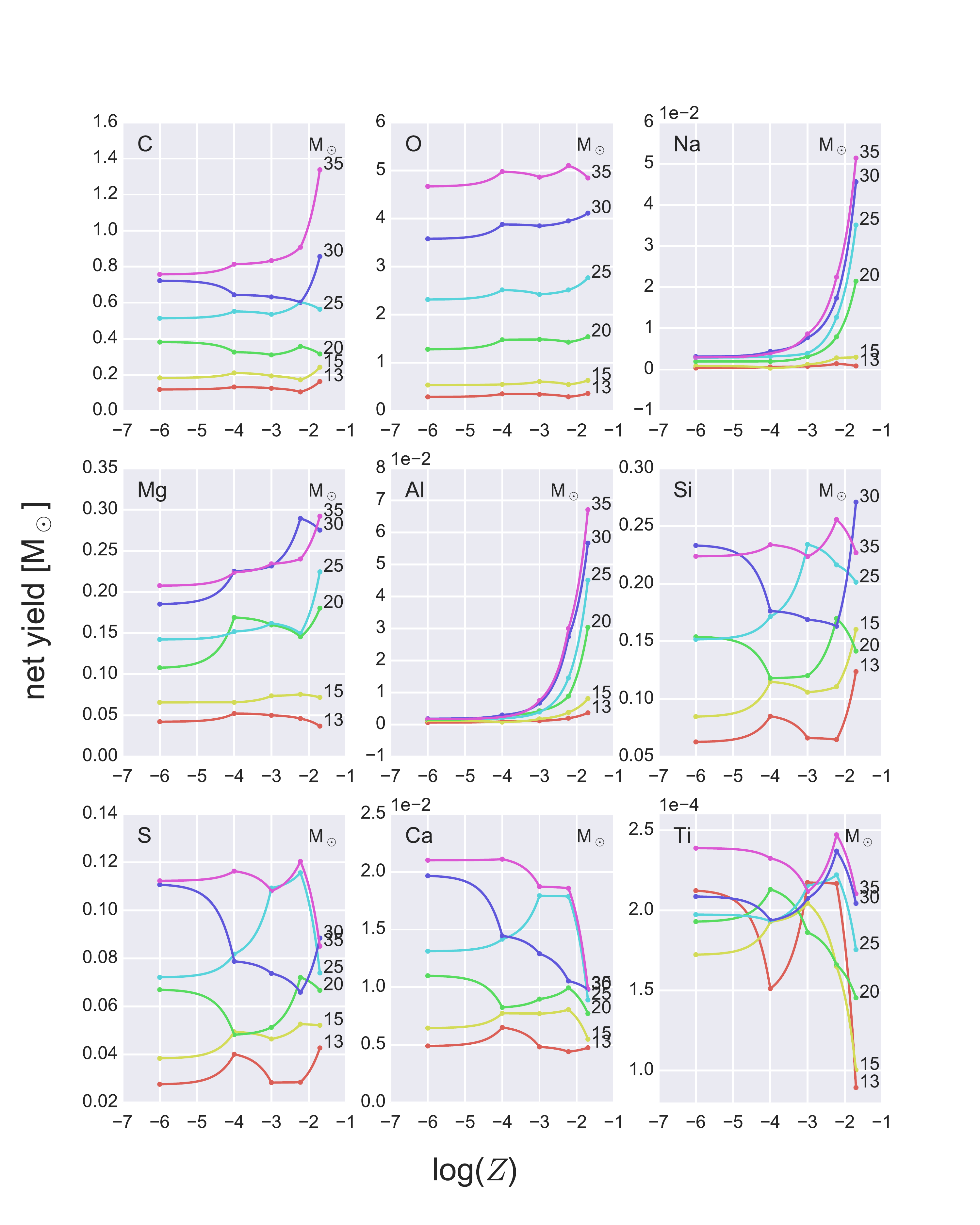}}

\caption{Net yields for the \citetalias{chieffi2004} and
\citetalias{limongi2006} CCSN yield set as a function of metallicity for various
elements.  The points show the yields of the masses and metallicities for the
models calculated by \citetalias{chieffi2004} and \citetalias{limongi2006}
between 13--35~\msun.  The lines show the linear interpolation in log
metallicity at fixed mass between the calculated CCSN models.}

\label{fig:cl04_yields_metallicity_mass}
\end{figure*}

\begin{figure}
\centerline{\includegraphics[width=9cm]
{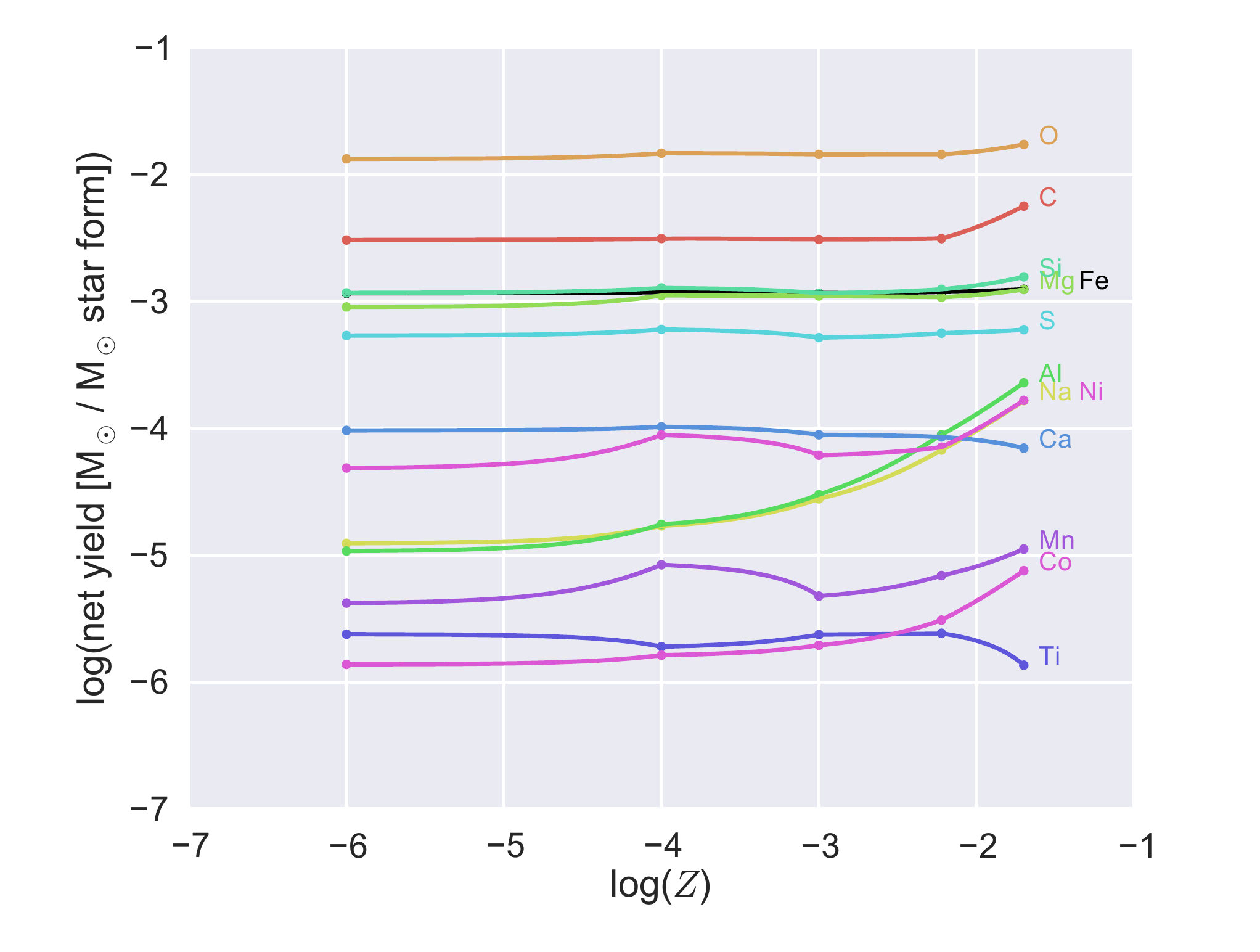}}

\caption{IMF integrated net yields for the \citetalias{chieffi2004} and
\citetalias{limongi2006} CCSN yield set as a function of metallicity.  The
points show the metallicities calculated by \citetalias{chieffi2004} and
\citetalias{limongi2006}, and the lines show the linearly interpolated yields.}

\label{fig:cl04_imf_int}
\end{figure}

Figures \ref{fig:cl04_yields_metallicity_mass} and \ref{fig:cl04_imf_int} show
how the \citetalias{chieffi2004} and \citetalias{limongi2006} CCSN net yields of
selected elements vary with metallicity. Figure
\ref{fig:cl04_yields_metallicity_mass} breaks down the yields by stellar mass
(for the masses calculated by \citetalias{chieffi2004}), and Figure
\ref{fig:cl04_imf_int} shows the CCSN yields from a stellar population by
convolving the yields with the Kroupa IMF. In Figure
\ref{fig:cl04_yields_metallicity_mass}, the points correspond to the masses and
metallicities of the models calculated by \citetalias{chieffi2004} and
\citetalias{limongi2006} (between 13--35~\msun), and the lines correspond to the
yields interpolated in log metallicity at fixed mass. In Figure
\ref{fig:cl04_imf_int}, the points indicate the metallicities of the models
calculated by \citetalias{chieffi2004} and \citetalias{limongi2006}, and the
lines indicate the (interpolated) IMF integrated yields.


\end{document}